\documentclass[12pt,tittlepage,a4paper]{report}
\usepackage{amsmath,amssymb}
\usepackage[latin1]{inputenc}
\usepackage{graphicx}
\renewcommand{\baselinestretch}{1.5}
\usepackage{anysize}
\usepackage[portuguese]{babel}
\usepackage{theorem}
\usepackage{latexsym}
\usepackage{amsfonts}
\usepackage{eufrak}
\usepackage{syntonly}
\usepackage[mathcal]{euscript}
\usepackage{pb-diagram}
\usepackage{epsf}

\def\a{\alpha}
\def\ab{\overline{\alpha}}
\def\b{\beta}
\def\bb{\overline\b}
\def\g{\gamma}
\def\gb{\overline\g}
\def\d{\delta}
\def\db{\overline\d}
\def\ep{\epsilon}
\def\eb{\overline\ep}
\def\r{\rho}
\def\rb{\overline\r}
\def\s{\sigma}
\def\sb{\overline\s}
\def\l{\lambda}
\def\lt{\widetilde\l}
\def\lh{\widehat\lambda}
\def\lth{\widehat{\lt}}
\def\om{\omega}
\def\ot{\widetilde\om}
\def\oh{\widehat\om}
\def\oth{\widehat{\ot}}
\def\O{\Omega}
\def\Ob{\overline\O}
\def\Ot{\widetilde\O}
\def\dt{\widetilde d}
\def\dth{\widehat{\dt}}
\def\t{\theta}
\def\tt{\widetilde\t}

\def\G{\Gamma}
\def\F{\Phi}
\def\L{\Lambda}
\def\Lt{\widetilde\L}
\def\D{\Delta}
\def\N{\nabla}
\def\Nb{\overline\nabla}
\def\Nt{\widetilde\N}
\def\Pb{\overline\Pi}
\def\p{\partial}
\def\pb{\overline\partial}
\def\dh{\widehat d}
\def\Jb{\overline J}
\def\Ct{\widetilde C}
\def\Qt{\widetilde Q}
\def\St{\widetilde S}
\def\jt{\widetilde j}
\def\Rt{\widetilde R}
\def\Tt{\widetilde T}
\def\Cb{\overline C}
\def\Db{\overline D}
\def\Hb{\overline H}
\def\Ib{\overline I}
\def\Jt{\widetilde J}
\def\Nn{\widetilde N}
\def\beq{\begin{equation}}
\def\eeq{\end{equation}}
\def\ber{\begin{eqnarray}}
\def\eer{\end{eqnarray}}
\def\bdm{\begin{displaymath}}
\def\edm{\end{displaymath}}

\marginsize{4.3cm}{0cm}{1cm}{3cm}

\usepackage{titlesec}

\titleformat{\chapter}[display] 
{\bfseries\Large} 
{
 \Large\chaptertitlename\ 
 \Large\thechapter} 
{8mm} 
{} 
[] 

\titleformat{\section}
{\large\bfseries}
{\thesection}{1em}{}
\titlespacing{\section}
{0pc}{3.2ex plus 0ex minus 0ex}{3.2ex minus .1ex}

\usepackage[latin1]{inputenc}
\usepackage[portuguese]{babel}

\def\bib{\bibitem}
\def\be{\begin{equation}}
\def\ee{\end{equation}}
\def\ba{\begin{eqnarray}}
\def\ea{\end{eqnarray}}
\def\'#1{\if#1i{\accent 19\i}\else{\accent 19 #1}\fi}

\def\ea{\'e }

\renewcommand{\baselinestretch}{1.2}
\begin{document}
\renewcommand{\contentsname}{Index}
\pagenumbering{arabic}
\renewcommand{\chaptername}{Chapter}
\renewcommand{\appendixname}{Ap\^endice}
\renewcommand{\bibname}{References}
\renewcommand{\figurename}{Figura}
\pagestyle{plain}
\def\baselinestretch{1.2}
\hoffset=-1.0 true cm
\voffset=-2 true cm
\topmargin=1.0cm
\thispagestyle{empty}
\def\thefootnote{\fnsymbol{footnote}}

\thicklines
\begin{picture}(370,60)(0,0)
\setlength{\unitlength}{1pt}
\put(40,53){\line(2,3){15}}
\put(40,53){\line(5,6){19}}
\put(40,53){\line(1,1){27}}
\put(40,53){\line(6,5){33}}
\put(40,53){\line(3,2){25}}
\put(40,53){\line(2,1){19}}
\put(40,53){\line(5,-6){17}}
\put(40,53){\line(1,-1){22}}
\put(40,53){\line(6,-5){30}}
\put(40,53){\line(3,-2){22}}
\put(40,53){\line(-2,1){15}}
\put(40,53){\line(-3,1){23}}
\put(40,53){\line(-4,1){26}}
\put(40,53){\line(-6,1){36}}
\put(40,53){\line(-1,0){40}}
\put(40,53){\line(-6,-1){32}}
\put(40,53){\line(-3,-1){20}}
\put(40,53){\line(-2,-1){10}}
\put(75,45){\Huge \bf IFT}
\put(180,56){\small \bf Instituto de F\'\i sica Te\'orica}
\put(165,42){\small \bf Universidade Estadual Paulista} 
\put(-25,2){\line(1,0){433}}
\put(-25,-2){\line(1,0){433}}
\end{picture}  


\vskip .3cm
\noindent
\hfill    IFT--D.004/2008\\

{TESE DE DOUTORAMENTO}

\vspace{3cm}
\begin{center}
{\large \bf }
Superstring Sigma Model Computations Using the Pure Spinor Formalism

\vspace{1.2cm}
Oscar Andr\'es Bedoya Delgado
\end{center}

\vskip 3cm
\hfill Advisor
\vskip 0.4cm
\hfill {\em }
Nathan Jacob Berkovits  
\vskip 0.4cm
\vskip 2.2cm
\vfill
\begin{center}
June 2008
\end{center}

\newpage

\pagenumbering{roman}

\begin{center}
{\Large \bf Aknowledgments}
\end{center}
\vskip 2.0cm
During the time of a Ph.D and my stay in S\~ao Paulo I met 
many people who I'd like to thank because they played and still do play
an important role in different aspects. This space is short, words are few and the memory
betrays.\\

Firstable, I thank CAPES and Funda\c c\~ao Instituto de F\'isica Teorica for 
unvaluable financial support.\\

Gracias a los {\it meninos}: Cristhian, Humberto, Alexis, Mamo y Juan;
que hicieron de la experiencia de compartir casa
algo extremadamente divertido y descomplicado. Gracias por incontables
momentos de distracci\'on mientras cocinabamos o en el caf\'e de la
tarde y por supuesto, por varias discusiones de F\'isica. Finalmente gracias por
la amistad y cuidarnos los unos a los otros. Agradezco a Hector por su acogida al llegar a S\~ao Paulo y su gran
apoyo desde que nos hicimos amigos. Tambi\'en agradezco a Boris y Mario quienes 
estuvieron a mi lado en la etapa inicial del doctorado y muy especialmente a
Aldo F. Rebolledo por su amistad y excelentes consejos. \\
Agrade\c co a meu caro amigo e {\it irm\~ao cient\'ifico} Carlos Mafra, por 
muitas discuss\~oes sobre teoria de cordas e varios outros assuntos, por
outras tantas ajudas com Linux e por ter sua porta e 
cora\c c\~ao aberto sempre que precisei. Tamb\'em agrade\c co a Pilar e aos
dois lhes desejo tudo do melhor.\\

Agrade\c co \`as pessoas com quem dividi sala, em especial ao Carlos Senise 
pelas discuss\~oes sobre supersimetria.\\
Gostaria tamb\'em de agradecer ao Osvaldo Chandia pela colabora\c c\~ao, e
ao Daniel e minha {\it irm\~a
cient\'ifica} Dafni por discuss\~oes. Para todas as pessoas que fizeram e
fazem parte do journal club em teoria de cordas, muito obrigado pelo
ensinado e pelas perguntas que me fizeram. \\

Agrade\c co de maneira muito especial ao Professor Nathan, por ter me
orientado, dado v\'arias
oportunidades e sobretudo, por compartilhar durante tudo este tempo sua maneira de pensar em F\'isica. \\

Agrade\c co a Dona Carmen e Seu Ivo, por serem t\~ao atenciosos comigo
e me fazer sentir acolhido.\\

Agrade\c co a {\it Mariana}, quem numa fase definitiva se tornou uma pessoa 
essencial. Me fez sentir renovado, me d\'a mais motivos
para me esfor\c car me enchendo de sonhos e me faz sentir mais vivo.\\

Finalmente agradezco a mis padres Oscar y Luzby, por haberse esforzado tanto
para ayudarme a llegar a este punto. Por toda la educaci\'on 
y ejemplos recibidos. Por su apoyo incondicional, comprensi\'on
y tambi\'en por el sacrificio de tener que estar separados
durante tantos a\~nos. A ellos dedico este humilde trabajo que a sus ojos es como un tesoro.


\newpage

\begin{center}
{\Large \bf Resumo}
\end{center}
\vskip 2.0cm

Nesta tese s\~ao apresentadas duas aplica\c c\~oes do 
modelo sigma para a supercorda usando o formalismo de 
espinores puros. A primera aplica\c c\~ao \'e o c\'alculo
da invari\^ancia conforme a um-loop para a supercorda tipo II, 
resultando em equa\c c\~oes de movimento no super-espa\c co 
para campos de fundo acoplados com a supercorda. A segunda aplica\c c\~ao 
est\'a relacionada com a invari\^ancia BRST da supercorda 
heter\'otica no n\'ivel qu\^antico, que permite encontrar 
corre\c c\~oes oriundas da teoria de supercordas para os v\'inculos
de super Yang-Mills/supergravidade em dez dimens\~oes.


\vskip 1.0cm
\noindent
{\bf Palavras Chaves}: Supersimetria; Supercordas; Modelo Sigma;
Corre\c c\~es de Chern-Simons.
\vskip 0.5cm
\noindent
{\bf \'Areas do conhecimento}: F\'isica de Part\'iculas e Campos.

\newpage

\begin{center}
{\Large \bf Abstract}
\end{center}
\vskip 2.0cm

In this thesis are presented two aplications of the sigma model for the
superstring in the pure spinor formulation. The first aplication
concerns the computation of the one-loop conformal invariance for the
type II superstring, resulting in equations of motion written in
superspace for the background fields coupled to the superstring. 
The second application is related to the BRST invariance of the
heterotic superstring at the quantum level, which allows to find
stringy corrections to the ten-dimensional super
Yang-Mills/supergravity constraints.


\vskip 0.5cm
\noindent
{\bf Areas of Knowledge}: Fields and Particles Physics.

\vfill \eject

\tableofcontents
\pagenumbering{arabic}
\chapter{Introduction}
The description of Physics in terms of fields dates back to the 19th century
and had as origin the study of the electric and magnetic phenomena. Since
then, the field language has seem appropriate to describe electromagnetism,
gravitation, and the remaining two type of interaction discovered in the
20th century; namely the weak and strong interactions. The Standard Model of
particle physics, which describes all but gravitational phenomena, is a
beautiful example of a {\it unified} description for various fundamental
interactions in terms of quantum fields. Nevertheless the Standard Model can be 
thought of as a built
theory, which can be adjusted if some minor changes are required by the
experiments. Furthermore, there are ingredients, in the philosophy of
constructing, that are put by hand instead of deduced from more fundamental
principles, for example, the way various particles are acomodated in the
standard model multiplets. In some way, the ability to
adjust such a theory also leaves the unsatisfactory taste of not having the
right core from where to extract it in a {\it unique} manner.\\

Although the gravitational field has a well established classical field
description, its quantum description has been elusive for quite long, as
well as its incorporation, together with the other three interactions, in a
single framework. Perhaps this first fact is an indication that the right
type of description has not been used. \\

An important step in the direction of a quantum theory of gravity has been
provided by precisely changing the type of description used in particle
physics, namely Quantum Field Theory.
String theory, which was {\it accidentally} discovered by studying an
apparently singular behavior of the mass and the spin of some heavy particles in the
late sixties; is a different proposal for describing particle physics. A string is a
one-dimensional object, which expand a two-dimensional surface as it
evolves in time, called the {\it worldsheet}. In its simplest
version, namely the bosonic string, the spectrum of particles is obtained by
quantizing the modes of vibration of closed and open strings. In the first
case, the massless sector of the spectrum contains a particle of spin two and
mass zero, which is the graviton. In the second case, the massless particle
of the spectrum, which has spin one is the photon. In this simplified model
one can already handle with gravity and electromagnetism using a single
framework. Actually this is not the first time that a single framework
contains gravity and a gauge field. This is the case of the Kaluza-Klein
theories, which appear as a compactification of a five-dimensional gravity
theory to four dimensions. In string theory the appearance of extra
dimensions is ``natural'', as explained below. In that
sense, string theory also has room for Kaluza-Klein theories.\\

String theory is a huge subject of study and it is not the aim of this
thesis to continue discussing their generalities. So in the following a description more
focused in the topic of interest will be given.

\section{Strings in a Generic Background}
It is known since the early  eighties that the coupling of strings to
a generic backgrounds puts restriction on it, namely, puts the background
on-shell. This equations of motion for the background can be computed
perturbatively by considering the quantum regime of the worldsheet symmetries.
In this section it will be discussed the bosonic string and superstring
in a generic background.

\subsection{Bosonic String Sigma Model}
In the simplest case, a bosonic string propagates in a Minkowski space-time. 
In such a case, the theory possess conformal symmetry at the worldsheet level. 
However, of primordial interest
in this thesis is to consider the case when the strings propagates in a 
curved space-time, which is described by coupling the bosonic string
to a generic space-time metric. Such a coupling is described by a non-linear sigma
model action 

\be S = {1\over {4\pi \a'}} \int d^2 \sigma \sqrt{g}g^{ab} \p_a X^m
\p_b X^n G_{mn}(X), \label{NLSM}\ee
where $X^m$ describe the coordinates of the string in $D$ dimensional 
space-time, $g_{ab}$
is a metric for the worldsheet, $\a'$ is proportional to the inverse of the
string tension and $G_{mn}$ is the space-time metric. This action is a 
direct generalization of the Polyakov action \cite{Polyakov} when the 
Minkowski $\eta_{mn} = diag(-1 , 1,{\ldots} 1)$ metric is 
replaced by the Riemannian metric. The interest of studying this type of
action is related with the information one can extract out of it. As well as
the preservation of the conformal symmetry at the quantum level indicates
that space-time should be $26$ dimensional in the Minkowski space case, in the
curved space case the preservation of this symmetry at the 
quantum level makes the space-time metric to satisfy the Einstein equations
\cite{Friedan80} \cite{GFM},
as will reviewed in detail. This is a way to obtain equations of motion for
space-time fields, which could help to know the structure of the string
effective action. Furthermore, perturbative methods can be used to compute
stringy corrections to space-time equations of motion, giving also hints
of string corrected effective actions: this requirement 
of conformal invariance can be computed perturbatively in the
string parameter $\a'$, so the Einstein equations can receive stringy
corrections. \\The space-time metric is associated to one of the massless
bosonic string states, namely the graviton. There are two more states at the massless level which
can be associated to an antisymmetric tensor, denoted by $B_{mn}$ and a scalar
field $\Phi$ known as dilaton.
In this way a generalized sigma model can be constructed, whose action in
the conformal gauge is given by 
\be S = {1\over {4\pi \a'}} \int d^2 \sigma  (\sqrt{g} g^{ab}\p_a X^m \p_b
X^n G_{mn} + \epsilon^{ab} \p_a X^m \p_b X^n B_{mn}) + 
{1\over{2\pi }} \int d^2 \sigma \sqrt{g} r^{(2)} \Phi (X), \label{GNLSM}\ee
where $\epsilon^{ab}$ is the purely antisymmetric tensor in two
dimensions and $r^{(2)}$ is the scalar curvature in two dimensions. The requirement 
of conformal invariance at the quantum level also puts the background field
on-shell. These equations of motion can be found as requirements for scale
invariance, i.e. computing the beta function for the generalized non-linear
sigma model (\ref{GNLSM}). 

The condition for conformal invariance can be
written as conditions for the stress-energy tensor being traceless: 
\be \langle T_a\,^a\rangle = \beta^G _{mn}g^{ab}\partial_a X^m
\partial_bX^n +\beta^B_{mn}\epsilon^{ab}\partial_aX^m \partial_b X^n + \beta^\Phi R^{(2)}, \label{trace}\ee
where
\ber \beta^G_{mn} &=& R_{mn}-\frac{1}{4} H_{mlr}H_n\,^{lr}+2\nabla_m\nabla_n \Phi, \label{betaG}\\ 
\beta^B_{mn} &=&-\frac{1}{2}\nabla^r H_{mnr} +\nabla_l\Phi H^l\,_{mn}, \label{betaB}\\
\beta^\Phi &=& -\frac{D-26}{12} + \alpha^\prime \left(R - \frac{H^2}{12}
+4\nabla^2 \Phi - 4 (\nabla\Phi)^2  \right),\label{betaPhi}\eer
and $H_{mnp}$ are the components of the three form $H = dB$. So, the theory is
conformal invariant if the beta-functions are zero. 
In the following it will be explained the procedure to compute this
$\beta$-functions, taking \cite{OABedoyaMaster} as reference.

\subsubsection{Covariant Background Field Expansion}

By making a perturbative expansion it will be found a diagramatic
expression for the terms in such expansion. With this goal, it will be
introduced the partition function 

\be Z_J \equiv e^{-W[J]}= \int [dX] exp\left(-( S[X]+X \cdot J  )\right), \label{funcional generador W} \ee 
which defines the functional generator $W[J]$ of connected diagrams, where 

\be X \cdot J = \int d^2 \sigma X^m J_m. \ee 
Variating respect of $J$, one defines the mean field

\be  X^m_0 \equiv \frac{\delta W}{\delta J_m} = \frac{1}{Z_J}\int [dX]X^m e^{S_J}, \ee
with 

\be S_J \equiv S[X]+X \cdot J.  \ee 
This mean field $ X^m_0$ will play the role of a background field; it will
be the field around which the perturbative expansion will be made. The
effective action is defined by 

\be \Gamma  \equiv W -  X_0 \cdot J. \ee 
From this equation, the current can be written as 

\be J_m = - \frac{\delta \Gamma}{\delta X_0^m}, \label{corrente}\ee
so, the effective action takes the form 

\be \Gamma = W + X_0 \cdot \frac{\delta \Gamma}{\delta X_0}, \label{acao efetiva}\ee
what allows to write $exp (-\Gamma)$ using (\ref{funcional generador W}),
(\ref{corrente}) and (\ref{acao efetiva}):

\be e^{-\Gamma}[X_0] = \int [d Y] \exp \left(-(S[X_0 + Y]-Y \cdot \frac{\delta \Gamma}{\delta X_0 })  \right),\label{definicion inicial} \ee 
where  $Y^m \equiv X^m - X^m_0 $. The field $Y^m$ will play the role of a
quantum field in the background field method. Insted of using the last
functional, it will be used 

\be \Omega[X_0] = \int [d Y] \exp \left(-(S[X_0 + Y]-S[X_0]-Y \cdot \frac{\delta \Gamma}{\delta X_0 })  \right). \label{partition function cap2} \ee 
This will be the generator of the 1PI diagrams. Subtracting $S[X_0]$ from
the exponential in (\ref{definicion inicial}), 
the expansions of the fields around $X_0$ will always contain the
quantum field $Y^m$. Making a Taylor expansion around $X_0$ a power series
in $Y^m$ will be obtained. Each term of such a expansion will not be invariant
under general coordinate transformations of spacetime, since $Y^m$ is a
subtraction of two coordinates, does not have such an invariance. That is why
it will be useful to find a system of coordinates in which the coordinate
invariance is manifest, such a coordinate system is denoted as normal
coordinate system. 

\paragraph{Normal Coordinate System}
Let $X_0^m$ be coordinates for a point $P_0$ in space-time and $X^m_0+
Y^m$ coordinates for a point $P$, it is possivel to find another
coordinate system by using a geodesic that joins both points. By considering
a parameter $t$ for the geodesic

\be \frac{d^2\lambda^m}{dt^2} + \Gamma^m_{np} \frac{d\lambda^n}{d
t}\frac{\lambda ^p}{dt} = 0 \label{geodesic}\ee
such that $\lambda^m(0)=X_0^m$ and $\lambda(1)=X_0^m+Y^m$. Defining
$\xi^m$ as  
\be \xi^m \equiv \frac{d \lambda^m}{dt}(0),\ee
it will be  a tangent vector to the geodesic in $P_0$, and as such,
will transform as a vector under a change of coordinates. So, any
geometrical object when expanded in Taylor series around $X_0^m$ will be a
diffeomorphism expression. 

\be  T_{m_1 ... m_i}(X_0+\xi) = \sum_{k=0}^\infty
\frac{1}{k!}\left(\frac{\partial}{\partial\xi^{n_1}}...\frac{\partial}{\partial\xi^{n_k}}\right)
T_{m_1...m_i}(X_0)\xi^{n_1}...\xi^{n_k}. \label{expansion coordenadas normales} \ee 
Supposing a solution for the geodesic equation in Taylor series
(\ref{geodesic})  
 
\be \lambda^m (t)  =
\sum_{k=0}^\infty\frac{1}{k!}\frac{d^k}{dt^k}\lambda^m(t)t^k ,\ee
then this solution, with the initial conditions already given has the form

\ber \lambda^m (t) & =&  X^m_0 + \xi^m t -\frac{1}{2}\Gamma^m_{n_1 n_2}
\xi^{n_1} \xi^{n_2}t^2
-\frac{1}{3!} \Gamma^m_{n_1 n_2 n_3}\xi^{n_1}\xi^{n_2}\xi^{n_3} t^3 - ... ,
\label{transformacion coordenadas normales} \\
  &=& X^m_0 + \xi^m t -\sum_{k=2}^\infty\frac{1}{k!}\Gamma^m_{n_1
  ...n_k}\xi^{n_1}...\xi^{n_k}t^k,\eer
where 
\ber \Gamma^m_{n_1 n_2 n_3}  &=&  \partial_{n_1} \Gamma^m _{n_2 n_3} -
\Gamma^l _{n_1 n_2}\Gamma^m_{l n_3}-\Gamma^l _{n_1 n_3}
\Gamma^m _{n_2 l}, \\
 &\equiv& \nabla_{n_1} \Gamma^m_{n_2 n_3}\label{derivada n=3}. \eer
The definition (\ref{derivada n=3}) is used recursively for defining
$\Gamma^m_{n_1 ...n_i}$ as the covariant derivative acting only in the lower
indices 

\be \Gamma^m _{n_1 ... n_i} \equiv \nabla_{n_1}\Gamma^m_{n_2 ... n_i}. \label{derivada conexiones} \ee
Finally, $\lambda(t=1)$ defines a coordinate transformation in which $Y^m$ is
written as a contravariant vector in this coordinate system 

\be \bar{\Gamma}^m_{(n_1...n_i)} = 0, \label{property}\ee
this defines the {\it normal coordinate system}. Here the bar notation
indicates that the expression is valid in a normal coordinate system.\\
Using such a coordinate system several expressions are simplified.
Christoffel symbols cancel, although not their derivatives, so the curvature
tensor in components is written as

\be \bar R^m\;_{nlp} = \partial_l \bar\Gamma^m_{np}-\partial_p\bar\Gamma^m_{nl}. \label{curvatura cc}\ee 
From (\ref{property}) when $i=3$ 

\be \partial_n \bar \Gamma^m_{l p} = - \partial_l \bar\Gamma^m_{pn } -
\partial_p \bar \Gamma^m_{ln}.\label{caso n=3}\ee 
Adding $2\partial_n \bar \Gamma^m_{pl}$ to both sides of (\ref{caso n=3})
and using the symmetry of the Christoffel symbols, one gets

\be \partial_{n_1}\bar \Gamma^m_{n_2 n_3} = \frac{1}{3}\left( \bar R^m \;
_{n_2 n_1 n_3}+\bar R^m \; _{n_3 n_1 n_2} \right). \label{conexiones 
como compcurvatura}\ee

In the following this result will be used to find an expansion in normal
coordinates for $G_{mn}$.

\subsubsection{Perturbative Expansion }

From (\ref{expansion coordenadas normales}) it can be found 
\ber \bar G_{mn}(X_0+\xi) &=& \bar G_{mn}(X_0) +\frac{\partial}{\partial
\xi^{l}}\bar G_{mn}(X_0)\xi^{l}+ \nonumber \\
&& + \frac{1}{2}\frac{\partial^2}{\partial \xi^{l}\partial\xi^p }
\bar G_{mn}(X_0) \xi^l \xi^p ... \, .\label{expansion G}\eer
As this expression is written in normal coordinates, then 

\be \frac{\partial}{\partial\xi^l }\bar G_{mn} =\nabla_l \bar G_{mn } ,\ee 
but if the metric $G_{mn}$ is covariantly constant, then this term does not
appear in the expansion. \\
The derivatives of the third term in (\ref{expansion G}) can be written as  

\be  \partial_l \partial_p \bar G_{mn} = \nabla_l \nabla_p \bar G_{mn} +
\partial_l \bar \Gamma^q_{pm}\bar G_{qn}+
\partial_l \bar \Gamma^q_{pn}\bar G_{mq}.\label{derivadas}\ee 
Symmetryzing (\ref{derivadas}) in the indices $l$ and $p$, and using (\ref{conexiones como compcurvatura})

\be \partial_{(l}\partial_{p)}\bar G_{mn} =\nabla_{(l}\nabla_{p)}\bar
G_{mn}+\frac{1}{3}[\bar R _{n(p l) m} + 
\bar R _{nm (l p)} + \bar R_{m (p l) n}+\bar R _{mn  (l p)}]. \ee
Replacing this expression in (\ref{expansion G}), and using the symmetries
of the curvatures tensor

\be R_{mnlp} = -R_{mnpl}, \,\, \,\,\,R_{mnlp} = R_{lpmn}, \label{simetrias tensor de Rieman}\ee 
one obtains
\be  \bar G_{mn}(X_0+Y) = \bar G_{mn}(X_0)+\frac{1}{3}\bar R_{mlpn} \xi^{l}
\xi^{p}+..., \label{expG}\ee
where again, the covariant derivatives of $G_{mn}$ are zero. Now, the
expansion (\ref{expG}) is written purely in tensorial terms and as such is
valid in any coordinate system. So, the bar notation is no longer necessary.
It will also be necessary to compute $\partial_a(X_0^m+Y^m)$. With this aim,
one computes (\ref{transformacion coordenadas normales}) in $t=1$ and take
derivatives :

\be \partial_a(X_0^m +Y^m) = \partial_a X_0^m + \partial_a \xi^m
-\frac{1}{2}\partial_n\bar \Gamma^m _{lp}\xi^l \xi^p 
\partial_a X_0^n - ... , \ee
and using again (\ref{conexiones como compcurvatura}),

\be \partial_a(X_0^m+Y^m) = \partial_a X_0^m + \nabla_a \xi^m
+\frac{1}{3}\bar R^m\;_{lpn}\partial_a X_0^n \xi^l \xi^p- ... , 
\label{expansion final derivadas} \ee
where 

\be \nabla_a \xi^m \equiv \partial_a \xi^m
+\Gamma^m_{lp}\xi^l\partial_a X_0^p \ee 
is the covariantization of $\partial_a\xi^m$. Then, (\ref{expansion
final derivadas}) can be written manifestly covariant. \\ Finally
multiplying (\ref{expG}) and (\ref{expansion final derivadas}) one finds 

\ber \label{expansion SG} S_G[X_0+Y] & = & S_G[X_0]+ \frac{1}{2\pi\alpha^{\prime}}\int_\Sigma
d^2\sigma \sqrt{|g|}g^{ab}G_{mn}\partial_a X_0^m \nabla_b \xi^n  \\ 
&& + \frac{1}{4\pi\alpha^{\prime}}\int_\Sigma d^2 \sigma \sqrt{|g|}g^{ab}
\left(G_{mn}\nabla_a\xi^m \nabla_b \xi^n+ R_{mlpn}\partial_a 
X_0^m \partial_b X_0^n \xi^l \xi^p \right) + {\ldots}, \nonumber  \eer
where $\Sigma$ denotes the two-dimensional manifold. In subsequent
chapters this notation will be dropped off. Now, for the antisymmetric tensor $B_{mn}$ one obtains 

\ber B_{mn}(X_0+Y) &=& B_{mn}(X_0)+\nabla_p B_{mn}(X_0)\xi^p 
+\frac{1}{2} ( \nabla_p \nabla_q B_{mn}(X_0) \nonumber  \\ && 
+ \frac{1}{3}R^l
\;_{pqm}B_{l n}(X_0)+\frac{1}{3}R^l\;_{pqn}B_{ml}(X_0)
) \xi^p \xi^q+... . \label{expansion B} \eer
From (\ref{expansion B}) and (\ref{expansion final derivadas}), it can be
found

\ber S_B [X_0+Y] & = &  S_B[X_0]+\frac{1}{2\pi\alpha^{\prime}}\int_\Sigma d^2 \sigma \epsilon^{ab}
(B_{mn}(X_0)\partial_a X_0^m \nabla_b\xi^n \nonumber \\  && +
\frac{1}{2}\nabla_l B_{mn} \partial_a 
X_0^m \partial_bX_0^n \xi^l )  \\ && + \frac{1}{4 \pi\alpha^{\prime}}\int_\Sigma d^2 
\sigma \epsilon^{ab}[ B_{mn}(X_0 )\nabla_a \xi^m \nabla_b \xi^n  +
2\nabla_l B_{mn}\partial_a 
X_0^m \xi^l \nabla_b \xi^n  \nonumber \\&& +\frac{1}{2}( \nabla_l \nabla_p
B_{mn} + 2B_{mq}
R^q\;_{lpn})\partial_a X_0^m \partial_b X_0^n \xi^l
\xi^p ]+... . \nonumber \eer
Nevertheless, note that the action for the antisymmetric tensor $S_B$ is
invariant under the following transformations   

\be B_{mn} \,\,\, \to \,\,\, B_{mn}+\partial_m \Lambda_n - \partial_n \Lambda_m, \label{gauge B}\ee
form some vector $\Lambda_m(X)$. So it is important that the terms in the
expansion have also this symmetry, and one look for a expansion in terms of
the field strength 
\be H_{m nl} \equiv \nabla_m B_{nl}+\nabla_n B_{lm}+\nabla_l B_{mn} = \nabla_{[m}B_{nl]},\ee
which is invariant under (\ref{gauge B}). The square bracket notation in
sub-indices means they are antisymmetrized. Up to surface terms, one finds
the following expression at second order in $\xi$

\be S_B[X_0+Y] = S_B[X_0]+\frac{1}{4\pi\alpha^\prime}\int_{\Sigma}d^2 \sigma \epsilon^{ab}\left(H_{lmn}\partial_a X_0^l \xi^m \nabla_b \xi^n  +\frac{1}{2}\nabla_l H_{mnp}\partial_a X_0^m \partial_b X_0^n \xi^l \xi^p \right)+ ... . \label{expansion SB} \ee
The expansion for $S_\Phi$ can easily be obtained 

\ber S_\Phi[X_0+Y] &=& S_\Phi[X_0]+\frac{1}{2\pi}\int_\Sigma d^2 \sigma
\sqrt{|g|}R^{(2)}\nabla_m\Phi(X_0)\xi^m \nonumber \\ && +
\frac{1}{4\pi}\int_\Sigma d^2 \sigma \sqrt{|g|}R^{(2)}\nabla_m\nabla_n\Phi(X_0)\xi^m \xi^n +...  . \label{expansion SD} \eer

If $X_0$ satisfy its classical equation of motion, then the linear term in
$\xi$ vanishes. To read the propagators, it will be useful to implement
an orthogonal frame, or vielbeins, for which $G_{mn} = e_m {}^i e_n
{}^j \eta_{ij}$. \\

\paragraph{Orthogonal Frame}
Denoting by $\{\hat{\partial}_m\}$ the vectors in a coordinate basis,
another basis can be written as a linear combination of them 

\be \hat{e}_i  =  e_i\,^m  \hat{\partial}_m, \ee 
with $i = 0,..., D-1$. For being orthonormal, it must satisfy 

\be G(\hat e_i, \hat e_j)  =  G_{mn}dX^m \otimes  dX^n (e_i \, ^l \hat
\partial_l, e_j \, ^p \hat \partial_p ) = \eta_{i j}, \ee
then

\be G_{mn} e_i \, ^m e_j \, ^n = \eta_{i j}. \label{base ortonormal}  \ee
Denoting by $e_m \, ^i$ the inverse elements of $e_i \, ^m$, they satisfy

\be e_i \, ^m e_m \, ^j = \delta_i \, ^j, \;\;\;\;\;\; e_m \, ^i e_i \, ^n = \delta_m \, ^n,  \ee
and from (\ref{base ortonormal}) one finds

\be \eta_{ij} e_m \, ^i e_n \, ^j = G_{mn}.  \ee
In this base, a conexion $\omega_m$ is introduced, with components
$\omega_m \, ^i \, _j$. It is defined by the condition that the covariant
derivative acting in the tetrad base is zero:

\be \nabla_m e_n \, ^i \equiv \partial_m e_n \, ^i - \Gamma^l _{m n}e_l \, ^i + \omega_m \, ^i \, _j e_n \, ^j = 0. \label{propiedades bases tetradas}\ee
Now, it is easy to see how the components of a vector are related among the
two basis

\be \xi = \xi^i \hat{e}_i = \xi^i e_i \, ^m \hat{\partial}_m = \xi^m \hat{\partial}_m, \ee
from which 

\be \xi^m = \xi^i e_i \, ^m. \label{relacion entre componentes} \ee
Therefore, using (\ref{base ortonormal}), (\ref{propiedades bases tetradas})
and (\ref{relacion entre componentes})

\ber  G_{mn}\nabla_a \xi^m \nabla_b \xi^n &=& \delta_{i j}e_m \, ^i e_n \, ^j \nabla_a \xi^m \nabla_b \xi^n, \\
&=& (\nabla_a \xi)^i(\nabla_b \xi)^i.   \eer
In this case, $\nabla_a$ denote that the derivative $\partial_a$ has been
covariantized, given by 

\be (\nabla_a \xi)^i = \partial_a \xi^i + \omega_m \, ^i \, _j \partial_a X^m \xi^j, \ee
and it is assumed that the derivative acting in $e_m \, ^i$ is zero.\\ 
With the help of the tetrad base, the following expansion for the
generalized bosonic non-linear sigma model is found

\ber S_\sigma[X_0+Y]&=& S_\sigma[X_0] +
\frac{1}{4\pi\alpha^{\prime}}\int_\Sigma d^2 \sigma \sqrt{|g|} g^{ab}
[(\nabla_a\xi)^i(\nabla_b \xi)^i \nonumber \\ && + R_{m i j n}(X_0)\partial_a X_0^m\partial_b X_0^n\xi^i \xi^j ] \nonumber \\ && +\frac{1}{3\pi\alpha^{\prime}}\int_{\Sigma}d^2 \sigma \sqrt{|g|} g^{ab}R_{m i j k}(X_0)\partial_a X_0^m (\nabla_b \xi)^k \xi^i \xi^j  \nonumber \\ && +\frac{1}{12\pi\alpha^{\prime}}\int_{\Sigma}d^2 \sigma \sqrt{|g|}g^{ab}R_{i k l j}(X_0) (\nabla_a \xi)^i (\nabla_b \xi)^j \xi^k \xi^l  \label{expansion modelo sigma} \\ && + \frac{1}{4\pi\alpha^\prime}\int_{\Sigma}d^2 \sigma \epsilon^{ab}[H_{m i j}(X_0)\partial_a X_0^m \xi^i (\nabla_b \xi)^j  \nonumber \\ && + \frac{1}{2}\nabla_i H_{mn j}(X_0)\partial_a X_0^m \partial_b X_0^n \xi^i \xi^j ] \nonumber \\ && +\frac{1}{4\pi}\int_\Sigma d^2 \sigma \sqrt{|g|} R^{(2)}\nabla_i\nabla_j\Phi(X_0)\xi^i \xi^j+ ... . \nonumber \eer

\subsubsection{One-loop conformal invariance}\label{plato fuerte}
In this section it will be studied the conditions for the energy-momentum
tensor being traceless. Noting that the functional generator 

\be \Omega = e^{-\Gamma}. \label{generating function}, \ee
depends only on the worldsheet metric (in a fixed gauge) and, as a
consequence of the diffeomorphism invariance satisfies

\be 0 = \int_\Sigma d^2 \sigma \frac{\delta \Gamma}{\delta g^{ab}}(\nabla^a v^b + \nabla^b v^a) ,  \ee
which in the conformal gauge and using coordinates $z= \sigma^1 + i\sigma^2$
and $\bar z= \sigma^1 - i\sigma^2$ takes the form

\be 0 = \int_\Sigma d^2 z \left( \frac{\delta \Gamma}{\delta g^{zz}}\nabla^z v^z + \frac{\delta \Gamma }{\delta g^{\bar z \bar z}}\nabla^{\bar z} v^{\bar z}
  -\frac{1}{2}\frac{\delta \Gamma}{\delta \omega}(\nabla_z v^z
  +\nabla_{\bar z} v^{\bar z}) \right),\ee
where $\omega$ is a conformal factor. Integrating by parts this expression

\ber  0 = \int_\Sigma d^2 z \sqrt{\hat g} & & [ \left( \nabla_z \left( \frac {1}{2\sqrt{\hat g}}
\frac{\delta \Gamma}{\delta \omega} \right) - \nabla^z \left(  \frac {1}{\sqrt{\hat g}}
\frac{\delta \Gamma}{\delta g^{zz}} \right)\right) v^z \nonumber \\ 
& & + \left( \nabla_{\bar z} \left( \frac {1}{2\sqrt{\hat g}}
\frac{\delta \Gamma}{\delta \omega} \right) - \nabla^{\bar z} \left(\frac{1}{\sqrt{\hat g}}
\frac{\delta \Gamma}{\delta g^{\bar z \bar z}} \right)\right) v^{\bar
z} ],  \eer
and as $v^z$ and $v^{\bar z} $ are arbitrary, the following set of equations
is obtained

\begin{eqnarray}  \nabla_z \left( \frac {1}{2\sqrt{\hat g}}
\frac{\delta \Gamma}{\delta \omega} \right) & = & \nabla^z \left(  \frac {1}{\sqrt{\hat g}}
\frac{\delta \Gamma}{\delta g^{zz}} \right) \label{conservacion cuantica} \\
\nabla_{\bar z} \left( \frac {1}{2\sqrt{\hat g}}
\frac{\delta \Gamma}{\delta \omega} \right)  & = &\nabla^{\bar z} \left(\frac{1}{\sqrt{\hat g}}
\frac{\delta \Gamma}{\delta g^{\bar z \bar z}} \right). \end{eqnarray}
These equations are the analog of the conservation of the classical
energy-momentum tensor. \\
It is possible to show that the right hand side are derivatives of the
expectation value of the $zz$ and $\bar z \bar z$ components of this tensor,
as explained in the following \\
From (\ref{generating function}),

\be \frac{\delta \Gamma}{\delta g^{zz}} = -\frac{1}{\Omega}\frac{\delta \Omega}{\delta g^{zz}} . 
\label{emt-gf}\ee
The measure element is chosen in such a way for not contributing to the
energy-momentum tensor. This allows to write

\be \nabla_z \left( \frac {4\pi}{\sqrt{\hat g}}
\frac{\delta \Gamma}{\delta g^{zz}} \right)  =  \nabla^z \langle T_{zz} \rangle , \label{conservacion 1} \ee
and in the same way

\be \nabla_{\bar z} \left( \frac {4\pi}{\sqrt{\hat g}}
\frac{\delta \Gamma}{\delta g^{\bar z \bar z}} \right)  =  \nabla^{\bar z} \langle T_{\bar z \bar z} \rangle . \label{conservacion 2} \ee
The left hand side of (\ref{conservacion cuantica}) will be identified with
the expectation value of the component $z \bar z$ of the energy-momentum.\\
In this way, independent of the metric $g$ component under consideration,
the variation of the effective respect of the metric is identified with the
energy-momentum tensor in the quantum regime:

\be \langle T_{ab}\rangle = \frac{4\pi}{\sqrt{|g|}}\frac{\delta \Gamma}{g_{ab}}, \label{energy momentum tensor} \ee
therefore, (\ref{conservacion cuantica}) takes the form of a conservation law

\be \nabla_z  \langle T_{\bar  z z}  \rangle + \nabla_{\bar z} \langle T_{zz} \rangle = 0. \label{conservacion}\ee

The idea is to use the value of $\langle T_{zz}\rangle$ computed at
$1-$loop, to compute the trace of the energy-momentum tensor. Initially can
be considered a flat worldsheet, and as discussed later, worldsheet
curvature effects will be taken into account. Using the notation $q = q^z$,
$\bar q = q^{\bar z}$, in momenta space the conservation law
(\ref{conservacion}) takes the form

\be \bar q \langle T_{z \bar z}(q) \rangle  + q \langle T_{zz}(q) \rangle  = 0   \label{conservacion en momentum space}.\ee
It will be first computed the contribution to $\langle T_{zz}\rangle$ coming
from the variation of the effective action containing the term $S_G$. From 
(\ref{energy momentum tensor}) and (\ref{emt-gf})

\be \langle T_{zz} \rangle =  \frac{1}{\Omega[X_0]}\frac{4\pi}{\sqrt{|g|}}\int [d\xi]exp\{-(S[X_0+\xi]-S[X_0])\}\frac{\delta(S_G[X_0+\xi]-S_G[X_0])}{\delta g^{zz}}. 
\label{comp ++ emt}\ee
The term in the exponential can be written as 

\be S_G[X_0+\xi]-S_G[X_0] = S_{Free}+S_{Int},\ee
with

\be S_{Free} = \frac{1}{4\pi\alpha^{\prime}}\int_\Sigma d^2 \sigma \sqrt{|g|}g^{ab}\partial_a \xi^i \partial_b \xi^i  \label{sliv}\ee
and

\be S_{Int} = -\frac{1}{4\pi \alpha^{\prime}}\int_\Sigma d^2 \sigma \sqrt{|g|}g^{ab}R_{i m j n}\partial_a X_0^m\partial_b X_0^n \xi^i \xi^j. \label{sint}\ee
Having chosen a tetrad basis allows to find an expression for the
propagator, which can be expressed diagrammatically as 

\be \epsfbox{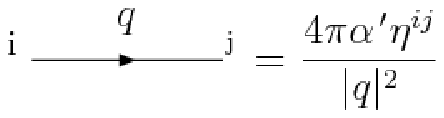}  \label{propagator}\ee
while from $S_{Int}$ can be found a vertex
\be \epsfbox{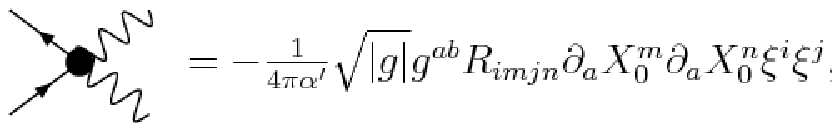}  \label{vertex}\ee
where the curved lines represent background fields.
Writing (\ref{comp ++ emt}) as

\be \langle T_{zz} \rangle = \frac{1}{\Omega[X_0]}\int [d\xi]e^{-(S_{Liv})}\left(\frac{1}{\alpha^{\prime}}\partial_z\xi^i \partial_z \xi^i + ... \right)e^{-S_{Int}} \ee 
and making an expansion of the exponential, the following diagram can be
formed

\be \epsfbox{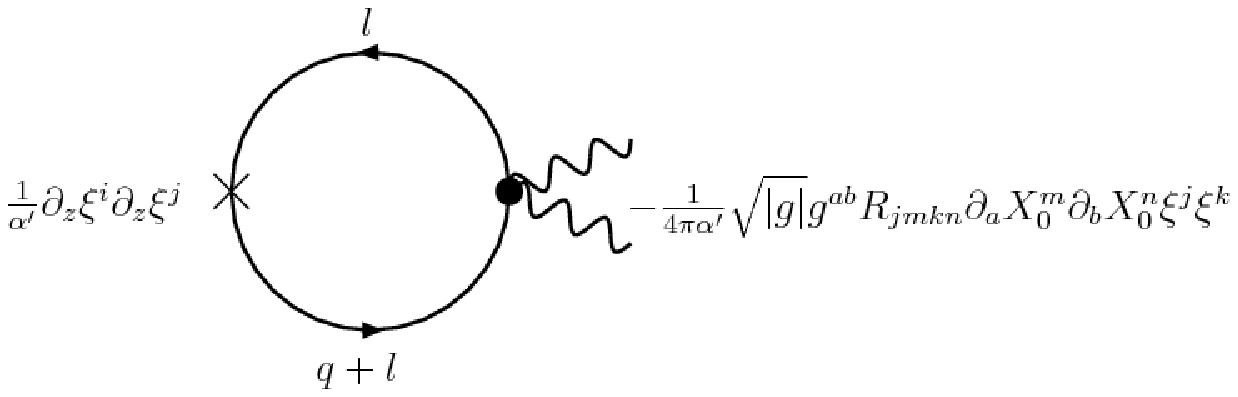}  \label{Pic0}\ee
where the cross represents an insertion of the energy-momentum tensor. This
diagram leads to 

\be \langle T_{zz} \rangle_{G} = \frac{1}{4}\int \frac{d^2 l}{2 \pi} \frac{\bar l(\bar l + 
\bar q)}{l^2 (l+q)^2}\{R_{mn}\partial_a X_0^m \partial^a X_0^n\}(q). \label{primera integral}\ee
In this equation the keys denote an expression in momentum space, but
independent of the momenta $l$.\\
To compute this integral in (\ref{primera integral}), one can use a Feynman
parameter $x$ to write it as a known integral, whose value could be found
using the dimensional regularization formulas \cite{Ramond}. Introducing the
parameter $x$, the integrand of  (\ref{primera integral}) is written as 

\be \frac{\bar l(\bar l+ \bar q)}{l^2 (l+q)^2} = \int_0^1 dx\frac{(l^1-il^2)^2+(l^1-il^2)(q^1-iq^2)}{[xl^2 + (1-x)(l+q)^2]^2}.  \ee 
Defining

\be k^a \equiv l^a+(1-x)q^a, \,\,\,  a=1,2 \label{definicion de k}\ee
then

\be \int \frac{dld\bar l}{2\pi}\frac{\bar l(\bar l+\bar q)}{l^2(l+q)^2} = \int_0^1 dx \int\frac{d^2k}{\pi}\frac{(k^1)^2-
(k^2)^2-2ik^1k^2-x(1-x)(q^1-iq^2)^2}{(k^2+\Delta)^2}, \label{calculo integral 1parcial} \ee
with 

\be \Delta \equiv x(1-x)q^2. \label{definicion de delta}\ee
In the integral (\ref{calculo integral 1parcial}) there are omitted linear
terms in $k$ that vanishes because of symmetry. Extending the
two-dimensional space to $d = 2+\epsilon$, (\ref{calculo integral 1parcial})
has a known form. The quadratic terms in $k$ cancel among them using  

\be \int d^d k \frac{k^a k^b}{(k^2+\Delta)^2} = \frac{\pi^\frac{d}{2}\delta^{ab}} {2}\Gamma(1-\frac{d}{2})\left(\frac{1}{\Delta} \right)^{1-\frac{d}{2}}. \ee
Now, using  

\be \int d^d k\frac{1}{(k^2+\Delta)^2} = \pi^{\frac{d}{2}}\Gamma(2-\frac{d}{2})\left(\frac{1}{\Delta}\right)^{2-\frac{d}{2}}\label{regularizacion sin potencias en el numerador}\ee
in the limit $\epsilon \to 0$, one obtains

\be \int \frac{d^2 l}{2 \pi} \frac{\bar l(\bar l + \bar q)}{l^2 (l+q)^2} = -\frac{\bar q}{q}. \label{Tzz de G}\ee
Substituting (\ref{Tzz de G}) in (\ref{conservacion en momentum space}) one
obtains an expression for the trace of the energy momentum tensor 
\be \langle T_{z\bar z}\rangle_{G} = \frac{1}{4}R_{mn}\partial_a X_0^m
\partial^a X_0^n ,\ee
which was the desired result. 

It can also be found an interaction with the first term in (\ref{expansion SB}). 
A diagram with only the linear term in the expansion of the exponential is
canceled by the antisymmetry of $H$, therefore, such a term should be
considered in quadratic order. The type of interaction is given by  

\be \epsfbox{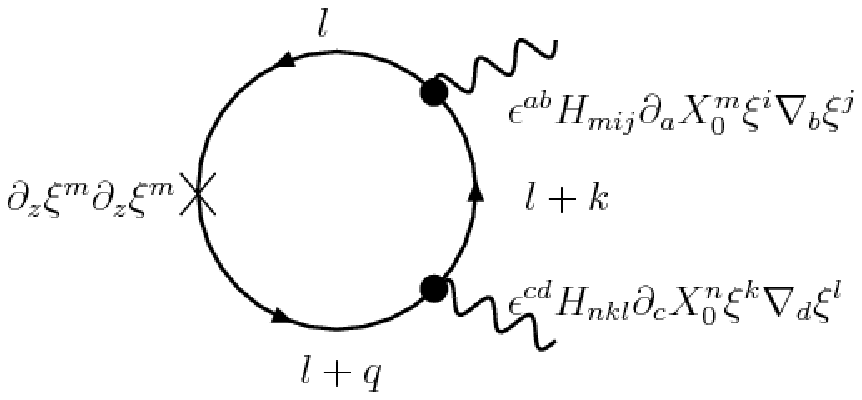}  \label{Pic02}\ee


from which one obtains

\be \langle T_{z z} \rangle_{H_2} = -\frac{1}{8\alpha^{\prime}}\frac{(4\pi\alpha^{\prime})^3}{(4\pi \alpha^{\prime})^2}\frac{\epsilon^{ab}}{\sqrt{|g|}}\frac{\epsilon^{cd}
}{\sqrt{|g|}}\partial_a X_0^m \partial_c X_0^n H_{m i j}H_{n l
k}\eta^{il}\eta^{j k} \label{emtas} \ee $$ \times \int \frac{dl d \bar l}{2 (2\pi)^2}
\frac{(l+q)_z(l+k)_b l_z l_d}{l^2(l+q)^2(l+k)^2}$$
Without lost of generality, one can make $k = 0$. Using 

\be l_b l_d = \frac{1}{2}g_{bd}l^2\ee
and 
\be \epsilon^{ab}\epsilon^{cd}g_{bd} = g^{ac},\ee
the equation (\ref{emtas}) is written 

\be \langle T_{zz}\rangle_{H_2} = -\frac{1}{64}H_{mn}^2 g^{ab}\partial_a X_0^m \partial_ bX_0^n \int \frac{dl d\bar l}{2\pi} \frac{\bar l (\bar l+ \bar q)}{l^2 (l+q)^2},    \ee
with
\be H_{mn}^2 \equiv H_{m l p}H_{n}\,^{lp}.\ee
Using the computed value for this integral (\ref{Tzz de G}) and having in
mind the existence of other four identical configurations to the diagram
(ref diagram 2) the following expression for $\langle T_{zz}\rangle$ can be
found
\be \langle T_{zz} \rangle_{H_2} =  \frac{1}{16} \frac{\bar q}{q}H_{mn}^2 g^{ab}\partial_a X_0^m \partial_ bX_0^n. \ee
Finally, using the expression for the energy-momentum conservation
(\ref{conservacion en momentum space}) can be found the expression for $\langle T_{z\bar z}\rangle$. \\
Up to now the results are 

\be \langle T_{z \bar z}\rangle  = \left(\frac{1}{4}R_{mn}-\frac{1}{16}H_{mn}\,^2 \right)\partial_a X_0^m \partial^a X_0^n + \left(-\frac{1}{8}\nabla^l H_{l m n}\right)\frac{\epsilon^{ab}}{\sqrt{|g|}}\partial_a X_0^m \partial_b X_0^n \label{1result}\ee
Until now the contributions coming from the dilaton were ignored by
choosing $g_{ab}=\delta_{ab}$. It seems that (\ref{1result}) would be all
the contributions at first order in  $\alpha^\prime$. Nevertheless, in a
flat worldsheet, variations of the action including the dilaton respect of
infinitesiaml variations of the metric, when evaluated in a flat worldsheet
are different from zero. Making this variation  

\be \delta S_\Phi  = -\frac{1}{2\pi}\int_\Sigma d^2 \sigma
\delta(\sqrt{|g|}R^{(2)})\Phi(X), \label{variacion de accion dilatonica} \ee
Palatini's identity can be used:

\be \delta R^{(2)}_{ab} = \nabla_c\delta \Gamma^c_{ab} - \nabla_b\delta \Gamma^c_{ac}  , \ee
where

\be \delta \Gamma^c_{ab} \equiv -g^{cd}\delta g_{de}\Gamma^e_{ab}+\frac{1}{2}g^{cd}(\delta g_{da,b}+\delta g_{db,a}-\delta g_{ab,d}), \label{definicion de delta Gamma}\ee
to write

\be \sqrt{|g|}g^{ab}\delta R^{(2)}_{ab} = \nabla_b(\sqrt{|g|}g^{ab}\delta \Gamma^c_{ac})-\nabla_c(\sqrt{|g|}g^{ab}\delta \Gamma^c_{ab}). \ee
Integrating by parts (\ref{variacion de accion dilatonica}) one finds

\ber \delta S_{\Phi} & = & \frac{1}{2 \pi} \int_{\Sigma}d^2\sigma\sqrt{|g|}\{g^{ab}\delta \Gamma^c_{ab}\partial_c \Phi(X)-g^{ab}\delta \Gamma^c_{ac}\partial_b \Phi(X)  \} \nonumber \\
&&+\frac{1}{2\pi} \int_{\Sigma}d^2 \sigma \{\delta
\sqrt{|g|}R^{(2)}\Phi(X)+\sqrt{|g|}\delta g^{ab}R^{(2)} \}. \label{delta
Phi} \eer
Replacing (\ref{definicion de delta Gamma}) in (\ref{delta Phi}),
integrating again by parts and making $g_{ab} = \delta_{ab}$, the
energy-momentum tensor can be computed

\be T^{dil}_{ab} = 2(\partial_a \partial_b -\delta_{ab} \Delta)\Phi(X).\ee
The simbol $\Delta$ is the Laplacian in the worldsheet. In coordinates $(z,\bar z)$, 

\be T^{dil}_{z \bar z} = -\frac{1}{2} \Delta \Phi(X(\sigma)), \label{trazo dilatonico} \ee
and the trace is different from zero, as expected by the lack of conformal
symmetry of $S_\Phi$. To compute the contributions of this trace to the
total conformal anomaly, one computes (\ref{trazo dilatonico}) in $X_0$,

\ber \Delta \Phi(X_0) & = & 2g^{z \bar z}\partial_z\partial_{\bar z}\Phi(X_0), \nonumber \\
&=& 2g^{z \bar z} \partial_z(\partial_m \Phi(X_0)\partial_{\bar z}X_0^m) \nonumber, \\
&=& \partial_a X_0^n \partial^a X_0^m \partial_n\partial_m \Phi(X_0)
+\Delta X_0^m \partial_m \Phi(X_0) . \label{DAlembertiano de Phi}  \eer
But as $X_0$ satisfy the classical equation of motion

\be \Delta X_0^m = -\Gamma^m_{nl}g^{ab}\partial_a X_0^n \partial^a X_0^l + \frac{1}{2}H^m \,_{nl} \\ \frac{\epsilon^{ab}}{\sqrt{|g|}}\partial_aX_0^n\partial_bX_0^l ,  \label{ecuacion de movimiento clasica para SG+SB}\ee
replacing (\ref{ecuacion de movimiento clasica para SG+SB}) in
(\ref{DAlembertiano de Phi}), one finds

\be \Delta \Phi(X_0) = \nabla_m \nabla_n \Phi(X_0)\partial_a X_0^m \partial^a X_0^n+\frac{1}{2}\nabla^m\Phi(X_0) H_{mnl}\frac{\epsilon^{ab}}{\sqrt{|g|}}\partial_a X_0^n \partial_b X_0^l ,  \ee
That is, the classical contributions coming from the variation of $S_\Phi$
are of the same order as the one-loop contributions coming from $\langle T_G
\rangle$ and $\langle T_H \rangle$. 
Therefore, the following partial result can be written

\be \langle T_a^a\rangle = \beta^G_{mn} \partial_a X_0^m \partial^a X_0^n + \beta^B_{mn}\frac{\epsilon^{ab}}{\sqrt{|g|}}\partial_a X_0^m \partial_b X_0^n ,  \ee
with

\ber \beta^G_{mn} &=&  R_{mn}(X_0)-\frac{1}{4}H_{ml p}H_n\,^{lp}(X_0)   +2\nabla_m \nabla_n \Phi(X_0), \\ 
\beta^B_{mn} &=& -\frac{1}{2}\nabla^l
H_{lmn}(X_0)+H_{lmn}\nabla^l \Phi(X_0). \eer
To find the remaining terms in the beta functions, some computations of two
point functions must be done, i.e two insertions of the energy momentum
tensor. To see that this is the case, one must remember that because of the 
diffeomorphism symmetry in two dimensions, a worldsheet metric can be written as a
scale factor times a flat metric. In this case, to state that a theory has
Weyl symmetry is equivalent to say that the energy-momentum tensor is
traceless, when computed using the metric $g_{ab} = e^{2\omega}\delta_{ab}$. 
This must be independent of the scale factor $\omega$ such that the result
is valid for any curved worldsheet. Then, at least the first variation with
respec to to $\omega$ must be zero. This first variation can be written as

\be \frac{\delta}{\delta\omega{(z^\prime)}}\langle T_{z\bar z}(z)\rangle_{e^{2\omega}\delta} = 
\frac{\delta \langle T_{z \bar z}(z)\rangle}{\delta g^{ab}}\frac{\delta g^{ab}}{\delta \omega}, \label{1v}\ee
and evaluating in  $\omega = 0$, the variation (\ref{1v}) is written in
terms of the two point function for the energy-momentum tensor

\be \frac{\delta}{\delta\omega{(z^\prime)}}\langle T_{z \bar z}(z)\rangle_{e^{2\omega}\delta} = 
-\frac{1}{\pi}\langle T_{z \bar z}(z)T_{z \bar z}(z^\prime)\rangle_{\delta}, \label{funcion de dos puntos} \ee
where 
\be \langle T _{z\bar z}(z)T_{z\bar z}(z^\prime) \rangle_{\delta}= \langle T^G_{z\bar z}(z)T^G_{z\bar z}(z^\prime)\rangle_\delta + 2\langle T^G_{z\bar z}(z)T^\Phi_{z\bar z}(z^\prime)\rangle_{\delta}+ \langle T^\Phi_{z \bar z}(z)T^\Phi_{z \bar z}(z^\prime)\rangle_\delta. \label{two point function} \ee
Integrating $\omega$ allows to write the trace of the energy-momentum tensor
in terms of the results obtained from the computation of the two point
function. The result will be something known: a term proportional to the
worldsheet curvature scalar $R^{(2)}$, whose constant of proportionality
contains the space-time dimension. Furthermore, some contributions fo the
fields $G$ and $H$ will be found. \\
Consider for the time being just the terms in the classical action with Weyl
symmetry: $S_G$ and $S_B$.
At the lowest order in $\alpha\,^\prime$ in the two point function, $\langle
T_{zz}T_{zz}\rangle$, the next graph is found 

\be \epsfbox{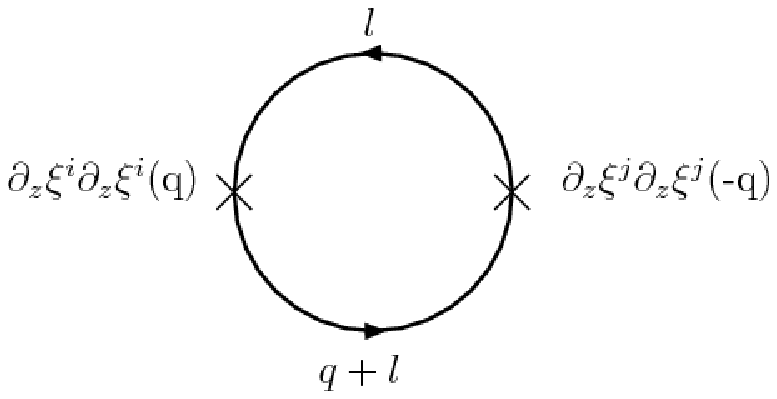}  \label{Pic02}\ee

from which is obtained

\be \langle T_{zz}(q)T_{zz}(-q)\rangle = \frac{D}{8}\int d^2 l \frac{{\bar l}^2(\bar l+ \bar q)^2}{l^2(l+q)^2} .\label{integralg3}\ee
To write the integral in (\ref{integralg3}) as an integral whose value is
known, a Feynman parameter $x$ is introduced. The two point function
(\ref{integralg3}) takes the form

\ber \langle T_{zz}(q)T_{zz}(-q)\rangle &=& \frac{D}{8}\int_0^1dx \int d^d k
\frac{(k^1-ik^2)^4+x^2(1-x)^2{\bar q}^4}{(k^2 + \Delta)^2}, \eer
with  $k$ and $\Delta$ given by (\ref{definicion de k}) and (\ref{definicion
de delta}) respectively. The terms with odd powers of $k$ are zero because
of symmetry, while the quadratic terms cancel among thems, as in
(\ref{calculo integral 1parcial}) . It is not difficult to prove that the
quadratic term in $k$ vanishes using 

\be \int d^d k \frac{k^a k^b k^c k^d}{(k^2+\Delta)^2} = \frac{\pi^{\frac{d}{2}}}{4}\Gamma\left(-\frac{d}{2}\right)\Delta^{\frac{d}{2}}(\delta^{ab}\delta^{cd}+\delta^{ac}\delta^{bd}+\delta^{ad}\delta^{bc}).\ee
Using (\ref{regularizacion sin potencias en el numerador}), one finds 

\be \langle T_{zz}(q)T_{zz}(-q)\rangle_{\delta} = \frac{\pi}{48}D\frac{\bar q^3}{q}. \ee
Using twice the energia-momentum tensor conservatoin $ q T_{zz}(q)+\bar q
T_{z \bar z}(q)= 0$, it can be found

\be \langle T_{z \bar z}(q)T_{z \bar z}(-q)\rangle_{\delta} = \frac{\pi}{48}D q\bar q.  \ee
Writting this equation in coordinate space

\be \langle T_{z \bar z}(z)T_{z \bar z}(0)\rangle_{\delta} = -\frac{\pi D}{12}\sqrt{|g|}\Delta \delta^{(2)}(z),  \ee
where it was used

\be \Delta \delta^{2}(\sigma) = -\frac{1}{\sqrt{|g|}}\int\frac{d^2 q}{ (2\pi)^2}\frac{q\bar q}{4}e^{iq \cdot z}.\ee
From (\ref{funcion de dos puntos}), an expression for $\langle T_{z \bar
z}\rangle$ can be found

\be  \langle T_{z \bar z}  \rangle_{e^{2\omega}\delta}  = \frac{D}{12}\sqrt{|g|}\Delta \omega.\ee
Using the following expression for the two-dimensional scalar curvature

\be R^{(2)} = -2\Delta \omega , \ee
which is valid in conformal gauge, one can write

\be \langle T_{z \, \bar z} \rangle_{e^{2\omega}\delta} = -\frac{D}{24}\sqrt{|g|} R^{(2)}. \label{funcion beta de phi a primer orden}\ee 
Multiplying both sides of (\ref{funcion beta de phi a primer orden}) by $2 g^{z\bar z}$

\be \langle T_a \, ^a \rangle = -\frac{D}{12}R^{(2)}.\ee
This expression is modified as $D \to D-26$ by considering the ghost fields
that appear when fixing the conformal gauge. 
\\ This last contribution to the trace of the energy-momentum tensor as a
different structure, since it is proportional to $R^{(2)}$ and neither to $\partial_a X_0^m\partial^a X_0^n$ 
nor $\epsilon^{ab}\partial_a X_0^m \partial_b X_0^n$. Moreover, in this
contribution do not appear terms containing $G$ neither $H$. 
To find their contributions proportional to $R^{(2)}$, it is necessary to go
to higher order terms in the expansions of $S_G[X]$ and $S_B[X]$. Those
contributions can be found computing the graphs \cite{Callan Thorlacius}

\be \epsfbox{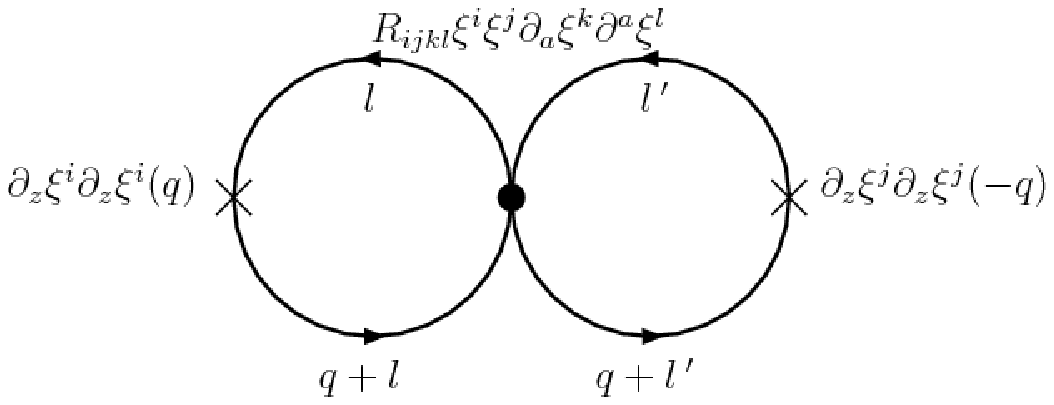}  \label{Pic03}\ee

\be \epsfbox{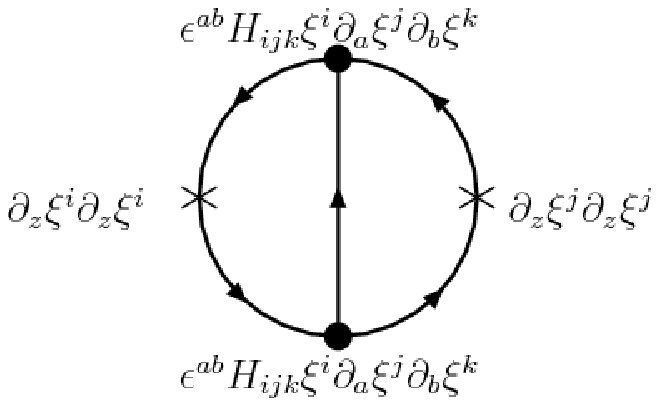}  \label{Pic04}\ee

\be \epsfbox{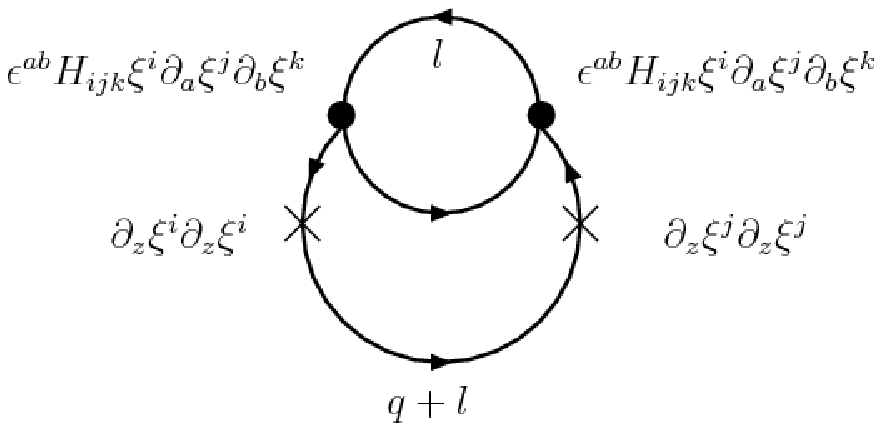}  \label{Pic05}\ee

These two loops computations are more complicated, since subdivergences can
appear. Nevertheless, according with \cite{Callan Thorlacius} their
contributions to the trace $\langle T_a \, ^a \rangle$ can be computed using
the same strategy of the conservation of the energy-momentum tensor.
They contribute

\be \langle T_a \, ^a \rangle = \alpha^\prime \left( R-\frac{H^2}{12} \right) R^{(2)}, \ee
where $R$ is the scalar curvature of space-time.\\
There are remaning contributions that can appear considering the two-point
functions $\langle T^{\Phi}_{z \bar z}(z)T^{\Phi}_{z\bar z}(z^\prime) 
\rangle$ and $\langle T^{G}_{z \bar z}(z)T^{\Phi}_{z\bar
z}(z^\prime)\rangle$. From the expansion for the action with $\Phi$ is not
difficult to find   

\ber T^\Phi_{z\bar z} &=&\frac{4\pi}{\sqrt{g}}\frac{\delta}{\delta g^{zz}}\left(- \frac{1}{2\pi}\int d^2 z \sqrt{g} R^{(2)}\nabla_i\Phi \xi^i \right)\nonumber\\  &=& 
2[\partial \bar \partial(\nabla_i \Phi)\xi^i +\partial (\nabla_i \Phi)\bar \partial \xi^i + \bar \partial(\nabla_i \Phi)\partial\xi^i +(\nabla_i \Phi)\partial \bar 
\partial \xi^i]. \eer
Taking just the last term in the last equation  
\be \langle T^\Phi_{z \bar z}(q)T^\Phi_{z \bar z}(-q)\rangle_{\delta} = \frac{1}{4}\nabla_i\Phi \nabla_j \Phi \langle \Delta \xi^i \Delta \xi^j \rangle_{\delta}
\ee which can be represented by the following diagram

\be \epsfbox{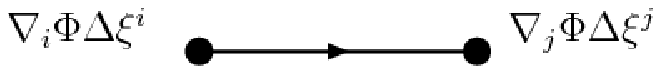}  \label{Pic06}\ee

from which it is obtained

\be \langle T^\Phi_{z \bar z}T^\Phi_{z \bar z}\rangle_{\delta} = \pi\alpha^\prime (\nabla \Phi)^2 q \bar q.\ee
Using the energy-momentum conservation twice, a contribution of
$-2\alpha^\prime(\nabla \Phi)^2R^{(2)}$ is obtained for the trace of the
energy-momentum tensor.

Nevertheless, the expansion for the energy-momentum tensor coming from
variating $S_\Phi$ also contributes at $\alpha^{\prime}$ order, so it is
necessary to compute $\langle T^G_{zz}T^\Phi_{zz}\rangle$. To compute
$T^\Phi_{zz}$ the same procedure that allowed to find contributions coming from the
variation of the action including $\Phi$ can be used. The following result
is found 

\ber T^\Phi_{zz} &=&\frac{4\pi}{\sqrt{g}}\frac{\delta}{\delta g^{zz}}\left(- \frac{1}{4\pi}\int d^2 z \sqrt{g} R^{(2)}\nabla_i\Phi\nabla_j\Phi \xi^i 
\xi^j\right)\nonumber\\  &=& -[\partial^2 (\nabla_i\nabla_j \Phi)\xi^i\xi^j + 4 \partial(\nabla_i\nabla_j\Phi)\partial \xi^i \xi^j+ 
2(\partial_i\partial_j\Phi)\xi^i \partial^2 \xi^j \nonumber \\ &
&\,\,\;\;\;\;+ 2 (\nabla_i \nabla_j \Phi)\partial\xi^i \partial \xi^j].
\label{TZZ phi}\eer
Out of these terms, only the last two will give a non-zero contribution. The
product of the last term with $T^G_{zz}$ is represented in the following
diagram 

\be \epsfbox{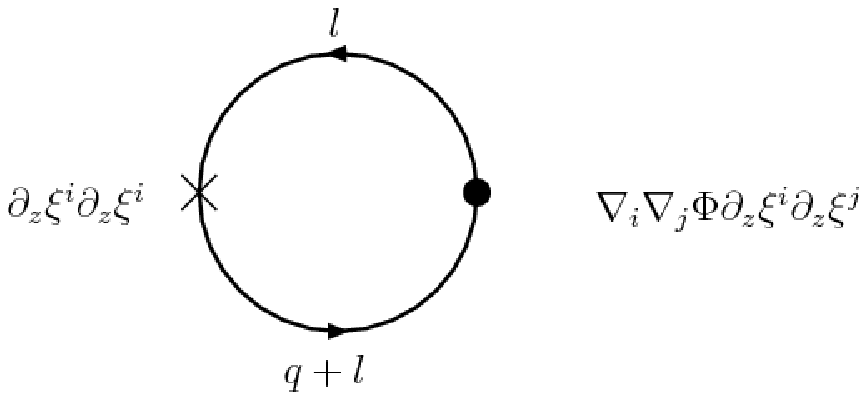}  \label{Pic07}\ee

while the term before the last in ($\ref{TZZ phi}$) with $T^G_{zz}$ can be
represented by a similar diagram. From the last diagram can it can be found

\be \langle T^G_{zz}T^\Phi_{zz}\rangle_{\delta} = -\frac{1}{2}\alpha^\prime \nabla^2 \Phi\int dl d \bar l \frac{{\bar l}^2(\bar l + \bar q)^2}{l^2(l+q)^2},\ee 
this integral is the same as in (\ref{integralg3}). Then, using the same
result one finds  

\be \langle T^G_{zz}T^\Phi_{zz}\rangle_{\delta} = -\frac{\pi \alpha^\prime}{6}\nabla^2 \Phi \frac{(\bar q)^3}{q} \label{firstresult}\ee
To compute the remaining contribution  

\be \int dl d\bar l\frac{{\bar l}^3(\bar l + \bar q )}{l^2(l+q)^2}\ee
has to be computed. The result is the double of (\ref{firstresult}) and
adding up everything, one finds  

\be \langle T_{z\bar z}\rangle_{e^{2\omega}\delta} = 2\alpha\,^\prime\sqrt{|g|}\nabla^2 \Phi R^{(2)}.\ee

Adding up all the computed contributions as indicated in (\ref{two point
function}) to the trace of the energy-momentum tensor it can be written as 

\be \langle T_a\,^a\rangle = \beta^G _{mn}g^{ab}\partial_a X_0^m \partial_bX_0^n +\beta^B_{mn}\epsilon^{ab}\partial_aX_0^m \partial_b X_0^n +
\beta^\Phi R^{(2)}, \label{trace}\ee
with 

\ber \beta^G_{mn} &=& R_{mn}-\frac{1}{4} H_{mlp}H_n\,^{lp}+2\nabla_m\nabla_n \Phi, \label{betaG}\\ 
\beta^B_{mn} &=&-\frac{1}{2}\nabla^p H_{mnp} +\nabla_l\Phi H^l\,_{mn}, \label{betaB}\\
\beta^\Phi &=& -\frac{D-26}{12} + \alpha^\prime \left(R - \frac{H^2}{12}
+4\nabla^2 \Phi - 4 (\nabla\Phi)^2  \right). \label{betaPhi}\eer

In the following, it will be shown that these beta functions can be
consistently set to zero, in such a way that the theory has no conformal
anomaly at one-loop.

\subsubsection{Consistency conditions of the Weyl invariance}
When the fields $B_{mn}$ and $\Phi$ are zero and $G_{mn}$ is the flat metric,
the coefficients $\beta^G_{mn}$, $\beta^B_{mn}$ in (\ref{trace}) are
zero, and $\beta^\Phi$ reduces to a number proportional to $D-26$. Nevertheless when the bosonic
string is coupled to space-time fields, there appear terms in the trace of
the energy-momentum tensor which do not have this property: they are
proportional to $\partial_a X^m \partial_b X^n$. Furthermore,
$\beta^\Phi$ appears as corrections to the $D-26$ term. 
In order to have an anomaly free theory, all of the three $\beta$ functions
must cancel, in a consistent way, where by consistent is meant to preserve
the property of $\beta^\Phi$ being a number. The term of order
$\alpha^\prime$ in $\beta^\Phi$ includes fields in spacetime, then in
principle would not be constant numbers. Nevertheless, as will be shown in
the following the conditions $\beta^G_{mn}=0$ and $\beta^B_{mn} = 0$
imply that the gradient of $\beta^\Phi$ is zero, therefore $\beta^\Phi$ is a
constant.\\ 
Using the Bianchi identity for the curvature tensor 

\be \nabla_{[l}R_{mn]p\eta} = 0, \label{Bianchi identity}\ee
is easy to see that the Ricci $R_{mn}$ tensor satisfy

\be \nabla^n R_{mn} = \frac{1}{2}\nabla_n R .\ee
From the definition of $H_{mnl}$ can be verified that it satisfies
the identity

\be \nabla_{[p}H_{mnl]} = 0,\ee
what allows to write 

\be \nabla^n (H_{mlp} H_{n}\,^{lp}) = H_{ml p}\nabla^n H_{n}\,^{lp}+ \frac{1}{6}\nabla_m H^2, \ee
Then, computing $\nabla^n \beta^G_{mn}$ is obtained

\be \nabla^n \beta^G_{mn} = \frac{1}{2}\nabla_{m}R - \frac{1}{24}\nabla_m H^2 -\frac{1}{4}H_m\,^{lp} \nabla^n H_{n l p} + 
2\nabla_m \nabla^2 \Phi + 2R_{mn}\nabla^n \Phi, \label{derivada de beta G}\ee
where it was used the definition of the curvature tensor
$[\nabla_m,\nabla_n]v^l = - R_{m n p}\,^l v^p$.
\\In terms of the beta functions $\beta^G_{mn}$ and $\beta^B_{mn}$
the equation (\ref{derivada de beta G}) is written as  

\be \nabla^n \beta^G_{mn} = \frac{1}{2}\nabla_{m}\left(R - \frac{1}{12}H^2 + 4 \nabla^2 \Phi - 4 (\nabla \Phi)^2 \right) + \frac{1}{2}\beta^B_{nl} 
H_m\,^{nl} + 2\beta^G_{mn}\nabla^n \Phi,\ee
or

\be \nabla^n \beta^G_{mn} = \frac{1}{2\alpha^{\prime}}\nabla_{m}\beta^\Phi + \frac{1}{2}\beta^B_{nl} H_m\,^{nl} + 
2\beta^G_{mn}\nabla^n \Phi,
\ee
which implies $\beta^\Phi$ is a constant if $\beta^G_{mn}$ and
$\beta^B_{mn}$ are zero. Therefore, when all the $\beta$ functions are
zero the theory has Weyl symmetry at the quantum level. \\

\subsubsection{Spacetime Effective Action }
The system of equations obtained by the vanishing of the $\beta$ functions
can be written in a more suggestive way.
From $\beta^\Phi=0$ it can be found

\be R = \frac{1}{12}H^2-4\nabla^2 \Phi-4 (\nabla \Phi)^2 ,\ee
with this and from $\beta^G_{mn} = 0$ one obtains

\be R_{mn} - \frac{1}{2}G_{mn}R = \Theta_{mn}, \label{Einstein equation} \ee
where

\be  \Theta_{mn} \equiv \frac{1}{4}\left(H_{mn}^2 -\frac{1}{6}G_{mn}H^2\right)-2 \nabla_m \nabla_n \Phi+ 2 G_{mn} \nabla^2 \Phi 
- 2 G_{mn}(\nabla \Phi)^2. \label{emtsp}\ee  
The equation (\ref{Einstein equation}), satisfied by the metric field,
is the Einstein equation in spacetime, with energy-momentum tensor
(\ref{emtsp}). This is a symmetric tensor, but its conservation must
be checked. In fact, applying the operator $\nabla^m$ to the equation  
(\ref{Einstein equation}), the left hand side is identically 
zero, as can be checked by using the Bianchi identity(\ref{Bianchi identity}). 
The right hand side can be written as 

\be \nabla^m \Theta_{mn} = \beta^B_{ml}H_{n}\,^{ml} - 2\beta^G_{mn}\nabla^m \Phi. \ee 
So the conditions for preserving the Weyl symmetry at the quantum level $\beta^G_{mn} = 0$, 
$\beta^B_{mn}=0$ guarantee the conservation for the matter
energy-momentum tensor.

Now is possible to find another two equations for the other fields in
spacetime.
By taking the trace of $\beta^G_{mn}$ 

\be G^{mn}\beta^G_{mn} = R - \frac{1}{4}H^2_{mn}+2\nabla^2 \Phi,
 \ee
but the condition $\beta^G_{mn} = 0$ together with $\beta^\Phi =
0$, allows to eliminate $R$ to write 

\be  \nabla^2 \Phi - 2 (\nabla \Phi)^2 = -\frac{1}{12}H^2. \label{dilaton equation} \ee
what constitutes the equation of motion for the dilaton.
Finally, the equation for the field $H$ will come from the condition $\beta^B_{mn}=0:$

\be \nabla^l H_{l mn} = 2 \nabla^l \Phi H_{lmn}. \label{antisimetric tensor equation}\ee
As is known, Einstein's equations can be derived as an equation of
motion of the Einstein-Hilbert action. It is tempting to think that
the three equations for the spacetime fields (\ref{Einstein equation}), (\ref{dilaton equation}) 
and(\ref{antisimetric tensor equation}) could be deduced from an
action principle. This work was done by Metsaev and Tseytlin
\cite{Metsaev Tseytlin}, where the following action was found

\be S = \int d^{26} X \sqrt{|G|}e^{-2\Phi} \left( R -\frac{1}{12} H^2 + 4 (\nabla \Phi)^2 \right). \label{spacetime action}\ee
By making the variation of (\ref{spacetime action}) with respect of
the fields $G$ $H$ and $\Phi$:

\ber  \delta S &=& \int d^D X \sqrt{|G|}e^{-2\Phi}\{\delta G^{mn}
[R_{mn}-\frac{1}{2}RG_{mn} + 2 \nabla_m \nabla_n \Phi - 2
G_{mn}\nabla^2 \Phi \nonumber \\ && + 2 G_{mn}(\nabla \Phi)^2 + \frac{1}{24} G_{mn}H^2 -\frac{1}{4}H^2_{mn}] -\frac{1}{2}\delta \Phi \beta^\Phi - \nonumber \\ && \frac{1}{6}H^{mnl}\delta H_{mnl} \}. \eer
The sector of derivatives of the field $G_{mn}$ in (\ref{spacetime
action}) does not have the form of the Einstein-Hibert action because
of the exponential of the field $\Phi$, but by making a transformation 
$G_{mn} = e^\omega \tilde{G}_{mn}$, (\ref{spacetime action})
takes the form

\ber S &=& \int d^D X \sqrt{|\tilde
G|}e^{-2\Phi}e^{\omega\frac{D-2}{2}} \{\tilde R - (D-1) \widetilde
{\nabla^2\omega} -\frac{1}{4}(D-1)(D-2)\widetilde{(\nabla \omega)^2} \nonumber \\ &&+ 4 \widetilde{(\nabla \Phi)^2} - \frac{1}{12}e^{-2\omega}\tilde H^2  \},\label{spacetime action two}\eer
where the notation with $\tilde{}$ indicates that the indices are
contracted with the metric $\tilde G$. The Einstein-Hilbert action
will be contained in (\ref{spacetime action two}) by choosing 

\be \omega = \frac{4\Phi}{D-2},\ee
in that way (\ref{spacetime action two}) can be written as

\be S = \int d^D X \sqrt{|\tilde G|} \left(\tilde R - \frac{4}{D-2}
\widetilde{(\nabla \Phi)^2} - \frac{1}{12}e^{\frac{8\Phi}{D-2}} \tilde H^2 
\right). \label{spacetime action III}\ee
In the action (\ref{spacetime action III}) can be identified a kinetic
term for the Dilaton and a Maxwell term for the antisymmetric field,
but with a coupling to the dilaton.\\


\subsection{Superstring Sigma model}
Phenomenologically, superstrings are more interesting than bosonic strings
since they contain fermions in their spectrum.
Before concentrating on the superstring sigma model, it will be given a 
brief description of the superstrings in Ramond-Neveu-Schwarz and
Green-Schwarz formalism \cite{BigJoe}, \cite{GSW}.\\

In the first half of the seventies {\it
supersymmetry} was discovered within string theory in an attempt to
construct a more realistic theory which could incorporate fermions in it's
spectrum. This more elaborated version of the string incorporating fermions is
known as the superstring, and differently from the bosonic string, which is
defined in 26 space-time dimensions to cancel the conformal anomaly, the
vanishing of the superconformal anomaly makes the superstring live in 10
space-time dimensions. The first formalism used for describing the
superstring dates back to that decade and is known a the RNS formalism standing for 
Ramond-Neveu-Schwarz. It has $ N=1 $ superconformal symmetry at the world-sheet 
level and is also space-time supersymmetric, although this feature is rather
involved. Nevertheless, the covariant quantization in this formalism is a
straightforward task. 

In the eighties superstring theory gained more interest. Green and Schwarz 
\cite{GS} showed that the theory is free of
gauge, gravitational and mixed anomalies by considering it's
low energy limit, which is $N=1$ $D=10$ super Yang-Mills theory coupled
to supergravity. Also in this decade Green and Schwarz found a new
formalism for the superstring, known as the GS formalism \cite{GSformalism}, which has manifest
space-time supersymmetry. It  has a new fermionic local symmetry at the
worldsheet level, known as {\it Kappa symmetry} \cite{SiegelKappa}, which is more involved than
the superconformal symmetry of the RNS formalism. For quantizing the GS
formalism, one has to use the light-cone gauge so the space-time symmetries
are no longer manifest. In this decade it was known the full set of
superstring theories: Type I, TypeIIA, Type IIB, Heterotic $SO(32)$ and
Heterotic $E_8 \times E_8$, which in the nineties were related to each
other using dualities \cite{Witten}.  \\

\subsubsection{Ramond-Neveu-Schwarz sigma model}
For the superstring, one can similarly consider the coupling to
background fields corresponding to massless states, giving further information
about the equations of motion for the background fields. Nevertheless,
because of the complicated supersymmetric space-time structure of the RNS formalism, 
only some sector of the possible couplings can be turned
on, namely, the NS-NS sector \cite{CFMP}  \cite{Sen}. In the conformal
gauge, the sigma model action for the
Heterotic string in the RNS formalism is
\ber \nonumber S &=& {1\over{2\pi\a'}} \int d^2 z d\t [{1\over 2}D{\bf X}^m  \pb {\bf
X}^n (G_{mn}({\bf X}) + B_{mn}({\bf X})) +
D {\bf X}^m  \Jb^I A_{m I}({\bf X}) + \t {\bf \bar\l} ^{\cal A} D {\bf \bar
\l}^{\cal A} ] \\ && + {1\over 2\pi}\int d^2 z r^{(2)} \Phi , ] \eer
where $D = \p_\t + \t \p_z$ is the $N=1$ $D=2$ supersymmetric derivative,
${\bf X}^m = X^m + \t \psi^m$ and $\Jb^I = {1\over 2} {\cal K}^I _{\cal A \cal
B} \bar \l ^{\cal A} \bar \l ^{\cal B}$ are the Heterotic string
currents that can be written in terms of the structure constants ${\cal K}^I 
_{\cal A \cal B}$ for the gauge group $E_8 \times E_8$ and right handed
fermions $\bar \l^{\cal A}$, with ${\cal A} = 1,{\ldots} ,32$. $A_m$ is the gauge potential. Besides the beta
functions already written, in the absence of the Kalb-Ramond and
Dilaton superfields, the check of conformal invariance allows to find
a beta function associated to the gauge field: $\b_{ n I}^A = \nabla^n F_{mn
I}$. The lack of manifest super-Poincar\'e invariance does not allow to
couple the RR sector, then using this formalism there are missing
equations of motion for the background fields. Therefore, the
sypersymmetrical aspects of those equations of motion are missed,
turning this formalism an inappropriate language for studying
supersymmetrical aspect such as dualities. The use of a manifestly space-time covariant formalism 
is in order. \\

\subsubsection{Green-Schwarz sigma model}
The GS formalism makes use of superspace in $10$ dimensions. For that
reason, the action written using superfields makes the supersymmetry
invariance manifest. Using the GS formalism, one can write a sigma model 
in a manifestly Super-Poincar\'e invariant form. The sigma model action is
given by \cite{GHMNT}

\be S = {1\over {2\pi\a'}}\int d^2 z \p Z^M \pb Z^N(G_{MN} + B_{MN})(Z) .  \ee
Here $Z^M$ stands for the $D=10$ $N=1$ superspace coordinates $(X^m, \t^\a)$,
with $m = 0,{\ldots} 9$ and $\a = 1,{\ldots} 16$. $G_{MN}$ and $B_{MN}$ are
superfields whose content is the supergravity multiplet. Some attempts have
been taken to compute the beta functions in this formalism, see \cite{GNZ}
\cite{GZ}. Nevertheless, its quantization in a manifest super-Poincar\'e
invariant way is an unsolved problem that rises difficulties in the
computation of the conformal anomaly. \\


Besides these two formalism, there exist a formalism suitable for
studying compactifications to four dimensions, known as the hybrid
formalism \cite{BerkovitsHF}. By using this description, the Heterotic
string and type II superstring beta functions were computed by studying the
$N=(2,0)$ and $N=(2,2)$
superconformal algebra respectively \cite{deBoerSkenderis} \cite{DNedel}
\cite{Nedel}. \\
Since one cannot couple all the background fields corresponding to  the
massless superstring states and covariantly quantize in ten dimensions neither with the RNS or 
GS formalism, one should look for a more convenient way of describing the
superstring. Fortunately, there exist a formulation in which the
super-Poincar\'e invariance is manifest and can be quantized covariantly.
This is the pure spinor formulation for the superstring \cite{Berkovits}, whose sigma model
for describing the Heterotic and type II superstrings \cite{BerkovitsUE} has been used to compute
the equations of motion for the backgrounds, giving respectively the
super Yang-Mills/supergravity equations of motion for the heterotic
case \cite{ChandiaHN} and  supergravity equations of motion for the
type II sigma model \cite{BedoyaChandia}. It is worth to note that
before pure spinors were used to describe superstrings, integrability
along pure spinor lines allowed to find the super Yang-Mills and
supergravity equations of motion in ten dimensions \cite{Howe} . Before discussing the pure spinor
sigma model, which will be done in chapter two and three, it will be useful to 
give a brief review of the pure spinor formalism in a flat background. For
detailed and pedagogical reviews, see \cite{BerkovitsICTP}
and \cite{Carlostesis}.

\section{Pure spinor formalism}
The Pure Spinor formalism has its roots in the Siegel approach for describing
the superstring \cite{siegel}. This approach had success for covariantly
quantizing the superparticle, but it could not be used to describe the
physical superstring spectrum. Nevertheless it had the advantage that all
the worldsheet fields are free, making trivial the computation of the OPE's.
Instead of describing the Siegel approach, the pure spinor description for
the Heterotic and type II superstrings will be given directly.

\subsection{Heterotic superstring in the Pure Spinor formalism}
The action for the heterotic superstring in the pure spinor formalism
 is given by 
\be S = {1\over{2\pi \a'}}\int d^2 z ({1\over 2} \p X^m \pb X_m
+ p_\a \pb \t^\a + \bar b \p \bar c) + S_\l + S_{\Jb} , \label{Heterotic}\ee
where the worldsheet variables $(X^m ,  \t^\a , p_\a)$, with $m = 0 {\ldots}
9$, $\a = 1{\ldots} 16$, describe the $N =1$ $D=10$ superspace. $p_\a$ is the
conjugate momentum to $\t^\a$. This formalism takes its name from the bosonic
spinor $\l^\a$, which is constrained to satisfy the pure spinor condition
$\l^\a (\g^m)_{\a\b} \l^\b =0$, where $\g^m$ are $16\times16$ symmetric
ten-dimensional gamma matrices. The pure spinor part of the action, denoted
by $S_\l$, is the action for a free
$\b$ $\g$ system, where the conjugate momentum to $\l^\a$ is denoted by
$\omega_\a$. $S_{\Jb}$ denotes the action for the heterotic right-moving currents
and $(\bar b,\bar c)$ are the right moving Virasoro ghosts. It is worth to note that the Lorentz currents $N^{ab} =
{1\over 2}\l \g^{ab}\omega$ and ghost number current $J = \l^\a \omega_\a $ satisfy
\be N^{mn}(y) N^{pq}(z) \rightarrow \a'{{\eta^{p[n} N^{m]q}(z) - \eta^{q[n}
N^{m]p}(z)}\over{y-z}} -3\a'^2{{\eta^{m[q}\eta^{p]n}}\over{(y-z)^2}},\ee
\be J(y) J(z) \rightarrow -{{4\a'^2}\over{(y-z)^2}} .\ee
It is worth to note that the $-3$ coefficient in the double pole, added with
a $+4$ coefficient for the double pole of the Lorentz curent in Siegel
approach $M^{mn} = {1\over 2} p \g^{mn}\t$, gives a $+1$ coefficient, which
is the same as in the Lorentz current of the RNS formalism.
These currents have OPEs with the pure spinors
\be N^{mn}(y) \l^\a (z) \rightarrow {1\over 2} \a'(\g^{mn})^\a{} _\b
{{\l^\b (z)}\over{y-z}} , \,\,\,\,\,\, J(y) \l^\a (z) \rightarrow \a'{{\l^\a
(z)}\over{y-z}},\ee
while the right-moving currents satisfy
\be \Jb^I (y)\Jb^J (z) \rightarrow \a'{{f^{IJ}{}_K \Jb^K
(z)}\over{\bar y - \bar z}} + \a'^2 {{\d^{IJ}}\over{(\bar y - \bar z)^2}}, \ee
where $f^{IJ}{}_K$ are the $E_8 \times E_8$ structure constants.
Physical states are defined as vertex operators in the cohomology of the
BRST charge\footnote {For a reference of BRST quantization of the
superstring and a proof of equivalence of the pure spinor formalism and GS
formalism, see \cite{DafniTese} and \cite{BerkovitsMarchioro}.} $Q = \oint dz \l^\a d_\a$ and $\bar Q = \oint (\bar c \bar T +
\bar c \pb \bar c \bar b)$, where $d_\a$ are the worldsheet
variables corresponding to $N=1$ $D=10$ space-time supersymmetric
derivatives and is given by 

\be d_\a = p_\a -{i \over 2} \g^m _{\a\b} \t^\b \p x_m + {1\over 8} \g^m
_{\a\b}(\g_m)_{\g\d} \t^\b \t^\g \p \t^\d. \ee

\subsection{Type II superstring in the Pure Spinor formalism}
The pure spinor closed string action in flat space-time is defined
by using the superspace coordinates $X^m$ with $m = 0, \dots, 9$ and the
conjugate pairs $( p_\a, \t^\a) , (\widetilde{p}_{\ab}, \tt^{\ab})$
with $(\a, \ab) = 1, \dots, 16$. For the type IIA superstring the
spinor indices $\a$ and $ \ab$ have the opposite chirality while for
the type IIB superstring they have the same chirality. In order to
define a conformal invariant system we need to include a pair of
pure spinor ghost variables $(\l^\a, \omega_\a)$ and $(\lt^{\ab},
\widetilde\omega_{\ab})$. These ghosts are constrained to satisfy the pure spinor
conditions $(\l \g^m \l) = (\lt \g^m \lt) = 0$, where $\g^m_{\a\b}$
and $\g^m_{\ab\bb}$ are the $16\times 16$ symmetric ten dimensional
gamma matrices. Because of the pure spinor conditions, $\omega$ and $\widetilde\omega$ are
defined up to $\d\omega = (\l\g^m) \L_m$ and $\d\ot = (\lt\g^m) \Lt_m$.
The quantization of the model is performed after the construction of
the BRST-like charges $Q = \oint \l^\a d_\a, \Qt = \oint \lt^{\ab}
\dt_{\ab}$, here $d_\a$ and $\dt_{\ab}$ are the world-sheet
variables corresponding to the $N=2 \,\, D=10$ space-time
supersymmetric derivatives and are supersymmetric combinations of
the space-time superspace coordinates of conformal weights $(1,0)$
and $(0,1)$ respectively. The action in flat space is a free action
involving the above fields, that is

\be  S = {1\over{2\pi\a'}} \int d^2 z ~ ({1\over 2} \p X^m \pb
X_m + p_\a \pb \t^\a + \widetilde{p}_{\ab} \p \tt^{\ab}) +
S_{pure},\label{actionflat}\ee where
$S_{pure}$ is the action for the pure spinor ghosts.

The left $N^{mn} = {1\over 2} \l \g^{mn}\omega$ and right-moving currents
$\widetilde N^{mn} = {1\over 2} \lt \g^{mn}\widetilde\omega$ satisfy the OPE's
\be N^{mn}(y) \l^\a (z) \rightarrow {1\over 2} \a'(\g^{mn})^\a{} _\b
{{\l^\b (z)}\over{y-z}} , \,\,\,\,\,\, \widetilde N^{mn}(y) \lt^{\ab} (z)
\rightarrow {1\over 2} \a'(\g^{mn})^{\ab}{} _{\bb}
{{\lt^{\bb} (z)}\over{\bar y-\bar z}}\ee

\be N^{mn}(y) N^{pq}(z) \rightarrow \a'{{\eta^{p[n} N^{m]q}(z) - \eta^{q[n}
N^{m]p}(z)}\over{y-z}} -3\a'^2{{\eta^{m[q}\eta^{p]n}}\over{(y-z)^2}},\ee

\be \widetilde N^{mn}(y) \widetilde N^{pq}(z) \rightarrow \a'{{\eta^{p[n}
\widetilde N^{m]q}(z) - \eta^{q[n}
\widetilde N^{m]p}(z)}\over{\bar y-\bar z}}
-3\a'^2{{\eta^{m[q}\eta^{p]n}}\over{(\bar y-\bar z)^2}},\ee \\

Having a covariantly quantized description for the superstring brings
important advantages. Scattering amplitudes have been computed up to two
loops, \cite{BerkovitsMulti}, \cite{BerkovitsAmpII}, \cite{AGV},
\cite{BerkovitsVallilo}, \cite{Mafra1loop}, \cite{BerkovitsMafraEquiv} and  
\cite{BerkovitsMafra}. Also it has been possible to study the superstring in a curved background more
properly, that means, including the full $D=10$ $N=1$ supermultiplet and
finding their equations of motion. In this area, is of importance to know
the effective field theories for the massless modes of the string. One
reason is to know the genuine effective stringy effects in the theory. A
second reason is that it would be possible to test duality conjectures
beyond the leading order and also would be interesting to know the effects
of the string corrections on the solutions of the supergravity equations of
motion. \\


As emphasized, these Ph.D thesis relays on the area of the non-linear sigma
model for the superstring in the pure spinor description. The coupling of
the pure spinor superstring to a generic background, including RR fields, was
given for the first time by Berkovits and Howe \cite{BerkovitsUE}, where
also a set of $D = 10$ $N=1$ super-Yang-Mills and supergravity constraints
were computed in the heterotic superstring case by studying the nilpotency of
the pure spinor BRST charge and the conservation of its respective BRST
current. Also they found a set $D=10$ $N=2$ supergravity
constraints analog considerations for the type II superstring in a
generic background. Both the heterotic and type II pure spinor superstrings 
in a generic background will be reviewed in chapter 2 and 3. For the open superstring in the pure spinor
description coupled to a background, it was shown that the classical
BRST invariance implies that the background fields satisfy the full
non-linear supersymmetric Born-Infeld equations of motion
\cite{BerkovitsPershin}. The one-loop beta functions for the heterotic superstring using the pure spinor formalism
were computed by Chandia and Vallilo \cite{ChandiaHN}. These authors also
show that the sYM/supergravity constraints makes the beta functions to be
zero, implying in conformal invariance at one-loop. In collaboration with
Chandia \cite{BedoyaChandia} the one-loop beta functions for the type II
superstring were computed, and also were verified the conformal
invariance of this theory by using the lowest order $D=10$ $N=2$
supergravity constraints. This will be developed in detail in chapter 3 of
this thesis. It is worth to note that using the pure spinor formalism,
the full superfield multiplet can be coupled to the superstring. So this
allowed to compute covariantly the equations of motion for the
background superfields, even the RR superfields in a manifestly covariant 
manner\footnote{For further studies of the pure spinor superstring in a
generic background see \cite{Kluson} and references therein. }.\\

There is one more study that can be performed using the superstring
sigma model using the pure spinor formalism. The Green-Schwarz mechanism 
demands an anomalous transformation of the the Kalb-Ramond superfield
\cite{GS}, 
which amounts to an $\a'$ order Chern-Smons modifications of the field-strength related to this
superfield. It will now be explained how to compute such $\a'$
corrections. Using the pure spinor sigma model it was shown at the lowest order in
$\a' $, that the BRST invariance puts the background fields on-shell
\cite{BerkovitsUE}. It is in the quantum regime of the BRST invariance
that it is expected to find the Chern-Simons
modifications. These Chern-Simons
modifications are of two types: Yang-Mills and Lorentz, as is known since the
Green-Schwarz mechanism for the cancellation of gauge, gravitational and mixed
anomalies in the framework of the ten-dimensional low-energy effective field 
theories \cite{GS}. Also Hull 
and Witten \cite{HullWitten} noted the appearance of those modifications to cancel the sigma
model gauge anomalies. Atick, Dhar and Ratra \cite{AtickDharRatra} gave further
evidence for the existence of the Yang-Mills Chern-Simons modification by making a Superspace
description of ${\cal N} =1$ supergravity coupled to ${\cal N }=1$ super
Yang-Mills. The Chern-Simons modifications were even noted in the
component formulation of supergravity in order to have a consistent
theory \cite{ChaplineManton}. Furthermore, integrability along pure
spinor lines allowed Howe \cite{HoweCS} to incorporate the
Chern-Simons corrections. By studying what conditions are imposed on the background
superfields by  the preservation of the BRST invariance properties at first order in
$\alpha'$ for the Heterotic sigma model in the pure spinor description, the Yang-Mills 
Chern-Simons modifications have been computed in as-yet unpublished work
\cite{Bedoya} which will be described in detail in chapter 4. The Lorentz Chern-Simons 
modifications constitute work in progress and will be discussed in
chapter 5. Both Yang-Mills and Lorentz
Chern-Simons modifications appear as stringy corrections to some of the
classical SYM/SUGRA constraints mentioned in the preceding paragraph. Besides the Chern-Simons 
modifications, other corrections are expected to preserve
supersymmetry. Being a manifestly supersymmetric, it seems promising that the
pure spinor sigma model would be useful to find a complete $\a'$
correction preserving supersymmetry in space-time. This will help to
settle an old debate found in the
literature, discussed in some works of Gates et al.
\cite{BellucciGates} , \cite{BellucciDepireuxGates} and 
\cite{GatesKissMerrell} on one hand and Bonora et al.
\cite{AuriaFreRaciti}, \cite{RacitiTivaZanon},
\cite{BonoraPastiTonin}, \cite{BonoraBregolaLechnerPastiTonin} and
\cite{Bonoraetal} where 
two sets of string corrected constraints cannot be related among them.\\
Recently a new set of supergravity constraints have been
introduced by Lechner and Tonin \cite{LechnerTonin} and it will be of 
interest to compare their $\a' $ corrections with those computed directly 
from the pure spinor superstring.

\chapter{Ten-Dimensional Supergravity Constraints from the Pure Spinor
Formalism for the Heterotic Superstring}

Before discussing the conformal anomaly and the gauge and Lorentz anomaly for the
superstring, it will be introduced the sigma-model type action in the 
pure spinor formalism \cite{BerkovitsUE} for the heterotic superstring. As in the bosonic string case \cite{BigJoeI}, 
the starting point for costructing a sigma-model action are the 
integrated vertex operators corresponding to the massless states. This
chapter is fully based on \cite{BerkovitsUE} and the pourpose of including
it in the thesis is to make the text more complete and set notation, instead
of being original in this topic. 

\section{Vertex Operators in the pure spinor formalism}
The massless supergravity and super Yang-Mills vertex operators are
respectively given by

\be V_{SG} = \int d^2 z [\p \t^\a A_{\a m}(x,\t)+\Pi^n A_{nm}(x,\t) + d_\a E_m
^\a (x\t) + N_{np} \O_m {}^{np}] \pb x^m, \label{VSG}\ee 

\be V_{sYM} = \int d^2 z [\p \t^\a A_{\a I}(x,\t)+\Pi^n A_{nI}(x,\t) + d_\a
W_I ^\a (x\t) + N_{np} U_I{}^{np}] \Jb^I,\label{VsYM}\ee 
where the last two terms in each vertex operators are present to make them
BRST invariant and $\Pi^m = \p x^m + {i\over 2}\t \g^m \p \t$. Note that any of this two vertex operators could be
constructed from the open string vertex operator

\be V_{open}  = \int d z [\p \t^\a A_\a (x,\t)+\Pi^n A_n(x,\t) + d_\a W^\a 
(x\t) + N_{np} U^{np}] \ee
multiplying either with $\int d \bar z \pb x^m$ or $\int d \bar z \Jb$. By computing the
conditions for BRST invariance of (\ref{VSG}) one finds

\be (\g_{npqrs})^{\a\b} D_\a A_{\b m} =0, \qquad \p^m (\p_m A_{\b n} - \p_n
A_{\b m})=0,  \label{LSGEOM}\ee
which are the linearized $N=1$ supergravity equations motions, and also
\be  A_{nm} = -{i\over 8} D_\a (\g_n )^{\a\b} A_{\b m}, \,\,\, E_{m}{}^\b = -
{i\over 10}(\g^n )^{\a\b} (D_\a A_{nm} - \p_n A_{\a m}), \label{LSGFS} \ee $$
\O_{m}{}^{np} = {1\over 8} D_\a (\g^{np})^\a {}_\b E_m{}^\b  = \p_{[n}A_{p
]m}, $$
which defines the linearized supergravity connections and field-strengths in
terms of $A_{\a m}$.
Similarly, the BRST invariance of (\ref{VsYM}) leads to the linearized $N=1$
super Yang-Mills equation of motion 
\be (\g_{npqrs})^{\a\b} D_\a A_{\b I} =0,  \label{LsYMEOM}\ee
as can be read-off from the condition $\l^\a \l^\b D_{\a} A_{\b I} =0$ using the pure spinor condition; and also to the definitions of the linearized super Yang-Mills connections and
field strengths in terms of $A_{\a I}$:

\be  A_{nI} = -{i\over 8} D_\a (\g_n )^{\a\b} A_{\b I}, \,\,\, W_{I}{}^\b = -
{i\over 10}(\g^n )^{\a\b} (D_\a A_{nI} - \p_n A_{\a I}), \label{LsTMFS} \ee 
$$ U_{I}{}^{np} = {1\over 8} D_\a (\g^{np})^\a {}_\b W^\b _I = \p_{[n}A_{p
]I}.$$

The on-shell graviton $h_{mn}$ is contained in the $(\g^n \t)_\a
h_{mn}(x)$ of $A_{\a m}(x,\t)$, while the on-shell gluon $a_{nI}$ is in the
$(\g^n \t)_\a a_{nI}(x)$ of $A_{\a I}(x,\t)$.

By considering the coupling of the superstring to a generic background,
(\ref{LSGEOM})-(\ref{LsTMFS}) will be generalized to covariant non-linear
equations.

\section{Heterotic Superstring in a Generic Background}

By adding the supergravity and super Yang-Mills vertex operators (\ref{VSG})
and (\ref{VsYM}) to the flat action (\ref{Heterotic}) and covariantizing
respect to $N = 1$ $D=10$ super-reparameterization invariance, one can
arrive to an action for the coupling of the Heterotic superstring to a
curved background. Also, one can consider the worldsheet fields $\p x^m$,
$\pb x^m$, $\p \t^\a$, $\pb \t^\a$, $d_\a$, $\Jb$ and $\l^\a \omega_\b$. Then, by
making products among them, one can write an expression which is classically invariant 
under worldsheet conformal transformations. The action 
is given by

\ber  S &=& {1\over {2\pi\a'}}\int d^2 z [{1\over 2}\p Z^M \pb Z^N(G_{NM}+
B_{NM}) + d_\a \pb Z^M E_{M}{}^\a + \p Z^M \Jb^I A_{MI}
\label{HeteroticCurved}  \\ && + d_\a \Jb^I W_I ^\a
+ \l^\a \omega_\b \Jb^I U_{I\a}{}^\b + \l^\a \omega_\b \pb Z^M \O_{M \a}{}^\b ]+ S_{FT} +
S_{ghost} + S_{\l} + S_{\Jb } . \nonumber \eer
In this notation $Z^M = (x^m , \t^\mu)$ are coordinates for the superspace.
Middle alphabet indices denote the curved superspace indices, while beginning
alphabet indices $a = (a,\a)$ denote the tangent superspace indices. The set
of background superfields is given by $G_{MN}$,  $B_{MN}$, $E_{M}{}^\a $, 
$A_{MI}$,  $ W_I^\a $,  $U_{I\a}{}^\b $, $\O_{M\a}{}^\b$ and $\Phi $. In
terms of the supervielbein $E_M {}^A$, $G_{MN}$ is given by $G_{MN} = E_{M}{}^a E_{N}{}^b
\eta_{ab}$. $B_{MN}$ is the two-form
Kalb-Ramond potential, and $\Phi$ is the dilaton. $A_{MI}$ is the super
Yang-Mills potential, while $W_I ^\a$ and $U_{I\a}{}^\b$ will be related to
the super Yang-Mills field strengths. $S_{FT}$ denotes the Fradkin-Tseytlin action $S_{FT} = {1\over{4\pi}}\int d^2 z \Phi(z) r^{(2)}$, where $r^{(2)}$ is the two-dimensional scalar curvature. Finally, $\O_{M\a}{}^\b$ is the sping
connection superfield. Because of the form that $\omega_\a$ appears in
(\ref{HeteroticCurved}) and the pure spinor condition, there is a gauge
invariance $\d \omega_\a = \L^a (\g_a \l)_\a$, so the background
superfields satisfy $(\g^{bcde})_\b {}^\a \O_{M\a}{}^\b = (\g^{bcde})_\b
{}^\a U_{I\a}{}^\b =0 $, which imply
\be \O_{M\a}{}^\b = \O_M ^{(s)}\d_\a {}^\b + {1\over 4}\O_M{}^{cd}
(\g_{cd})_\a {}^{\b} , \,\,\, U_{I\a}{}^\b = U_I ^{(s)} \d_\a {}^\b + {1\over
4} U_{I}{}^{cd} (\g_{cd})_\a {}^{\b}.  \ee

The action (\ref{HeteroticCurved}) is invariant under local gauge
transformations

\ber &&\nonumber  \d E_M {}^b = \eta_{cd}\L^{bc} E_M {}^d ,\,\, \d E_M {}^\a =
\Sigma_\b ^\a E_M {}^\b ,\,\, \d \O_{M\a}{}^\b = \p_M \Sigma_\a {}^\b + \Sigma _\a ^\g
\O_{M\g}{}^\b - \Sigma_\g ^\b \O_{M\a}{}^\g ,\,\, \\ &&
\d W_I ^\a = \Sigma_\g ^\a W_I ^\g ,\,\, \d U_{I \a}{}^\b  = \Sigma_\a ^\g
U_{I\g}{}^\b - \Sigma_\g ^\b U_{I\a}{}^\g, \,\, \d \l^\a = \Sigma_\g ^\a
\l^\g ,\,\, \d\omega_\a = - \Sigma_\a ^\g \omega_\g ,\eer
as well as under local shift transformations.

\be \d \O_\a ^{(s)} = 4 (\g_c)_{\a\b} h^{c\b} ,\,\, \d\O_\a {}^{bc} = 2
(\g^{[b})_{\a\b} h^{c]\b}, \,\, \d d_\a = -\d \O_{\a\b}{}^\g \l^\b \omega_\g
,\,\, \d U_{I\a}{}^\b = W_I ^\g \d\O_{\g\a}{}^\b , \ee 
where the transformation of $\O_{\a\b}{}^\g$ has been chosen in such a way for not to change the pure spinor BRST current.

\subsection{Heterotic Nilpotency Constraints}
The constraints found by requiring that the BRST charge remains nilpotent
when the string is coupled to a curved background can be found either by
using canonical commutation relations \cite{BerkovitsUE}, computing directly
twice the BRST variation on various worldsheet fields in (\ref{HeteroticCurved})
\cite{ChandiaIX} or by
a tree level computation, as explained in chapter 4. In this section we use
the first approach, with a commutator algebra
\be [P_M , Z^N] = \d_M {}^N , \,\,\, [\omega_\alpha ,\l^\b] = \d_\a ^\b
,\,\, [\Jb^I , \Jb^J] = f^{IJ}{}_K\Jb^K , \label{PBAlgebra}\ee
where the canonical momentum is defined as usual $P_M = \d L / \d (\p_0
Z^M)$. By computing this momentum one finds

\be d_\a = E_{\a}{}^M (P_M + {1\over 2}(\p Z^N - \pb Z^N)B_{NM}
-\l^\d \omega_\b \O_{M\d}{}^\b -\Jb^I A_{MI}).\ee

Then, one can use the commutators algebra (\ref{PBAlgebra}) to find 
\be \{Q , Q\} = \oint \l^\a \l^\b [T_{\a\b}{}^C D_C - {1\over 2}(\p Z^N -
\pb Z^N)H_{\a\b N} - \l^\g \omega_\d R_{\a\b\g}{}^\d - \Jb^I F_{\a\b I}],
\label{HetNilpotencyT}\ee
where $D_C = E_C {}^M (P_M - \l^\a \omega_\b \O_{M \a}{}^\b -\Jb^I A_{MI}).$

From (\ref{HetNilpotencyT}) one can read the nilpotency constraints
\be \l^\a \l^\b T_{\a\b}{}^C = \l^\a \l^\b H_{\a\b C} = \l^\a \l^\b \l^\g
R_{\a\b\g}{}^\d = \l^\a \l^\b F_{\a\b I} =0.\ee

\subsection{Heterotic Holomorphicity Constraints}
In this subsection it will be computed the conditions for $\pb(\l^\a d_\a )=0$ at 
the lowest order. Again there are three possible ways to compute this
constraints. One is by using the classical equations of motion derived for
the worldsheet fields in (\ref{HeteroticCurved}) \cite{BerkovitsUE}, by
computing directly the BRST variation of this action \cite{ChandiaIX} or by
computing tree level diagrams, as will be shown in chapter four. In this
chapter it will be followed the first approach.

By variating $\l^\a$ and $\omega_\a$ in (\ref{HeteroticCurved}) one obtains
respectively
\be \pb \omega_\a = -(\pb Z^M \O_{M\a}{}^\b + \Jb^I U_{I\a}{}^\b)\omega_\b, 
\,\,\, \pb \l^\a = (\pb Z^M \O_{M\b}{}^\a + \Jb^I U_{I\b}{}^\a)\l^\b.
\label{EOMlo}\ee

The equations of motion for the right-moving Heterotic currents can be found
by using bosonization. The result is 

\be \p \Jb^I = f^{IJ}{}_K \Jb^K (\p Z^M A_{MJ} + d_\a W_J ^\a + \l^\a
\omega_\b U_{J\a}{}^\b).\label{EOMJ}\ee
Finally, by computing the variation of (\ref{HeteroticCurved}) with respect of superspace 
coordinates $Z^M$ one finds
\be \pb d_\a = E_{\a}{}^P [(\p_{[P}E_{M]}{}^a)E_N {}^b \eta_{ab} + \p_{[P}
E_{N]}{}^a E_M {}^b \eta_{ab} - {1\over 2} H_{PMN}) \p Z^M \pb Z^N \label{EOMZ}\ee
$$ +2(\p_{[P} E_{N]}{}^\b) d_\b + \p_{[P}\O_{N]\g}{}^\b \l^\g \omega_\b) \pb
Z^N -\O_{P\g}{}^\b \pb (\l^\g \omega_\b) - A_{PI}\p \Jb^I $$
$$ + (2\p_{[P}A_{M]I}\p Z^M + \p_P W_I ^\b d_\b + \p_P U_{I\g}{}^\b \l^\g
\omega_\b)\Jb^I ] .$$
So, by using (\ref{EOMlo}), (
\ref{EOMJ}) and (\ref{EOMZ}) one finds the that
derivative in the $\bar z$ direction of the BRST current is

\be \pb(\l^\a d_\a) = \l^\a [\Pi^b \Pb^c (T_{\a bc}+ T_{\a cb} - H_{\a bc})]
+ {1\over 2} \Pi^\b \Pb^c (T_{\a\b c} - H_{\a\b c}) + d_\b \Pb^c T_{\a
c}{}^\b + \l^\b \omega_\g \Pb^c R_{\a c\b}{}^\g ]\ee
$$ + \l^\a [\Pi^b \Jb^I (F_{\a bI} + {1\over 2}W_I ^\b (T_{\a\b b} - H_{\a\b
b}))+ \Pi^\g \Jb^I(F_{\a\g I} + {1\over 2}W_I ^\b H_{\a\g\b})] $$
$$+ \l^\a [d_\b \Jb^I(\nabla_\a W_I ^\b - T_{\a\g}{}^\b W_I ^\g - U_{I\a}{}^\b )
+ \l^\g \omega_\d \Jb^I(\nabla_\a U_{I\g}{}^\d - R_{\a\b\g}{}^\d) ], $$
where $\Pi^A = \p Z^M E_M {}^A$, $\Pb^A = \pb Z^M E_M {}^A$ and $T_{AB
c}\equiv T_{AB}{}^d \eta_{dc} $.

Since $\Pb^\a$ is related to $\Jb^I$ through $\Pb^\a = - \Jb^I W_I^\a$ by
using the equation of motion for the worldsheet field $d_\a$ in
(\ref{HeteroticCurved}), we arrive at the following set of constraints for
holomorphicity of the BRST current at the lowest order in $\a'$
\be T_{\a (bc)} = - H_{\a bc} = T_{\a\b}{}^c - H_{\a\b}{}^c = T_{c\a}{}^\b
 = 0, \,\, \l^\a \l^\b R_{d\a\b}{}^\g =0, \,\, F_{\a\b I} = - {1\over 2}W_I^\g
 H_{\g\a\b}, \ee 
 $$ F_{\a bI} = - W_I^\g T_{\g\a b} , \,\,  \N_\a W_I^\b -
 T_{\a\g}{}^\b W_I^\g = U_{I\a}{}^\b , \,\, \l^\a \l^\b (\N_\a U_{I\b}{}^\g +
 R_{\a\g\b}{}^\d W_I ^\g) =0 .$$

It will be explained in chapter 4 how to compute those constraints
perturbatively in $\a'$.

In the following chapter, it will be discussed the pure spinor sigma model 
for the type II superstring.

\chapter{One-loop Conformal Invariance of the Type II Pure Spinor
Superstring in a Curved Background }

Having gained experience with the Heterotic sigma model, the type II
sigma model will be introduced and the conditions for conformal
invariance will be computed. At the end of the chapter it will be 
shown how the classical constraints imply in the equations of motion
for the background.

\section{Classical Considerations}
In a curved background, the pure spinor sigma model action for the
type II superstring is obtained by adding to the flat action of
(\ref{actionflat}) the integrated vertex operator for supergravity
massless states and then covariantizing respect to ten dimensional
$N = 2$ super-reparameterization invariance. The result of doing
this is

\be S = {1\over{2\pi\a'}} \int d^2 z ~ ({1\over 2} \Pi^a \Pb^b
\eta_{ab} + {1\over 2} \Pi^A \Pb^B B_{BA} + d_\a  \Pb^\a + \dt_{\ab}
\Pi^{\ab} + (\l^\a \omega_\b) \Ob_\a{}^\b + (\lt^{\ab} \widetilde{\omega}_{\bb})
\Ot_{\ab}{}^{\bb} \label{action} \ee $$  + d_\a \dt_{\bb} P^{\a\bb} + (\l^\a \omega_\b)
\dt_{\gb} C_\a{}^{\b\gb} + (\lt^{\ab} \widetilde\omega_{\bb}) d_\g
\Ct_{\ab}{}^{\bb\g} + (\l^\a \omega_\b)  (\lt^{\ab} \widetilde\omega_{\bb})
S_{\a\ab}{}^{\b\bb} )+ S_{pure} + S_{FT},$$
where $\Pi^A = \p Z^M
E_M{}^A, \Pb^A = \pb Z^M E_M{}^A$ with $E_M{}^A$ the supervielbein
and $Z^M$ are the curved superspace coordinates, $B_{BA}$ is the
super two-form potential. The connections appears as $\Ob_\a{}^\b =
\pb Z^M \O_{M\a}{}^\b = \Pb^A \O_{A\a}{}^\b$ and $\Ot_{\ab}{}^{\bb}
= \p Z^M \Ot_{M\ab}{}^{\bb} = \Pi^A \Ot_{A\ab}{}^{\bb}$. They are
independent since the action of (\ref{action}) has two independent Lorentz
symmetry transformations. One acts on the $\a$-type indices and
the other acts on the $\ab$-type indices. $S_{pure}$ is the
action for the pure spinor ghosts and is the same as in the flat
space case of (\ref{actionflat}).

As was shown in \cite{BerkovitsUE}, the gravitini and the dilatini
fields  are described by the lowest $\t$-components of the
superfields $C_{\a}{}^{\b\gb}$ and $\Ct_{\ab}{}^{\bb\g}$, while the
Ramond-Ramond field strengths are in the superfield $P^{\a\bb}$. The
dilaton is the theta independent part of the superfield $\Phi$ which
defines the Fradkin-Tseytlin term

\be  S_{FT} = {1\over{2\pi}} \int d^2z ~ r ~ \Phi, \label{ft} \ee where $r$ is
the world-sheet curvature. Because of the pure spinor constraints, the
superfields in \ref{action} cannot be arbitrary. In fact, because of the gauge invariances $\d \omega_\a = \L^a (\g_a \l)_\a$ and $\d \widetilde\omega_{\ab} = \bar \L^a (\g_a \lt)_{\ab}$ one can find 

\be \O_{A\a}{}^\b = \O_A \d_\a{}^\b + {1\over 4} \O_{Aab}
(\g^{ab})_\a{}^\b,\quad \Ot_{A\ab}{}^{\bb} = \Ot_A \d_{\ab}{}^{\bb}
+ {1\over 4} \Ot_{Aab} (\g^{ab})_{\ab}{}^{\bb},\label{fields} \ee
\be 
C_\a{}^{\b\gb} = C^{\gb} \d_\a{}^\b + {1\over 4} C_{ab}{}^{\gb}
(\g^{ab})_\a{}^\b,\quad \Ct_{\ab}{}^{\bb\g} = \Ct^\g
\d_{\ab}{}^{\bb} + {1\over 4} \Ct_{ab}{}^\g
(\g^{ab})_{\ab}{}^{\bb},\ee
\be
S_{\a\ab}{}^{\b\bb} = S \d_\a{}^\b \d_{\ab}{}^{\bb} + {1\over 4}
S_{ab} (\g^{ab})_\a{}^\b \d_{\ab}{}^{\bb} + {1\over 4} \St_{ab}
(\g^{ab})_{\ab}{}^{\bb} \d_\a{}^\b + {1\over{16}} S_{abcd}
(\g^{ab})_\a{}^\b (\g^{cd})_{\ab}{}^{\bb}.\ee

The action of (\ref{action}) is BRST invariant if the background fields
satisfy suitable constraints. As was shown in \cite{BerkovitsUE}, these
constraints imply that the background field satisfy the type II
supergravity equations. The BRST invariance is obtained by requiring
that the BRST currents $j_B = \l^\a d_\a$ and $\jt_B = \lt^{\ab}
\dt_{\ab}$ are conserved. Besides, the BRST charges $Q = \oint j_B$
and $\Qt = \oint \jt_B$ are nilpotent and anticommute. Let us review
these properties now.

\subsection{Nilpotency}

As was shown in \cite{BerkovitsUE} (see also
\cite{ChandiaIX}), nilpotency is obtained
after defining momentum variables in (\ref{action}) and then using the
canonical Poisson brackets. The only momentum variable that does
not appear in (\ref{action}) is the conjugate momentum of $Z^M$ which is
defined as $P_M = (2\pi\a') \d S / \d (\p_0 Z^ M)$ where $\p_0 =
{1\over 2} ( \p + \pb )$. It is not difficult to see that $\omega_\a$ is the
conjugate momentum to $\l^\a$ and that $\ot_{\ab}$ is the one for
$\lt^{\ab}$. Nilpotence of $Q$ determines the constraints

\be \l^\a \l^\b H_{\a\b A} = \l^\a \l^\b \l^\g R_{\a\b\g}{}^\d
 = \l^\a \l^\b \Rt_{\a\b\gb}{}^{\db} = 0, \label{qdos} \ee
\be 
\l^\a \l^\b T_{\a\b}{}^a = \l^\a \l^\b T_{\a\b}{}^\g = \l^\a \l^\b
T_{\a\b}{}^{\gb} = 0,\ee 
where $H=dB$, the torsion $T_{AB}{}^\a$
and $R_{AB\g}{}^\d$ are the torsion and the curvature constructed
using $\O_{A\b}{}^\g$ as connection. Similarly, $T_{AB}{}^{\gb}$
and $\Rt_{AB\gb}{}^{\db}$ are the torsion and the curvature using
$\Ot_{A\bb}{}^{\gb}$ as connection.

The nilpotence of the BRST charge $\Qt$ leads to the constraints

\be\lt^{\ab} \lt^{\bb} H_{\ab\bb A} = \lt^{\ab}
\lt^{\bb} R_{\ab\bb\g}{}^\d = \lt^{\ab} \lt^{\bb} \lt^{\gb}
\Rt_{\ab\bb\gb}{}^{\db} = 0, \label{qtildos} \ee
\be
\lt^{\ab} \lt^{\bb} T_{\ab\bb}{}^a = \lt^{\ab} \lt^{\bb}
T_{\ab\bb}{}^\g = \lt^{\ab} \lt^{\bb} \Tt_{\ab\bb}{}^{\gb} = 0. \ee
Finally, the anticommutation between $Q$ and $\Qt$ determines

\be H_{\a\bb A} = T_{\a\bb}{}^a = T_{\a\bb}{}^\g =
T_{\a\bb}{}^{\gb} = \l^\a \l^\b R_{\gb\a\b}{}^\d = \lt^{\ab}
\lt^{\bb} \Rt_{\g\ab\bb}{}^{\db} = 0. \label{qqtil} \ee

Note that given the decomposition (\ref{fields}) for the connections, we
can respectively write

\be R_{DC\a}{}^\b = R_{DC}\d_\a{}^\b + {1\over 4}
R_{DCef} (\g^{ef})_\a{}^\b, \ee
\be \Rt_{DC\ab}{}^{\bb} = \Rt_{DC}\d_{\ab}{}^{\bb} + {1\over
4}\Rt_{DCef}(\g^{ef})_{\ab}{}^{\bb}. \label{decurvatures} \ee

\subsection{Holomorphicity}

The holomorphicity of $j_B$ and the antiholomorphicity of $\jt_B$
constraints are determined after the use of the equations of motion
derived from the action (\ref{action}). The equation for the pure spinor
ghosts are

\be \Nb\l^\a + \l^\b ( \dt_{\gb} C_\b{}^{\a\gb} +
\lt^{\ab} \widetilde\omega_{\bb} S_{\b\ab}{}^{\a\bb} ) = 0,\quad \Nb\omega_\a - (
\dt_{\gb} C_\a{}^{\b\gb} + \lt^{\ab} \widetilde\omega_{\bb} S_{\a\ab}{}^{\b\bb}
)\omega_\b = 0, \label{pureseom} \ee and

\be \N\lt^{\ab} + \lt^{\bb} ( d_\g \Ct_{\bb}{}^{\ab\g} +
\l^\a \omega_\b S_{\a\bb}{}^{\b\ab} ) = 0,\quad \N\widetilde\omega_{\ab }- ( d_\g
\Ct_{\ab}{}^{\bb\g} + \l^\a \omega_\b S_{\a\ab}{}^{\b\bb} )\widetilde\omega_{\bb} =
0, \label{gheom} \ee 
where $\N$ is a covariant derivative which acts with $\O$ or
$\Ot$ connections according to the index structure of the fields it
is acting on. For example,

$$
\N P^{\a\bb} = \p P^{\a\bb} + P^{\g\bb} \O_\g{}^\a + P^{\a\gb}
\Ot_{\gb}{}^{\bb}.$$ The variations respect to $d_\a$ and
$\dt_{\ab}$ provide the equations

\be \Pb^\a + \dt_{\bb} P^{\a\bb} + \lt^{\ab} \widetilde\omega_{\bb}
\Ct_{\ab}{}^{\bb\a} = 0,\quad \Pi^{\ab} - d_\b P^{\b\ab} + \l^\a
\omega_\b C_\a{}^{\b\ab} = 0. \label{pieom} \ee
The most difficult equations to obtain
are those coming from the variation of the superspace coordinates.
Let us define $\s^A = \d Z^M E_M{}^A$, then it is not difficult to
obtain

$$
\d \Pi^A = \p \s^A - \s^B \Pi^C E_B{}^M E_C{}^N \p_{[N} E_{M]}{}^A
(-1)^{C(B+M)}.$$ Here we can express this variation in terms of
the connection $\O$ . In fact,

$$
\d \Pi^A = \N \s^A - \s^B \Pi^C ( T_{CB}{}^A + \O_{BC}{}^A
(-1)^{BC} ). $$
There is a point about our notation for the torsion that we should make
clear. Using tangent superspace indices, the torsion can be written
as
\be T_{BC}{}^A = -E_B{}^N (\p_N E_C{}^M)E_M{}^A +(-)^{BC}E_C{}^N
(\p_N E_B{}^M)E_M{}^A +\O_{BC}{}^A -(-)^{BC}\O_{CB}{}^A .\label{TorsionI} \ee
In our notation, $T_{BC}{}^\a$ will mean that the connection in (\ref{TorsionI})
is $\O_{C\b}{}^\a$ while $T_{BC}{}^{\ab}$ means that the connection in (\ref{TorsionI})
 is $\Ot_{C\bb}{}^{\ab}$. Since we also have two connections with bosonic
tangent space index $\O_{Cb}{}^a$ and $\Ot_{Cb}{}^a$, we use $T_{BC}{}^a$ to
denote the torsion when we use the first and $\Tt_{BC}{}^a$ to denote the
torsion when we use the second.

We vary the action (\ref{action}) under these transformations and, after
using the equations (\ref{gheom}), (\ref{pieom}) and some of the nilpotence
constraints, we obtain

\ber \Nb d_\a &=& -{1\over 2} \Pi^a \Pb^b ( T_{\a(ba)} + H_{\a ba} ) +
{1\over 2} \Pi^\b \Pb^a ( T_{\b\a a} - H_{\b\a a} ) - d_\b \Pb^a
T_{a\a}{}^\b \label{deom} \\  && - \dt_{\bb} \Pi^a ( T_{a\a}{}^{\bb} + {1\over 2}
P^{\g\bb} ( T_{\g\a a} + H_{\g\a a} ) ) + \l^\b \omega_\g \Pb^a
R_{a\a\b}{}^\g \nonumber \\ && + \lt^{\bb} \widetilde\omega_{\gb} \Pi^a ( \Rt_{a\a\bb}{}^{\gb} -
{1\over 2} \Ct_{\bb}{}^{\gb\d} ( T_{\d\a a} + H_{\d\a a} ) ) -
\dt_{\bb} \Pi^\g ( T_{\g\a}{}^{\bb} + {1\over 2} P^{\d\bb} H_{\d\g\a} )
\nonumber \\ &&+
\l^\b \omega_\g \Pb^{\db} R_{\db\a\b}{}^\g + \lt^{\bb} \widetilde\omega_{\gb} \Pi^\d
( \Rt_{\d\a\bb}{}^{\gb} + {1\over 2} \Ct_{\bb}{}^{\gb\r} H_{\r\d\a} )
\nonumber \\ &&
+ d_\b \dt_{\gb} ( P^{\d\gb} T_{\d\a}{}^\b - \N_\a P^{\b\gb} ) +
\lt^{\bb} \widetilde\omega_{\gb} d_\d ( \N_\a \Ct_{\bb}{}^{\gb\d} +
\Ct_{\bb}{}^{\gb\r} T_{\r\a}{}^\d + P^{\d\rb}
\Rt_{\rb\a\bb}{}^{\gb} ) \nonumber \\ &&
+\l^\b \omega_\g \dt_{\db} ( \N_\a C_\b{}^{\g\db} - P^{\r\db}
R_{\r\a\b}{}^\g ) - \l^\b \omega_\g \lt^{\db} \widetilde\omega_{\rb} ( \N_\a
S_{\b\db}{}^{\g\rb} + C_\b{}^{\g\sb} \Rt_{\sb\a\db}{}^{\rb} \nonumber \\&&+
\Ct_{\db}{}^{\rb\s} R_{\s\a\b}{}^\g ), \nonumber \eer and

\ber \N \dt_{\ab} &=& -{1\over 2} \Pi^a \Pb^b ( T_{\ab(ba)} + H_{\ab
ba} ) + {1\over 2} \Pi^a \Pb^{\bb} ( T_{\bb\ab a} + H_{\bb\ab a} ) -
\dt_{\bb} \Pi^a T_{a\ab}{}^{\bb} \label{dteom} \\ && - d_\b \Pb^a (
T_{a\ab}{}^\b - {1\over 2} P^{\b\gb} ( T_{\gb\ab a} - H_{\gb\ab a} ) ) +
\lt^{\bb} \widetilde\omega_{\gb} \Pi^a \Rt_{a\ab\bb}{}^{\gb} \nonumber \\ &&+ \l^\b \omega_\g
\Pb^a ( R_{a\ab\b}{}^\g - {1\over 2} C_\b{}^{\g\db} ( T_{\db\ab a} -
H_{\db\ab a} ) ) - d_\b \Pb^{\gb} ( T_{\gb\ab}{}^\b + {1\over 2}
P^{\b\db} H_{\gb\db\ab} ) \nonumber \\ && + \lt^{\bb} \widetilde\omega_{\gb} \Pi^\d
\Rt_{\d\ab\bb}{}^{\gb} + \l^\b \omega_\g \Pb^{\db} ( R_{\db\ab\b}{}^\g
+ {1\over 2} C_\b{}^{\g\rb} H_{\db\rb\ab} ) \nonumber \\ &&
+ d_\b \dt_{\gb} ( P^{\b\db} T_{\db\ab}{}^{\gb} - \N_{\ab}
P^{\b\gb} ) + \l^\b \omega_\g \dt_{\db} ( \N_{\ab} C_\b{}^{\g\db} +
C_\b{}^{\g\rb} \Tt_{\rb\ab}{}^{\db} - P^{\r\db} R_{\r\ab\b}{}^\g
) \nonumber \\ && 
+\lt^{\bb} \widetilde\omega_{\gb} d_\d ( \N_{\ab} \Ct_{\bb}{}^{\gb\d} + P^{\d\rb}
\Rt_{\rb\ab\bb}{}^{\gb} ) - \l^\b \omega_\g \lt^{\db} \widetilde\omega_{\rb} (
\N_{\ab} S_{\b\db}{}^{\g\rb} + C_\b{}^{\g\sb}
\Rt_{\sb\ab\db}{}^{\rb} \nonumber \\ &&+ \Ct_{\db}{}^{\rb\s} R_{\s\ab\b}{}^\g ).
\nonumber \eer
From these equations, (\ref{pureseom}), (\ref{gheom}) and also two equations
in (\ref{qqtil}) we obtain the holomorphicity constraints. In fact, $\Nb
j_B = 0$ implies

\ber  \nonumber &&T_{\a(ab)} = H_{\a ab} = T_{\a\b a} - H_{\a\b a} =
T_{a\a}{}^\b = T_{a\a}{}^{\bb} + P^{\g\bb} T_{\g\a a} = \l^\a
\l^\b R_{a\a\b}{}^\g = 0,  \\ && 
\Rt_{a\a\bb}{}^{\gb} - \Ct_{\bb}{}^{\gb\d} T_{\d\a a} =
T_{\g\a}{}^{\bb} + {1\over 2} P^{\d\bb} H_{\d\g\a} =
\Rt_{\d\a\bb}{}^{\gb} + {1\over 2} \Ct_{\bb}{}^{\gb\r} H_{\r\d\a} = 0,
\label{pjb}\\ &&
P^{\d\gb} T_{\d\a}{}^\b - \N_\a P^{\b\gb} - C_\a{}^{\b\gb} =  
\N_\a \Ct_{\bb}{}^{\gb\d} + \Ct_{\bb}{}^{\gb\r} T_{\r\a}{}^\d +
P^{\d\rb} \Rt_{\rb\a\bb}{}^{\gb} - S_{\a\bb}{}^{\d\gb} = 0, \nonumber \\ && \l^\a \l^\b
( \N_\a C_\b{}^{\g\db} - P^{\r\db} R_{\r\a\b}{}^\g ) = \l^\a
\l^\b ( \N_\a S_{\b\db}{}^{\g\rb} + C_\b{}^{\g\sb}
\Rt_{\sb\a\db}{}^{\rb} + \Ct_{\db}{}^{\rb\s} R_{\s\a\b}{}^\g ) =
0, \nonumber \eer
and $\N \jt_B = 0$ implies

\ber \nonumber && T_{\ab(ab)} = H_{\ab ab} = T_{\ab\bb a} + H_{\ab\bb a} =
T_{a\ab}{}^{\bb} = T_{a\ab}{}^\b - P^{\b\gb} T_{\gb\ab a} =
\lt^{\ab} \lt^{\bb} \Rt_{a\ab\bb}{}^{\gb} = 0, \\ &&
R_{a\ab\b}{}^\g - C_\b{}^{\g\db} T_{\db\ab a} = T_{\gb\ab}{}^\b +
{1\over 2} P^{\b\db} H_{\gb\db\ab} = R_{\db\ab\b}{}^\g + {1\over 2} C_\b{}^{\g\rb}
H_{\db\rb\ab} = 0 \label{pjtb} \\ &&  P^{\b\db} T_{\db\ab}{}^{\gb} - \N_{\ab} P^{\b\gb}
+ \Ct_{\ab}{}^{\gb\b} =  \N_{\ab} C_\b{}^{\g\db} + C_\b{}^{\g\rb} T_{\rb\ab}{}^{\db} -
P^{\r\db} R_{\r\ab\b}{}^\g - S_{\b\ab}{}^{\g\db} =0 \nonumber \\ &&  \lt^{\ab}
\lt^{\bb} ( \N_{\ab} \Ct_{\bb}{}^{\gb\d} + P^{\d\rb}
\Rt_{\rb\ab\bb}{}^{\gb} ) = 0,
\lt^{\ab} \lt^{\bb} ( \N_{\ab} S_{\d\bb}{}^{\r\gb} + C_\d{}^{\r\sb}
\Rt_{\sb\ab\bb}{}^{\gb} + \Ct_{\bb}{}^{\gb\s} R_{\s\ab\d}{}^\r ) =
0. \nonumber \eer

\subsection{Solving the Bianchi identities}

We can gauge-fix some of the torsion components and determine others
through the use of Bianchi identities. It is not necessary but it
will simplify the computation of the one-loop beta functions. As in
\cite{BerkovitsUE}, we can set $H_{\a\b\g} = H_{\a\b\gb} = H_{\a\bb\gb} =
H_{\ab\bb\gb} = 0$ since there is no such ten-dimensional
superfields satisfying the nilpotency constrains of $Q$ and $\widetilde Q$. We can 
use the Lorentz rotations to gauge fix $T_{\a\b}{}^a = \g^a_{\a\b}$ and
$T_{\ab\bb}{}^a = \g^a_{\ab\bb}$, therefore the above constraints
imply $H_{\a\b a} = (\g_a)_{\a\b}$ and $H_{\ab\bb a} =
-(\g_a)_{\ab\bb}$. We can use the shift symmetry of the action
(\ref{action})

$$
\d d_\a = \d \O_{\a\b}{}^\g \l^\b \omega_\g, \quad \d \dt_{\ab} = \d
\Ot_{\ab\bb}{}^{\gb} \lt^{\bb} \widetilde\omega_{\gb},\quad \d C_\a{}^{\b\gb} =
P^{\d\gb} \d \O_{\d\a}{}^\b,\quad \d \Ct_{\ab}{}^{\bb\g} = -
P^{\g\db} \d \Ot_{\db\ab}{}^{\bb},
$$
$$
\d S_{\a\bb}{}^{\g\db} = C_\a{}^{\g\rb} \d \Ot_{\rb\bb}{}^{\db} +
\Ct_{\bb}{}^{\db\r} \d \O_{\r\a}{}^\g,$$ to gauge-fix $T_{\a\b}{}^\g
= T_{\ab\bb}{}^{\gb} = 0$.

The Bianchi identity for the torsion is

\be (\N T)_{ABC}{}^D \equiv \N_{[A} T_{BC]}{}^D +
T_{[AB}{}^E T_{EC]}{}^D - R_{[ABC]}{}^D = 0, \label{bianchis} \ee 
where brackets in (\ref{bianchis}) mean (anti-)symmetrization
respect to the $ABC$ indices. The curvature will be $R$ or $\Rt$
if the upper index $D$ is $\d$ or $\db$ respectively. When
$D = d$, we use the notation $(\N T)_{ABC}{}^d$ or $(\N \Tt)_{ABC}{}^d$,
if we use the connection $\O_{Bc}{}^a$ or $\Ot_{Bc}{}^a$;
then the curvatures in each case will be $R$ or $\Rt$.

The Bianchi identity $(\N T)_{\a\b\g}{}^a = 0$ implies $T_{\a ab} =
2 (\g_{ab})_\a{}^\b \O_\b$. Similarly, the Bianchi identity $(\N
\Tt)_{\ab\bb\gb}{}^a = 0$ implies $\Tt_{\ab ab} = 2
(\g_{ab})_{\ab}{}^{\bb} \Ot_{\bb}$. The Bianchi identity $(\N
T)_{\a\bb\gb}{}^a = 0$ implies $\Ot_\a = \Tt_{\a a}{}^b = 0$.
Similarly, the Bianchi identity $(\N T)_{\ab\b\g}{}^a = 0$ implies
$\O_{\ab} = T_{\ab a}{}^b = 0$. It is not difficult to show that the
constraints $T_{a\a}{}^\a = T_{a\ab}{}^{\ab} = 0$ imply $\O_a =
\Ot_a = 0$.

We can write two sets of Bianchi identities for $H$ depending on
what is the connection we choose in the covariant derivative. Note
that the components of the superfield $H$ do not depend on such
choice. The Bianchi identities come from
$\N H = 0$ and $\Nt H = 0$ and it is not difficult to check that
both sets are equivalent. Let us write only one of
them

\be (\N H)_{ABCD} \equiv \N_{[A} H_{BCD]} + {3\over 2}
T_{[AB}{}^E H_{ECD]} = 0. \label{dhzero} \ee
There is one more Bianchi identity
involving a derivative of the curvature 
\be (\N R)_{ABCD}{}^E
\equiv \N R_{[ABC]D}{}^E + T_{[AB}{}^F R_{F C]D}{}^E=0. \label{drs} \ee

The identities $(\N H)_{\a\b\g\d}, (\N
H)_{\a\b\g\db}, (\N H)_{\a\b\gb\db}, (\N H)_{\a\bb\gb\db}, (\N
H)_{\ab\bb\gb\db}$ are easily satisfied if we recall the identities
for gamma matrices $\g^a_{(\a\b} (\g_a)_{\g)\d} = \g^a_{(\ab\bb}
(\g_a)_{\gb)\db} = 0$. The identities $(\N H)_{a\a\b\g}, (\N
H)_{a\a\b\gb}, (\N H)_{a\a\bb\gb}, (\N H)_{a\ab\b\gb}$ are satisfied
after using the dimension-${1\over 2}$ constraints. The identity $(\N
H)_{ab\a\b} = 0$ implies $T_{abc} + H_{abc} = 0$ and the identity
$(\Nt H)_{ab\ab\bb} = 0$ implies $\Tt_{abc} - H_{abc} = 0$. The
identity $(\N H)_{ab\a\bb} = 0$ is satisfied if we use the
constraints involving the superfield $P^{\a\bb}$ in the first lines
of (\ref{pjb}) and (\ref{pjtb}).

\subsection{The remaining equation of motion}

In the computation of the one-loop beta function we will need to
know the equation of motion for $\Pi^a$ and $\Pb^a$. Since we know
that the difference $\N \Pb^a - \Nb \Pi^a$ is given by the torsion
components, then we only need to determine $\N \Pb^a + \Nb \Pi^a$
which is determined by the varying the action respect to $\s^a = \d
Z^M E_M{}^a$. To make life simpler we will write this equation using
the above results for torsion and $H$ components. The equation turns
out to be

\ber  \nonumber {1\over 2} ( \Nt \Pb_a + \Nb \Pi_a ) &=& {1\over 2} \Pi^b \Pb^c H_{cba} -
{1\over 2} \Pi^\a \Pb^b T_{\a ab}  + d_\a \Pb^b T_{ab}{}^\a  + \l^\a \omega_\b
\Pb^b R_{ab\a}{}^\b \\ \nonumber && + {1\over 2}
\dt_{\ab} \Pi^\b T_{a\b}{}^{\ab} 
 + \dt_{\ab} \Pi^b ( T_{ab}{}^{\ab}+ {1\over 2}
P^{\b\ab} T_{\b ab} ) \\ \nonumber && + \lt^{\ab} \widetilde\omega_{\bb} \Pi^b (
\Rt_{ab\ab}{}^{\bb} + {1\over 2} \Ct_{\ab}{}^{\bb\g} T_{\g ab} )  \\ \nonumber && + {1\over 2} \lt^{\ab} 
\widetilde\omega_{\bb} \Pi^\g \Rt_{a\g\ab}{}^{\bb}  + {1\over 2} d_\a \Pb^{\bb} 
T_{a\bb}{}^\a + {1\over 2} 
\l^\a \omega_\b \Pb^{\gb} R_{a\gb\a}{}^\b + d_\a \dt_{\bb} \N_a
P^{\a\bb} \\ \nonumber &&+ \l^\a \omega_\b \dt_{\gb} ( \N_a C_\a{}^{\b\gb} - P^{\d\gb}
R_{a\d\a}{}^\b )+ \lt^{\ab} \widetilde\omega_{\bb} d_\g ( \N_a
\Ct_{\ab}{}^{\bb\g} + P^{\g\db} \Rt_{a\db \ab}{}^{\bb} ) \\  &&+ \l^\a
\omega_\b \lt^{\gb} \widetilde\omega_{\db} ( \N_a S_{\a\gb}{}^{\b\db} -
\Ct_{\gb}{}^{\db\r} R_{a\r\a}{}^\b - C_\a{}^{\b\rb}
\Rt_{a\rb\gb}{}^{\db} ). \label{piaeom} \eer

\subsection{Ghost number conservation}

As it was shown in \cite{BerkovitsUE}, the vanishing of the ghost number
anomaly determines that the spinorial derivatives of the dilaton
superfield $\Phi$ are proportional to the scale connection $\O$. This
relation is crucial to cancel the beta function in heterotic string
case \cite{ChandiaHN} and will be equally essential in our computation.
Let us recall how this relation is obtained. Consider the coupling
between ghost number currents and the connections in the action
(\ref{action}). Namely

$$
{1\over{2\pi\a'}} \int d^2z ~ (J \Ob + \Jt \O ).$$ The BRST variation
on this term contains the term

$$
- {1\over{2\pi\a'}} \int d^2z ~ (\pb J \l^\a \O_\a + \p \Jt \l^\a
\Ot_\a ).$$ The anomaly in the ghost number current conservation turns
out to be proportional to the two dimensional Ricci scalar, as noted
by dimensional grounds. The proportionality can be determined by
performing a Weyl transformation, around the flat world-sheet, of
the anomaly equation. In this way, the triple-pole in the OPE
between the current and the corresponding stress tensor yields

\be\N_\a \Phi = 4 \O_\a,\quad \N_{\ab} \Phi = 4 \Ot_{\ab},
\label{dilat} \ee
which will be used in section 5 to cancel the UV divergent part of
the effective action.

\section{Covariant Background Field Expansion}

We use the method explained in \cite{deBoerSkenderis} and \cite{ChandiaHN}. 
Here, we need to define a
straight-line geodesic which joins a point in superspace to
neighbor ones and allows us to perform an expansion in superspace.
It is given by $Y^A$ which satisfies the geodesic equation $\D Y^A =
Y^B \N_B Y^A = 0$. The connection we choose to define this covariant
derivative has the non-vanishing components $\O_{Aa}{}^b,
\O_{A\a}{}^\b$ and $\Ot_{A\ab}{}^{\bb}$. These same connections
are defined in the action (\ref{action}). In this way, the covariant
expansions of the different objects in (\ref{action}) are determined by

\be \D \Pi^A = \N Y^A - Y^B \Pi^C T_{CB}{}^A,\quad \D
\Ob_\a{}^\b = -Y^A \Pb^B R_{BA\a}{}^\b, \quad \D \Ot_{\ab}{}^{\bb} =
-Y^A \Pi^B \Rt_{BA\ab}{}^{\bb}. \label{covexps}\ee Any superfield $\Psi$ is 
expanded as $\D \Psi = Y^A \N_A \Psi$.

As in \cite{ChandiaHN}, we see that $d_{\a}, \dt_{\ab}$ and the pure
spinor ghosts are treated as fundamental fields, then we expand them
according to

\ber && d_{\a} = d_{\a 0} + \dh_{\a},\quad \l^{\a} = \l_0^{\a} +
\lh^{\a}, \quad \omega_{\a} = \omega_{\a 0} + \oh_{\a},\\ &&
\dt_{\ab} = \dt_{\ab0} + \dth_{\ab},\quad \lt^{\ab} = \lt_0^{\ab} +
\lth^{\ab},\quad \widetilde\omega_{\ab} = \widetilde\omega_{\ab 0} + \oth_{\ab 0}, \label{fund} \eer 
where the subindex $0$ means the background value of the corresponding
field which will dropped in the subsequent discussion.

The quadratic part of the expansion of (\ref{action}), excluding the
Fradkin-Tseytlin term, has the form

\ber S_2 &&= S_p + {1\over{2\pi\a'}} \int d^2z ~ (Y^A Y^B E_{BA}
+ Y^A \Nb Y^B C_{BA} + Y^A \N Y^B \Cb_{BA} \\ \nonumber &&+ \dh_\a Y^A \Db_A{}^\a 
 + \dth_{\ab} Y^A D_A{}^{\ab}   + (\lh^\a \oh_\b) \Hb_\a{}^\b +
(\lth^{\ab} \oth_{\bb}) H_{\ab}{}^{\bb} + (\lh^\a \omega_\b + \l^\a
\oh_\b) Y^A \Ib_{A\a}{}^\b \\ \nonumber &&+ (\lth^{\ab} \widetilde\omega_{\bb} + \lt^{\ab}
\oth_{\bb}) Y^A I_{A\ab}{}^{\bb} + \dh_\a \dth_{\bb} P^{\a\bb}
 + (\lh^\a \omega_\b + \l^\a \oh_\b) \dth_{\gb} C_\a{}^{\b\gb} + \\
 \nonumber &&
(\lth^{\ab} \widetilde\omega_{\bb} + \lt^{\ab} \oth_{\bb}) \dh_\g
\Ct_{\ab}{}^{\bb\g} + (\lh^\a \omega_\b + \l^\a \oh_\b) (\lth^{\gb}
\widetilde\omega_{\db} + \lt^{\gb} \oth_{\db}) S_{\a\gb}{}^{\b\db} ) ,\label{sdos} \eer
where
$E_{BA}, C_{BA}, \dots$ are background superfields given by

\ber  \nonumber E_{BA} &=& {1\over 4} \Pi^C \Pb^D ( T_{CB}{}^E H_{EDA}
(-1)^{D(C+B)} - T_{DB}{}^E H_{ECA} (-1)^{BC}  + \N_B
H_{DCA}(-1)^{B(C+D)} \\ \nonumber && + 2 T_{CB}{}^a T_{DAa} (-1)^{D(C+B)} )-
{1\over 4} \Pi^{(a} \Pb^{C)} ( R_{CBAa} - T_{CB}{}^D T_{DAa} +
\N_B T_{CAa} (-1)^{BC} ) \\ \label{eba} &&+ {1\over 2} d_\a \Pb^C (-1)^{A+B} ( -R_{CBA}{}^\a
+ T_{CB}{}^D T_{DA}{}^\a - \N_B T_{CA}{}^\a (-1)^{BC} )\\ \nonumber && + {1\over 2}
\dt_{\ab} \Pi^C (-1)^{A+B} ( -R_{CBA}{}^{\ab} + T_{CB}{}^D
T_{DA}{}^{\ab} - \N_B T_{CA}{}^{\ab} (-1)^{BC} ) \\ \nonumber &&+ {1\over 2} \l^\a \omega_\b
\Pb^C ( T_{CB}{}^D R_{DA\a}{}^\b - \N_B R_{CA\a}{}^\b (-1)^{BC}
) \\ \nonumber  &&+ {1\over 2} \lt^{\ab} \widetilde\omega_{\bb} \Pi^C ( T_{CB}{}^D R_{DA\ab}{}^{\bb} -
\N_B R_{CA\ab}{}^{\bb} (-1)^{BC} ) + {1\over 2} d_\a \dt_{\bb} \N_B \N_A
P^{\a\bb} \\ \nonumber && + {1\over 2} \l^\a \omega_\b \dt_{\gb} \N_B \N_A C_\a{}^{\b\gb}
(-1)^{A+B} + {1\over 2} \lt^{\ab} \widetilde\omega_{\bb} d_\g \N_B \N_A
\Ct_{\ab}{}^{\bb\g} (-1)^{A+B} \\ \nonumber && + {1\over 2} \l^\a \omega_\b \lt^{\gb} \widetilde\omega_{\db}
\N_B \N_A S_{\a\gb}{}^{\b\db},  \eer

\be  C_{BA} = -{1\over 4} \Pi^a T_{BAa} - {1\over 2} \Pi^C T_{CAa}
\d^a_B - {1\over 4} \Pi^A H_{CBA} - {1\over 2} d_\a T_{BA}{}^\a (-1)^{A+B}
- {1\over 2} \l^\a \omega_\b R_{BA\a}{}^\b, \label{cba} \ee

\be  \Cb_{BA} = -{1\over 4} \Pi^b T_{BAa} - {1\over 2} \Pb^C T_{CAa}
\d^a_B + {1\over 4} \Pi^A H_{CBA} - {1\over 2} \dt_{\ab} \Tt_{BA}{}^{\ab}
(-1)^{A+B} - {1\over 2} \lt^{\ab} \widetilde\omega_{\bb} \Rt_{BA\ab}{}^{\bb}, \label{ccba} \ee

\be \Db_A{}^\a = - \Pb^B T_{BA}{}^\a + \dt_{\bb} \N_A
P^{\a\bb} (-1)^A + \lt^{\bb} \widetilde\omega_{\gb} \N_A \Ct_{\bb}{}^{\gb\a},
\label{dbaa} \ee

\be D_A{}^{\ab} = -\Pi^B \Tt_{BA}{}^{\ab} - d_\b \N_A
P^{\b\ab} (-1)^A + \l^\b \omega_\g \N_A C_\b{}^{\g\ab}, \label{daab} \ee

\be \Hb_\a{}^\b = \Ob_\a{}^\b + \dt_{\gb} C_\a{}^{\b\gb}
\lt^{\gb} \widetilde\omega_{\db} S_{\a\gb}{}^{\b\db}, \label{hbab} \ee

\be H_{\ab}{}^{\bb} = \Ot_{\ab}{}^{\bb} + d_\g
\Ct_{\ab}{}^{\bb\g} + \l^\g \omega_\d S_{\g\ab}{}^{\d\bb},\label{habbb} \ee

\be \Ib_{A\a}{}^\b = -\Pb^A R_{BA\a}{}^\b + \dt_{\gb} \N_A
C_\a{}^{\b\gb} (-1)^A + \lt^{\gb} \widetilde\omega_{\db} \N_A
S_{\a\gb}{}^{\b\gb}, \label{ibaab} \ee

\be I_{A\ab}{}^{\bb} = -\Pi^B \Rt_{BA\ab}{}^{\bb} + d_\g
\N_A \Ct_{\ab}{}^{\bb\g} (-1)^A + \l^\g \omega_\d \N_A
S_{\g\ab}{}^{\d\bb}.\label{iaabbb} \ee 
In (\ref{sdos}) $S_p$ provides the propagators for
the quantum fields and is given by

\be  S_p = {1\over{2\pi\a'}} \int d^2z ~ ({1\over 2}\N Y^a \Nb Y_a +
\dh_\a \Nb Y^\a + \dth_{\ab} \N Y^{\ab}) + {\cal L}_{pure}, \label{sp} \ee 
where ${\cal L}_{pure}$ is the Lagrangian for the pure spinor ghosts.

\section{The one-loop UV divergent Part of the Effective Action}

The effective action is given by

\be e^{-S_{eff}} = \int D{\cal Q} ~ e^{-S},\label{seff} \ee where ${\cal Q}$
represents the quantum fluctuations.

To compute the one-loop beta functions we need to expand (\ref{action}) up
to second order in the quantum fields. In this way, we will obtain
the UV divergent part\cite{Friedan80} of the effective action, $S_\L$. Here $\L$ is
UV scale. Note that the Fradkin-Tseytlin term is evaluated on a sphere
with metric $\L dz d{\bar z}$. Finally, the complete UV divergent
part of the effective action becomes

\be  S_\L + {1\over 2\pi} \int d^2 z ~ ( \N \Pb^A \N_A \Phi +
\Pb^A \Pi^B \N_B \N_A \Phi ) \log\L . \label{sdiv} \ee

The computation of $S_\L$ is performed by contracting the quantum
fields. From (\ref{sp}) we read

\be  Y^a (z,{\bar z}) Y^b (w,{\bar w}) \to -\a' \eta^{ab}
\log |z-w|^2 \ee \be \dh_\a(z) Y^\b(w) \to
{\a'\d_\a{}^\b\over{(z-w)}},\quad \dth_{\ab}({\bar z}) Y^{\bb}({\bar
w}) \to {\a'\d_{\ab}^{\bb}\over{({\bar z}-{\bar w})}}. \label{props} \ee 
For the pure spinor ghosts we note that, because of (\ref{fields}), they enter in the
combinations

$$
N^{ab} = {1\over 2} (\l \g^{ab} \omega),\quad J = \l^\a \omega_\a,\quad \Nn^{ab} =
{1\over 2} (\lt \g^{ab} \ot), \quad \Jt = \lt^{\ab} \ot_{\ab}.$$ We can
expand each of these combinations as $J + J_1 + J_2$, similarly for $\Jt$,
$N^{ab}$ and $\Nn^{ab}$. As in
\cite{ChandiaHN}, the only relevant OPE's involving the pure spinor ghosts
and contributing to $S_\L$ are

\be N_1^{ab}(z) N_1^{cd}(w) \to {1\over(z-w)} ( - \eta^{a[c}
N^{d]b}(w) + \eta^{b[c} N^{d]a}(w) ), \label{nope}\ee

\be  \Nn_1^{ab}({\bar z}) \Nn_1^{cd}({\bar w}) \to {1\over({\bar
z}-{\bar w})} ( - \eta^{a[c} \Nn^{d]b}({\bar w}) + \eta^{b[c}
\Nn^{d]a}({\bar w}) ). \label{nnope} \ee

The one-loop contributions to $S_\L$ come from self-contraction of
$Y^A$'s in the term with $E_{BA}$ in (\ref{sdos}) and a series of double
contractions in (\ref{sdos}). These come from products between the term
involving $C_{BA}$ with the one involving $\Cb_{BA}$, $C_{BA}$ with
$\Db_{A}{}^{\bb}$, $\Cb_{BA}$ with $D_{A}{}^{\b}$, $\Db_{A}{}^{\bb}$
with $D_{A}{}^{\b}$, $E_{BA}$ with $P^{\a\bb}$, $\Ib_{C\a}{}^{\b}$
with $C_{\a}{}^{\b\gb}$, $I_{C\ab}{}^{\bb}$ with
$\Ct_{\ab}{}^{\bb\g}$ and $S_{\a\gb}{}^{\b\db}$ with itself. After
adding up all these contributions, the one-loop UV divergent part of
the effective action is proportional to

\be \int d^2z ~   [ - \eta^{ab} E_{ba} + \eta^{a[c} \eta^{d]b}
C_{ba} \Cb_{dc} + \eta^{ab} C_{[a\a]} \Db_b{}^\a + \eta^{ab}
\Cb_{[a\ab]} D_b{}^{\ab} + \Db_{\ab}{}^\b D_\b{}^{\ab} + E_{[\a\bb]}
P^{\a\bb} \label{uv} \ee $$  + N^{ab} \Ib_{\ab a}{}^c C_{cb}{}^{\ab} + \Nn^{ab}
I_{\a a}{}^c \Ct_{cb}{}^\a + {1\over 2} N^{ab} \Nn^{cd} S_a{}^e{}_c{}^f
S_{bedf} + \N \Pb^A \N_A \Phi + \Pb^A \Pi^B \N_B \N_A \Phi ]
\log\L, $$ where we used the expressions (\ref{fields}).

Now it will be shown that (\ref{uv}) vanishes as consequence of the
classical BRST constraints.

\section{One-loop Conformal Invariance}

To write the equations derived from the vanishing of (\ref{uv}), we need to
determine $\N\Pb^A$ from the classical equations of motion from
(\ref{action}). In order to do this, we need to know

\be \Nb \Pi^A - \N \Pb^A = \Pi^B \Pb^C T_{CB}{}^A. \label{diff} \ee Note that
we are using here the connection $\O_A{}^B$ to calculate the
covariant derivatives and the torsion components.

The equation for $\N\Pb_a$ is

\ber \N \Pb_a &=& \Pi^b \Pb^c T_{abc} - \Pi^\a \Pb^b T_{\a ab} +
\dt_{\ab} \Pi^b T_{ab}{}^{\ab} + d_\a \Pb^b T_{ab}{}^\a +
\lt^{\ab} \widetilde\omega_{\bb} \Pi^b \Rt_{ab\ab}{}^{\bb} \\ \nonumber &&+ \l^\a \omega_\b
\Pb^b R_{ab\a}{}^\b 
 + \dt_{\ab} \Pi^\b T_{a\b}{}^{\ab} + \lt^{\ab}
\widetilde\omega_{\bb} \Pi^\g \Rt_{a\g\ab}{}^{\bb} + d_\a \dt_{\bb} \N_a
P^{\a\bb} \\ \nonumber &&+ \l^\a \omega_\b \dt_{\gb} ( \N_a C_\a{}^{\b\gb} - P^{\d\gb}
R_{a\d\a}{}^\b ) + \lt^{\ab} \widetilde\omega_{\bb} d_\g ( \N_a
\Ct_{\ab}{}^{\bb\g} + P^{\g\db} \Rt_{a\db \ab}{}^{\bb} ) \\ \nonumber &&+ \l^\a
\omega_\b \lt^{\gb} \widetilde\omega_{\db} ( \N_a S_{\a\gb}{}^{\b\db} -
\Ct_{\gb}{}^{\db\r} R_{a\r\a}{}^\b - C_\a{}^{\b\rb}
\Rt_{a\rb\gb}{}^{\db} ). \label{npba} \eer Now we compute the equation for $\Pb^\a$.
We start by noting that this world-sheet field is determined from
the equation of motion (\ref{pieom}), then

$$
\N\Pb^\a = - \N ( \dt_{\bb} P^{\a\bb} + \lt^{\bb} \widetilde\omega_{\gb}
\Ct_{\bb}{}^{\gb\a} ).$$

Remember that the covariant derivative on $P^{\a\bb}$ and
$\Ct_{\ab}{}^{\bb\g}$ acts with $\O_\a{}^\b$ on $\a$-indices and with
$\Ot_{\ab}{}^{\bb}$ on ${\ab}$-indices. Now we can use the equations
(\ref{gheom}) and (\ref{dteom}) to obtain

\ber  \N \Pb^\a &=& d_\b \dt_{\gb} ( \Ct_{\db}{}^{\gb\b}
P^{\a\db} + P^{\b\db} \N_{\db} P^{\a\gb} ) + \l^\b \omega_\g \dt_{\db} (
- S_{\b\rb}{}^{\g\db} P^{\a\rb} + C_\b{}^{\g\rb} \N_{\rb} P^{\a\db}
) \\ \nonumber  && - \dt_{\bb} \Pi^a \N_a P^{\a\bb}
 - \dt_{\bb} \Pi^\g \N_\g
P^{\a\bb} + \lt^{\bb} \widetilde\omega_{\gb} d_\d ( \Ct_{\bb}{}^{\rb\d}
\Ct_{\rb}{}^{\gb\a} - \Ct_{\rb}{}^{\gb\d} \Ct_{\bb}{}^{\rb\a} -
P^{\d\rb} \N_{\rb} \Ct_{\bb}{}^{\gb\a}\\ \nonumber &&
-P^{\a\db}(\N_{\db}\Ct_{\bb}{}^{\gb\d}+P^{\d\eb}\Rt_{\eb\db\bb}{}^{\gb})
) + \l^\b \omega_\g \lt^{\db} \widetilde\omega_{\rb} ( S_{\b\db}{}^{\g\sb}
\Ct_{\sb}{}^{\rb\a} - S_{\b\sb}{}^{\g\rb} \Ct_{\db}{}^{\sb\a} \\ \nonumber &&+
C_\b{}^{\g\sb} \N_{\sb} \Ct_{\db}{}^{\rb\a}+P^{\a\eb}(\N_{\eb}
S_{\b\db}{}^{\g\rb} + C_\b{}^{\g\sb} \Rt_{\sb\eb\db}{}^{\rb} +
\Ct_{\db}{}^{\rb\s} R_{\s\eb\b}{}^\g )) \\ \nonumber &&- \lt^{\bb} \widetilde\omega_{\gb} \Pi^a
(\N_a \Ct_{\bb}{}^{\gb\a} +\Rt_{a\db\bb}{}^{\gb}P^{\a\db}) -
\lt^{\bb} \widetilde\omega_{\gb} \Pi^\d S_{\d\bb}{}^{\a\gb}. \label{npbaa} \eer To obtain the
equation for $\Pb^{\ab}$ we can use (\ref{diff}). After all this we get

\ber \N \Pb^{\ab} &=& d_\b
\dt_{\gb} ( C_\d{}^{\b\gb} P^{\d\ab} - P^{\d\gb} \N_\d P^{\b\ab} )
+ \lt^{\bb} \widetilde\omega_{\gb} d_\d ( S_{\r\bb}{}^{\d\gb} P^{\r\ab} -
\Ct_{\bb}{}^{\gb\r} \N_\r P^{\d\ab} ) \label{npbba} \\ \nonumber &&+ d_\b \Pb^a \N_a
P^{\b\ab} + d_\b \Pb^{\gb} \N_{\gb} P^{\b\ab} + \l^\b \omega_\g
\dt_{\db} ( C_\b{}^{\r\db} C_\r{}^{\g\ab} - C_\r{}^{\g\db}C_\b{}^{\r\ab} +
P^{\r\db} \N_\r C_\b{}^{\g\ab}
\\ \nonumber &&+P^{\d\ab}( \N_\d C_\b{}^{\g\db} - P^{\epsilon\db} R_{\epsilon\d\b}{}^\g )) + \l^\b
\omega_\g \lt^{\db}
\widetilde\omega_{\rb} ( S_{\b\db}{}^{\s\rb} C_\s{}^{\g\ab} - S_{\s\db}{}^{\g\rb}
C_\b{}^{\s\ab} \\ \nonumber &&+ \Ct_{\db}{}^{\rb\s} \N_\s C_\b{}^{\g\ab}
+P^{\a\eb}(\N_{\eb} S_{\b\db}{}^{\g\rb} + C_\b{}^{\g\gb}
\Rt_{\gb\eb\db}{}^{\rb} + \Ct_{\db}{}^{\rb\s} R_{\s\eb\b}{}^\g)) \\
\nonumber && -
\l^\b \omega_\g \Pb^a (\N_a C_\b{}^{\g\ab}
-R_{a\epsilon\b}{}^{\g}P^{\epsilon\ab}) 
 - \l^\b \omega_\g \Pb^{\db} S_{\b\db}{}^{\g\ab} + \Pi^a \Pb^b
T_{ab}{}^{\ab} - \Pi^\b \Pb^a T_{a\b}{}^{\ab} \\ \nonumber && - \dt_{\bb} \Pi^a
P^{\g\bb} T_{a\g}{}^{\ab} - \lt^{\bb} \widetilde\omega_{\gb} \Pi^a
\Ct_{\bb}{}^{\gb\d} T_{a\d}{}^{\ab}.  \eer

\subsection{Beta functions}

Now we can obtain the equations for the background fields implied
by the vanishing of the beta functions. These are the background
dependent expressions for the conformal weights $(1,1)$
independent couplings in (\ref{uv}). That is, all the independent
combinations formed from the products between $ (
\Pi^a, \Pi^\a, d_\a, \l^\a \omega_\b ) $ and $ ( \Pb^a, \Pb^{\ab},
\dt_{\ab}, \lt^{\ab} \ot_{\bb} ) $ because $\Pi^{\ab}$ and
$\Pb^\a$ are determined from the equations of motion (\ref{pieom}).
Let
us first concentrate on the beta functions coming from the couplings
to $\Pi^A \Pb^B$,  $d_\a \Pb^B$ and $\Pi^A \dt_{\bb}$ fields.
After using the results for the expansion (\ref{eba})-(\ref{iaabbb}) and the
equations (\ref{npba})-(\ref{npbba}) in (\ref{uv}), the couplings $\Pi^\a \Pb^{\bb},
\Pi^\a \Pb^{b}, \Pi^a \Pb^{\bb}$ and $\Pi^a \Pb^b$ lead respectively
to a first set of equations

\be T_{c\bb}{}^\d T_{\d\a}{}^c-T_{c\a}{}^{\db}T_{\db
\bb}{}^c + 4\N_{\a}\N_{\bb}\F =0, \label{pipidos} \ee

\be \N^d T_{\a db}+R_{\a deb}\eta^{de}+T_{bc}{}^{\d}
T_{\d\a}{}^c +4\N_b \N_\a \F =0,\label{pipitres} \ee

\be R_{\bb dea}\eta^{de}+T_{ac}{}^{\db} T_{\db\bb}{}^c
-T_{c\bb}{}^\d T_{\d a}{}^c +4\N_a\N_{\bb} \F =0, \label{pipicuatro} \ee

\be \eta^{cd} ( R_{acdb} + R_{bcda} ) - \N^c T_{abc} +
T_{c(a}{}^\a T_{b)\a}{}^c + 8 T_{a\a}{}^{\bb} T_{b\bb}{}^\a + 4
T_{ab}{}^c \N_c \Phi \label{pipiuno} \ee $$ + 4 T_{ab}{}^{\ab}\N_{\ab}\F + 4 \N_a \N_b \Phi
= 0. $$
We wrote them by increasing their dimensions, that is, if
$X^a$ has dimension $-1$ and each $\theta ^\a$, $\tt^{\ab}$ have
dimension $-{1\over 2}$, then the first has dimension $1$, the
second and third dimension ${3\over 2}$ and the fourth dimension
$2$. The couplings to $d_\a \Pb^{\bb}$, $\Pi^\a \dt_{\bb}$,
$d_\a \Pb^b$ and $\Pi^a \dt_{\bb}$ lead respectively to a second set of
equations

\be \N^c T_{c\bb}{}^{\a} - 2 \N_{\bb} P^{\a\gb} \N_{\gb} \F
+ 2 P^{\a\gb} \N_{\gb} \N_{\bb} \F = 0, \label{dalPbbb} \ee

\be \N^c T_{c\a}{}^{\bb} + 2 \N_{\a} P^{\g\bb} \N_{\g} \F
- 2 P^{\g\bb} \N_\g \N_\a \F = 0, \label{Pialdtbb} \ee

\be \N^c T_{cb}{}^{\a} - T_{cd}{}^{\a} T_{b}{}^{cd} + (T_{\d
b}{}^c T_{c\gb}{}^{\a} - R_{b\gb\d}{}^{\a}) P^{\d\gb} +
T_{b\gb}{}^{\d}( 3 \N_{\d} P^{\a\gb} - 2 P^{\a\gb}\N_{\d} \F )
+ 2 T_{bc}{}^{\a} \N^c \F \label{dalPbb} \ee $$ - 2 \N_{b} P^{\a\gb}
\N_{\gb} \F = 0, $$

\be  \N^c T_{ca}{}^{\bb} - 2 T_{cd}{}^{\bb} T_{a}{}^{cd} +
P^{\g\bb} T_{\a}{}^{de} T_{ade} + \Rt_{a\g\db}{}^{\bb} P^{\g\db}  -
T_{a\g}{}^{\db} ( 3 \N_{\db} P^{\g\bb} - 2 P^{\g\bb}\N_{\db} \F )
\label{Piadtbb} \ee $$  + 2 T_{ac}{}^{\bb} \N^c \F + 2 \N_{a} P^{\g\bb} \N_{\g} \F = 0.
 \label{Piadtbb} $$
The first two with have dimension 2 and the second two have dimension ${5\over 2}$.
Now we will prove that these equations are implied by the classical
BRST constraints, the Bianchi identities (\ref{bianchis}) and the
relations (\ref{dilat}).

Firstly, it is important to know the expression for the scale
curvature in terms of the scale connection. This are found to be

\ber  \nonumber && R_{\a\b} = \N_{(\a } \O_{\b )},  \quad R_{\a\bb} =
\N_{\bb} \O_\a ,\quad R_{\ab\bb} =0, \\ && R_{ab} = T_{ab}{}^\g
\O_\g ,\quad R_{a\b} = \N_a \O_\b ,\quad R_{a\bb} = T_{a\bb}{}^\g
\O_\g. \label{curcon}  \eer

\ber \nonumber  && \Rt_{\ab\bb} = \N_{(\ab }\Ot_{\bb )},  \quad
\Rt_{\a\bb} = \N_{\a} \Ot_{\bb} ,\quad \Rt_{\a\b} =0, \\ && \Rt_{ab}
= T_{ab}{}^{\gb} \Ot_{\gb} ,\quad \Rt_{a\bb} =  \N_a \Ot_{\bb}
,\quad \Rt_{a\b} = T_{a\b}{}^{\gb} \Ot_{\gb}. \label{curcont} \eer

Secondly, let us write some expressions useful for later use. We
note that the Bianchi identity $(\N T)_{\a ab}{}^c = 0$,  using
(\ref{curcon})  can be written as

\be R_{\a [ab]c} = \N_{\a} T_{abc} -2 (\g_{c[a})_{\a}{}^\b
R_{b]\b} + (\g_{c})_{\a\b}T_{ab}{}^\b - T_{\a dc} T_{ab}{}^d - T_{\a
[a}{}^d T_{b]d c}, \label{Ralbcd} \ee now, we can use the identity

\be 2 R_{\a abc} = R_{\a [ab]c}+ R_{\a [ca]b} -R_{\a [bc]a},
\label{idzero} \ee
and the Bianchi identity $(\N H)_{\a abc} =0$ to write (\ref{Ralbcd}) as

\be R_{\a abc} = T_{a[b}{}^\b (\g_{c]})_{\b\a} - 2
(\g_{bc})_{\a}{}^\b R_{a\b}. \label{idI} \ee An identical procedure starting with
$(\N \Tt)_{\ab ab}{}^c =0$ allows us to find

\be \Rt_{\ab abc}= T_{a[b}{}^{\bb} (\g_{c]})_{\ab\bb} - 2
(\g_{bc})_{\ab}{}^{\bb}\Rt_{a\bb}. \label{idIp} \ee
Then, replacing (\ref{idI}) and (\ref{idIp})
respectively in $(\N T)_{a\a\b}{}^\b = 0$ and $( \N
 T )_{a\ab\bb}{}^{\bb} =0,$ we find

\be \g^b _{\a\b}T_{ba}{}^\b = 8 R_{a\a},\quad \g^b_{\ab\bb}
T_{ba}{}^{\bb} = 8 \Rt_{a\ab}. \label{idII} \ee

We have enough information to show that the equations (\ref{pipidos}) ,
(\ref{pipitres}) and (\ref{pipicuatro}) are satisfied. From the Bianchi identity
$(\N T)_{\a\b\gb}{}^\b =0$ we obtain

\be T_{\a\b}{}^d T_{d\gb}{}^\b = 17 R_{\a\gb} + {1\over 4}
R_{\gb\b cd} (\g^{cd})_{\a}{}^{\b}.\label{BiII} \ee 
Since we need an expression for
$R_{\gb\b cd}$, we can use $(\N T)_{\a\bb a}{}^b=0$, finding

\be R_{\gb\b cd} = 2 ( \g_{cd})_{\b}{}^\d \N_{\gb}\O_\d +
T_{c\gb}{}^\epsilon (\g_d)_{\epsilon\b} + T_{c\b}{}^{\eb} (\g_d)_{\eb\gb}.
\label{BiXXVIII} \ee
Replacing (\ref{BiXXVIII}) in (\ref{BiII}) ,using the second equation in
(\ref{curcon}) , $\N_\a \F= 4\O_\a$ and the constraints coming from
holomorphicity-antiholomorphicity of the BRST current
$T_{a\b}{}^{\gb} = -(\g_a)_{\b\d}P^{\d\gb}$, $T_{a\bb}{}^\g =
(\g_a)_{\bb\db}P^{\g\db}$ we can verify the equation (\ref{pipidos}).

To verify (\ref{pipitres}) and (\ref{pipicuatro}), we must contract the $a$ and
$b$ indices using $\eta^{ab}$ in (\ref{idI}) and (\ref{idIp}), and use
(\ref{idII})
together with the relations (\ref{dilat}).

For deriving the remaining equation of the first set, the coupling to
$\Pi^a \Pb^b$, it is useful to find an expression for $R_{abcd}$,
which can be found from the Bianchi identity $(\N T)_{ab\a}{}^\b$

\be R_{abcd} = -{1\over 8} (\g_{cd})_\b{}^\a ( \N_\a
T_{ab}{}^\b  - T_{\a[a}{}^e T_{b]e}{}^\b - T_{\a[a}{}^{\gb}
T_{b]\gb}{}^\b ), \label{rabcd} \ee
from this equation we construct $\eta^{cd} ( R_{acdb} +
R_{bcda} )$:

\ber \eta^{cd} ( R_{acdb} + R_{bcda} ) &=& -{1\over 8 }\eta^{cd}
[(\g_{db})_{\b}{}^\a \N_\a T_{ac}{}^\b + (\g_{da})_{\b}{}^\a \N_\a
T_{bc}{}^\b ] \\ \nonumber && +{1\over 8} \eta^{cd}[ (\g_{db})_{\b}{}^\a T_{\a
[a}{}^e T_{c]e}{}^\b  + (\g_{da})_{\b}{}^\a T_{\a [b}{}^e
T_{c]e}{}^\b ] \\ \nonumber && +{1\over 8} \eta^{cd} [(\g_{db})_{\b}{}^\a
T_{\a [a}{}^{\eb} T_{c]\eb}{}^\b  + (\g_{da})_{\b}{}^\a T_{\a
[b}{}^{\eb} T_{c]\eb}{}^\b]. \label{Rsab} \eer 
Let us consider the right hand side of
(\ref{Rsab}) line by line. We can use (\ref{idII}) , the Bianchi identity $(\N
R)_{\a a\d \b}{}^\g$ to write

\be (\g_b)^{\d\a} \N_{\a} R_{a\d} = - 2 \N_a \N_b \F - 2
T_{ab}{}^\g \N_\g \F - 2 (\g_b \g_{ae})^{\d\b}\O_\b R^{e}{}_\d -
(\g_a \g_b)_{\b}{}^\d P^{\b\eb} R_{\eb\d}, \label{BiRaladl} \ee
and the beta function with dimension $1$ (\ref{pipidos}) to find
the following expression for the first line in the right hand side
of (\ref{Rsab})

\be -4 \N_b \N_a \F + 2 T_{ab}{}^C \N_C \F - 4
\eta_{ab}(\g^e)^{\d\b} \O_\b R_{e\d} + 4 (\g_b)^{\d\b}\O_\b R_{a\d}
+ 4 (\g_a)^{\d\b}\O_\b R_{b\d} + {1\over 4} \eta_{ab} \eta^{cd}
T_{c\b}{}^{\db} T_{d\db}^{\b}. \label{fstRsab} \ee 
Finding an expression for the second
line is a matter of gamma matrices algebra, once we use (\ref{curcon}) .
For this line we find ${1\over 4}\eta_{ab}T_{\b cd} T^{cd\b} -
{3\over 4}T_{c(a}{}^\b T_{b)\b}{}^c $. Using $T_{a\b}{}^{\gb} =
-(\g_a)_{\b\d}P^{\d\gb}$ and some gamma matrices algebra, it is
straightforward to find $T_{\b (a}{}^{\gb} T_{b)\gb}{}^b -{1\over
4}\eta_{ab}\eta^{cd}T_{d\b}{}^{\gb}T_{c\gb}{}^\b$ for the third
line. So, adding the results for the three lines and using (\ref{idII}) we
find

\be \eta^{cd} ( R_{acdb} + R_{bcda} ) = - 4 \N_b \N_a \F -
T_{c(a}{}^\b T_{b)\b}{}^c + 2 T_{ab}{}^E \N_E \F + T_{\b
(a}{}^{\gb}T_{b)\gb}{}^\b ,\label{RsabII} \ee 
which contains some of the terms in
(\ref{pipiuno}) . It is also needed to use $(\N T)_{abc}{}^c=0$ in order
to generate the term $\N^c T_{abc}$. This Bianchi identity gives

\be \N^c T_{abc} -T_{c[a}{}^e T_{b]e}{}^c -T_{c[a}{}^\epsilon
 T_{b]\epsilon}{}^c -\eta^{cd}(R_{acdb}-R_{bcda})=0 . \label{BiXXVII} \ee 
Finding an expression for $\eta^{cd}(R_{acdb}-R_{bcda})$ is not difficult
following the description given to compute (\ref{RsabII}) . After we
compute it and replace it in (\ref{BiXXVII}) we find
\be \N^c T_{abc} +T_{\b [a}{}^{\db} T_{b]\db}{}^\b -2T_{ab}{}^c 
\N_c \F+2T_{ab}{}^\g \N_\g \F -2T_{ab}{}^{\gb}\N_{\gb}\F =0.
\label{BiXXVIIp} \ee
Combining
(\ref{RsabII}) and (\ref{BiXXVIIp}) gives the desired beta function equation
(\ref{pipiuno}).

A similar procedure, but with more steps, is performed to prove the
equations of the second group. To probe (\ref{dalPbbb}) one can start by
computing $\{\N_{\ab},\N_{\bb}\}P^{\g\bb} = -\g^c _{\gb\bb}\N_c
P^{\g\bb}+\Rt_{\ab\bb\db}{}^{\bb}P^{\g\db }$. Then we split the
curvature as a scale curvature plus a Lorentz curvature. For the
latter, use $(\N \Tt)_{\ab\bb c}{}^d =0$ to obtain

\be \Rt_{\ab\bb cd}(\g^{cd})_{\db}{}^{\bb} = - 180 \N_{\ab}
\Ot_{\db} + (\g^{cd})_{\db}{}^{\bb} \N_{\bb} \Tt_{\ab cd} + 16
\Tt_{\db}{}^{cd} \Tt_{\ab cd} + (\g^{cd} \g^e)_{\db\ab} \Tt_{ecd},
\label{Rabbbcd} \ee
so on one hand we will have

\ber \nonumber \{\N_{\ab},\N_{\bb}\}P^{\g\bb} &=& - \N^c T_{c\ab}{}^\g +
\Rt_{\ab\db} P^{\g\db} - 45 \N_{\ab}\Ot_{\db}P^{\g\db} + {1\over 4}
(\g^{cd})_{\db}{}^{\bb} \N_{\bb} \Tt_{\ab cd} P^{\g\db} \\ &&
 - 4 \Tt_{\ab cd} \Tt_{\db}{}^{cd} P^{\g\db} + {1\over
4} (\g^{cd}\g^e)_{\db\ab} \Tt_{ecd} P^{\g\db}. \label{DDPI} \eer 
On the other hand, we can use $\N_{\ab}P^{\b\gb} = \Ct_{\ab}{}^{\gb\b}$, $\Ct^{\g} = -
P^{\g\db} \Ot_{\db}$  and \\$\Ct_{cd}{}^\g = 1/10 (\g^a)^{\g\a}
\Rt_{a\a cd}$, which come from antiholomorphicity of the BRST
current, to write

\be \{\N_{\ab},\N_{\bb}\}P^{\g\bb} = - 17 \N_{\ab} P^{\g\db}
\Ot_{\db} - 17 P^{\g\db} \N_{\ab} \Ot_{\db} + {1\over 40}
(\g^{a})^{\g\a} \N_{\bb} ( \Rt_{a\a cd}(\g^{cd})_{\ab}{}^{\bb}
).\label{DDPII} \ee
Using $(\N \Tt)_{\a bcd}=0$ and $(\Nt H)_{\a bcd}=0$ it is
straightforward to find

\be (\g^a)^{\g\a}\Rt_{a\a cd} = 10 T_{cd}{}^\g - 10 P^{\g\eb}
\Tt_{\eb cd}. \label{gRt} \ee
Since there is a derivative acting on this terms in
(\ref{DDPII}) , we make use of $(\N \Tt)_{\bb cd}{}^\g =0$ to find

\be (\g^{cd})_{\ab}{}^{\bb} \N_{\bb} T_{cd}{}^\g = - 18
\N^d T_{d\ab}{}^\g + (\g^{cd}\g^e)_{\ab\db} \Tt_{ecd} P^{\g\db} + 16
\Tt_{\ab cd} T_{dc}{}^\g. \label{BiTtbbcdg}\ee
We can now replace the last two equations
in (\ref{DDPII}) and equate it to (\ref{DDPI}) . The identity

\be (\g^{ab})_{(\ab}{}^{\bb}(\g_{ab})_{\gb )}{}^{\db} = - 10
s\d_{(\ab}{}^{\bb} \d_{\gb )}{}^{\db} + 8
(\g^a)_{\ab\gb}(\g_a)^{\bb\db}, \label{idIII} \ee
which can be proved using
$(\g^a)_{(\ab\bb}(\g_a)_{\gb )\db}=0$, will be of help to find
(\ref{dalPbbb}). A completely analog procedure allows us to arrive to
(\ref{Pialdtbb}).

To prove (\ref{dalPbb}) we make use of the Bianchi identities $(\N R)_{\a
ab\b}{}^\g =0$, $(\N T)_{c\a\b}{}^\g=0$ and the identity
$(\g_a)^{\a\b} R_{\a\b\g}{}^\d = - 2 (\g_a)^{\a\b} R_{\g\a\b}{}^\d$,
which follows from $(\N T)_{\a\b\g}{}^\d =0$, to arrive to

\ber && \nonumber (\g)^{\a\b} ( \N_{\a} R_{ab\b}{}^\g - 2 T_{\a [a}{}^e
 R_{b]e\b}{}^\g - T_{\a[a}{}^{\eb}R_{b]\eb\b}{}^\g) -
8 T_{b}{}^{ac} T_{ac}{}^\g + 8 \N^a T_{ab}{}^\g + 2 T_{ab}{}^\g \N^a
\F \\ &&
- {1\over 8} (\g)^{\a\b}(\g^{cd})_{\epsilon}{}^\g R_{\a\b cd} 
 T_{ab}{}^\epsilon + T_{ab}{}^{\eb} (\g^a)^{\a\b}R_{\eb\a\b}{}^\g =0.
\label{idIV} \eer
The last term in this equation is zero as can easily seen using $(\N
T)_{\eb\a\b}{}^\g =0$. The first term can be worked out using
(\ref{rabcd}) and $(\N T)_{a\eb\b}{}^\g =0$, the curvature in the first
term of the second line can be rewritten using $(\N T)_{\a\b
a}{}^b=0$. The use of $(\N T)_{cdb}{}^\d =0$ will be also needed to
generate (\ref{dalPbb}). Again, an analog procedure will allow at arrive to
(\ref{Piadtbb}).

So far, we concentrated on a specific set of beta functions. The
remaining ones can be classified in a third and fourth sets. The
third set involves first order derivatives of the curvatures. We
present it again as the dimension increases.

At dimension $5/2$ we find respectively from the couplings to $J
\Pb^{\bb}$, $\Pi^\a \Jt$, $N^{ac}\Pb^{\bb}$ and  $\Pi^\a \Nn^{bc}$

\be \N^a R_{a\bb} + \N_{(\eb} R_{\bb) \d } P^{\d\eb} + 2
(\N_{\bb}C^{\ab} - R_{\bb\g} P^{\g\ab}) \N_{\ab} \F + 2 C^{\ab}
\N_{\ab} \N_{\bb} \F = 0, \label{JPbbb}\ee

\be \N^b \Rt_{b\a} - \N_{(\d} \Rt_{\a)\eb} P^{\d\eb} + 2
(\N_{\a} \Ct^{\b} + \Rt_{\a\gb} P^{\b\gb})\N_{\b}\F + 2 \Ct^\b \N_\b
\N_\a \F = 0, \label{PialJt} \ee

\be \N^d R_{d\bb ac} + \N_{(\eb} R_{\bb\d) ac} P^{\d\eb} +
2 (\N_{\bb} C_{ac}{}^{\ab} - R_{\bb\g ac} P^{\g\ab}) \N_{\ab} \F + 2
C_{ac}{}^{\db} \N_{\db} \N_{\bb} \F = 0, \label{NacPbbb} \ee

\be \N^d \Rt_{d\a bc} - \N_{(\d}\Rt_{\a )\eb bc} P^{\d\eb}
+ 2 (\N_{\a} \Ct_{bc}{}^{\g} + \Rt_{\a\db bc} P^{\g\db}) \N_{\g} \F
+ 2 \Ct_{bc}{}^\g \N_\g \N_\a \F = 0. \label{PialNtbc}  \ee
While at dimension $3$ we
find respectively from the couplings to $J\Pb^b$, $\Pi^a \Jt$,
$N^{ac}\Pb^b$ and $\Pi^a \Nn^{bc}$

\ber && \nonumber \N^a R_{ab} - T_{ba}{}^c R^{a}{}_c + T_{ba}{}^{\g}
R^{a}{}_{\g} + 3 T_{b\gb}{}^{\a} \N_{\a} C^{\gb} + 2 R_{bc} \N^c \F
+ 2 R_{b\ab} P^{\g\ab}\N_{\g} \F \\ && + 2 (\N_b C^{\ab} - R_{b\g}
P^{\g\ab}) \N_{\ab} \F + P^{\d\eb} (\N_{\eb} R_{\d b} + T_{b\d}{}^c
R_{\eb c} + T_{b\eb}{}^{\g} R_{\d\g})=0, \label{JPbb} \eer

\ber && \nonumber \N^b \Rt_{ba} + T_{ab}{}^{\gb} \Rt^{b}{}_{\gb} + T_{abc}
\Ct^\d T_{\d}{}^{bc} + 3 T_{a\g}{}^{\bb} \N_{\bb} \Ct^{\g} + 2
\Rt_{ab} \N^b \F - 2 \Rt_{a\g} P^{\g\bb} \N_{\bb} \F \\ &&
 + 2(\N_a \Ct^{\b} + \Rt_{a\gb} P^{\b\gb}) \N_{\b} \F
- P^{\d\eb} (\N_{\d} \Rt_{\eb a} + T_{\d a}{}^c \Rt_{\eb c} +
T_{a\d}{}^{\gb} \Rt_{\eb\gb})=0, \label{PiaJt} \eer

\be \N^d R_{dbac} - T_{b}{}^{de} R_{deac} + T_{b}{}^{d\epsilon}
R_{d\epsilon ac} + 3 T_{b\db}{}^{\g}\N_{\g}C_{ac}{}^{\db}+2R_{bdac}\N^d \F
+2R_{b\db ac} P^{\epsilon\db} \N_{\epsilon} \F \label{NacPbb} \ee $$ + 2 (\N_b C_{ac}{}^{\db} -
R_{b\epsilon ac} P^{\epsilon\db}) \N_{\db} \F  +2 R_{b\db ea}C_{c}{}^{e\db} +
P^{\d\eb} (\N_{\eb} R_{\d bac} + T_{b\d}{}^f R_{\eb fac} +
T_{b\eb}{}^{\g} R_{\d\g ac})=0,$$

\be \N^d \Rt_{dabc} + T_{a}{}^{d\eb} \Rt_{d\eb bc} +
T_{adf} \Ct_{bc}{}^\epsilon T_{\epsilon}{}^{df} + 3 T_{a\d}{}^{\eb}
\N_{\eb}\Ct_{bc}{}^{\d} + 2 \Rt_{adbc} \N^d \F - 2 \Rt_{a\d bc}
P^{\d\eb} \N_{\eb} \F \label{PiaNtbc} \ee $$ + 2 (\N_a \Ct_{bc}{}^{\d} + \Rt_{a\eb
bc} P^{\d\eb}) \N_{\d} \F + 2 \Rt_{a\d eb} \Ct_{c}{}^{e\d} -
P^{\d\eb} (\N_{\d} \Rt_{\eb abc} + T_{a\d}{}^f  \Rt_{\eb fbc} +
T_{a\d}{}^{\gb} \Rt_{\eb\gb bc})=0. $$

The fourth set involves second order derivatives of the
background fields $P^{\a\bb}$, $C_{\a}{}^{\b\gb}$,
$\Ct_{\ab}{}^{\bb\g}$ and $S_{\a\bb}{}^{\g\db}$. There is an
equation at dimension $3$, coming from the coupling to $d_\a
\dt_{\bb}$

\ber && \N^2 P^{\a\bb} - 2 P^{\g\db} S_{\g\db}{}^{\a\bb} +
T_{de}{}^{\a} T^{de\bb} - 2 \N_{\gb} P^{\d\bb} \N_{\d} P^{\a\gb} - 2
\N_c P^{\a\bb} \N^c \F \\ \nonumber && - 2 (P^{\g\db} \N_{\db} P^{\a\bb} +
P^{\a\db} \N_{\db} P^{\g\bb}) \N_{\g} \F + 2 (P^{\d\gb} \N_{\d}
P^{\a\bb} + P^{\d\bb} \N_{\d} P^{\a\gb}) \N_{\gb} \F =0.\label{daldtbb} \eer
At 
dimension $7/2$ we find respectively from the couplings to
$J\dt_{\bb}$, $d_\a \Jt$, $N^{ac}\dt_{\bb}$ and $d_\a \Nn^{bc}$

\be  \N^2 C^{\bb} - P^{\a\gb} \N_{[\a} \N_{\gb ]} C^{\bb} -
T_{ac}{}^{\bb} R^{ac} + 2 R_{\g}{}^a \N_a P^{\g\bb} + 2 \N_{\gb}
P^{\a\bb} \N_\a C^{\gb} - C^{\ab} \Rt_{\ab\d\eb}{}^{\bb} P^{\d\eb}
\label{Jdbb} \ee $$ + P^{\a\bb} (\N_c R_\a{}^c - \N_{[\d}R_{\gb ]\a} P^{\d\gb}) - 2 (\N_a
C^{\bb} - P^{\g\bb} R_{a\g}) \N^a \F - 2 (P^{\a\bb} \N_\a C^{\gb} +
P^{\a\gb} \N_\a C^{\bb} $$ $$
+ P^{\a\bb} R_{\a\g} P^{\g\gb}) \N_{\gb} \F + 2( S P^{\a\bb} + {1\over
4} \St_{cd} (\g^{cd})_{\eb}{}^{\bb} P^{\a\eb} - C^{\gb} \N_{\gb}
P^{\a\bb}) \N_{\a} \F =0,$$

\be  \N^2 \Ct^{\a} - P^{\b\gb} \N_{[\b}\N_{\gb ]} \Ct^{\a} -
T_{bc}{}^{\a} \Rt^{bc} - 2 \Rt_{\gb}{}^b \N_b P^{\a\gb} - 2 \N_{\gb}
\Ct^\b \N_\b P^{\a\gb} + \Ct^{\b} R_{\b\eb\d}{}^{\a} P^{\d\eb} \label{daJt}
\ee $$ - P^{\a\bb} (\N_c \Rt_{\bb}{}^c + \N_{[\d} \Rt_{\gb
]\bb} P^{\d\gb}) - 2 (\N_b \Ct^{\a} + P^{\a\gb} \Rt_{b\gb}) \N^b \F
+ 2 (P^{\a\bb} \N_{\bb} \Ct^{\g} + P^{\g\bb} \N_{\bb} \Ct^{\a}$$ $$ + P^{\a\eb} \Rt_{\eb\gb} P^{\g\gb}) \N_{\g} \F - 2 (S P^{\a\bb} + {1\over
4} S_{cd} (\g^{cd})_{\epsilon}{}^{\a} P^{\epsilon\bb} - \Ct^{\g} \N_{\g}
P^{\a\bb}) \N_{\bb} \F =0, $$ 

$$  \N^2 C_{ac}{}^{\bb} - P^{\d\eb} \N_{[\d}\N_{\eb
]}C_{ac}{}^{\bb} - R_{deac} T^{de\bb} - 2 R_{d\epsilon ac} \N^d P^{\epsilon\bb}
+ 2 \N_{\db} P^{\epsilon\bb} \N_{\epsilon} C_{ac}{}^{\db} - C_{ac}{}^{\gb}
\Rt_{\gb\d\eb}{}^{\bb} P^{\d\eb} $$ $$ - P^{\b\bb} (\N^d R_{d\b ac}
- \N_{[\d}R_{\eb ]\b ac} P^{\d\eb} + 2 R_{\b\db ea} C_{c}{}^{e\db})
+ 2 \N_{\db} C_{ea}{}^{\bb} C_{c}{}^{e\db} - 2 (\N_d C_{ac}{}^{\bb}
- P^{\epsilon\bb} R_{d\epsilon ac}) \N^d \F
$$ $$ -2 (C_{ac}{}^{\db} \N_{\db} P^{\g\bb}- S_{ac} P^{\g\bb} -
{1\over 4} S_{acbd}(\g^{bd})_{\db}{}^{\bb} P^{\g\db}) \N_\g \F $$ \be
- 2 (P^{\g\bb} \N_\g C_{ac}{}^{\db} + P^{\g\db} \N_\g C_{ac}{}^{\bb}
- P^{\epsilon\bb} R_{\epsilon\g ac} P^{\g\db}) \N_{\db} \F = 0,
\label{Nacdbb} \ee

$$ \N^2 \Ct_{bc}{}^{\a} - P^{\d\eb} \N_{[\d}\N_{\eb ]}
\Ct_{bc}{}^{\a} - \Rt_{debc} T^{de\a} + 2 \Rt_{d\eb bc} \N^d
P^{\a\eb} - 2 \N_{\db} \Ct_{bc}{}^\epsilon \N_\epsilon P^{\a\db} +
\Ct_{bc}{}^{\g} R_{\g\eb\d}{}^{\a} P^{\d\eb} $$ $$ - P^{\a\bb} (\N^d
\Rt_{d\bb bc} - \N_{[\d} \Rt_{\eb ]\bb bc} P^{\d\eb} + 2 \Rt_{\bb\d
eb} \Ct_{c}{}^{e\d}) + 2 \N_{\d} \Ct_{eb}{}^{\a} \Ct_{c}{}^{e\d} - 2
(\N_d \Ct_{bc}{}^{\a} + P^{\epsilon\bb} \Rt_{d\bb bc}) \N^d \F
$$ $$ + 2 (\Ct_{bc}{}^{\d} \N_{\d} P^{\a\gb} - \St_{bc} P^{\a\gb}
- {1\over 4} S_{adbc} (\g^{ad})_{\d}{}^{\a} P^{\d\gb})\N_{\gb} \F $$\be
 + 2 (P^{\a\bb} \N_{\bb} \Ct_{bc}{}^{\d} + P^{\d\bb} \N_{\bb}
\Ct_{bc}{}^{\a} + P^{\a\eb} \Rt_{\eb\gb bc} P^{\d\gb}) \N_{\d}
\F=0. \label{daNtbc} \ee

Finally, at dimension $4$ we find from the couplings to  $J\Jt$,
$J\Nn^{ac}$, $N^{ab}\Jt$ and $N^{ab}\Nn^{cd}$ respectively

$$ \N^2 S - P^{\d\eb} \N_{[\d} \N_{\eb ]} S - R^{ab} \Rt_{ab}
+ 2 \Rt_{a\bb} \N^a C^{\bb} + 2 R_{a\b} \N^a \Ct^\b - 2 \N_{\ab}
\Ct^\b \N_\b C^{\ab} $$ $$ - \Ct^\b (\N^a R_{a\b} - P^{\d\eb}
\N_{[\d}R_{\eb ]\b}) - C^{\bb} (\N^a \Rt_{a\bb} - P^{\d\eb} \N_{[\d}
\Rt_{\eb ]\bb}) + 2 (\Ct^\a R_{b\a} + C^{\ab} \Rt_{b\ab}) \N^b \F
$$ $$ - 2(C^{\ab} \N_{\ab} \Ct^\b
+ P^{\b\ab}(\N_{\ab} S + C^{\gb} \Rt_{\gb\ab} + \Ct^\g R_{\g\ab}))
\N_\b \F $$ \be - 2 (\Ct^{\a} \N_{\a} C^{\bb}  - P^{\a\bb} (\N_\a S 
 + C^{\gb} \Rt_{\gb\a} + \Ct^\g R_{\g\a})) \N_{\bb} \F =0, \label{JJt} \ee

$$ \N^2 \St_{ac} - P^{\d\eb} \N_{[\d} \N_{\eb ]} \St_{ac} -
R^{ed} \Rt_{edac} + 2 \Rt_{b\db ac} \N^b C^{\db} + 2 R_{b\d} \N^b
\Ct_{ac}{}^\d - 2 \N_{\bb} \Ct_{ac}{}^\d \N_\d C^{\bb} $$ $$- 2 \N_\d
\St_{ba} \Ct_{c}{}^{b\d}  - C^{\bb} (\N^d \Rt_{d\bb ac} -
P^{\d\eb} \N_{[\d} \Rt_{\eb ]\bb} + 2 \Rt_{\bb \d ea}
\Ct_{c}{}^{e\d}) - \Ct_{ac}{}^{\b} (\N^d R_{d\b} - P^{\d\eb} \N_{[\d}
R_{\eb ]\b})$$ \be  +2 (\Ct_{ac}{}^\b R_{d\b} + C^{\bb} \Rt_{d\bb ac})
\N^d \F - 2 C^{\db} \N_{\db} \Ct_{ac}{}^\g \N_\g \F + 4 \St_{ab}
\Ct_{c}{}^{b\g} \N_\g \F - 2 \Ct_{ac}{}^{\b} \N_{\b} C^{\gb} \N_{\gb}
\F =0,\label{JNtac} \ee

$$ \N^2 S_{ab} - P^{\d\eb} \N_{[\d} \N_{\eb ]} S_{ab} -
\Rt^{cd} R_{cdab} + 2 R_{c\d ab} \N^c \Ct^{\d} + 2 \Rt_{c\db} \N^c
C_{ab}{}^{\db} -2 \N_{\gb} \Ct^\d \N_\d C_{ab}{}^{\gb} $$ $$- 2 \N_{\db}
S_{ca} C_{b}{}^{c\db} - \Ct^{\g} (\N^d R_{d\g ab} - P^{\d\eb}
\N_{[\d} R_{\eb ]\g ab} + 2 R_{\g \db ea} \Ct_{b}{}^{e\db}) -
C_{ab}{}^{\gb} (\N^d \Rt_{d\gb} - P^{\d\eb} \N_{[\d} R_{\eb ]\gb})$$ \be
 + 2 (\Ct^\g R_{d\g ab} + C_{ab}{}^{\gb} \Rt_{d\gb ab}) \N^d \F
- 2 C_{ab}{}^{\gb} \N_{\gb} \Ct^\d \N_\d \F + 4 S_{ac}C_{b}{}^{c\db}
\N_{\db} \F - 2 \Ct^{\g} \N_{\g} C_{ab}{}^{\db} \N_{\db} \F =0,
\label{NabJt} \ee

$$ \N^2 S_{abcd} - P^{\d\eb} \N_{[\d} \N_{\eb ]} S_{abcd} -
\Rt^{ef}{}_{cd} R_{efab} + 2 \Rt_{f\eb cd} \N^f C_{ab}{}^{\eb} + 2
R_{f\epsilon ab} \N^f \Ct_{cd}{}^\epsilon $$ $$ - 2 \N_{\eb} \Ct_{cd}{}^\g \N_\g
C_{ab}{}^{\eb}
 + 2\N_{\eb} S_{afcd} C_{b}{}^{f\eb} + 2 \N_\epsilon
S_{abcd} \Ct_{d}{}^{f\epsilon} - C_{ab}{}^{\eb} (\N^e \Rt_{e\eb cd} -
P^{\d\gb} \N_{[\d} \Rt_{\gb ]\eb cd} $$ $$+ 2 \Rt_{\eb \d ec}
\Ct_{d}{}^{e\d})  - \Ct_{cd}{}^{\epsilon} (\N^e R_{e\epsilon ab}
- P^{\d\gb} \N_{[\d} R_{\gb ]\epsilon ab}) + 2 (\Ct_{cd}{}^\epsilon R_{e\epsilon ab}
+ C_{ab}{}^{\eb} \Rt_{e\eb cd}) \N^e \F $$ \be - 2 C_{ab}{}^{\gb} \N_{\gb}
\Ct_{cd}{}^\epsilon \N_\epsilon \F + 4 S_{abcf} \Ct_{d}{}^{f\epsilon}
\N_\epsilon \F  - 2
\Ct_{cd}{}^{\g} \N_{\g} C_{ab}{}^{\eb} \N_{\eb} \F + 4
S_{afcd}C_{b}{}^{f\eb}\N_{eb}\F=0.\label{NabNcd} \ee

Since the Bianchi identities allow to write the curvature components
in terms of the torsion components, we expect that the beta
functions of the third set will be implied by the eight beta
functions already proven, i.e first and second set. In the same way
we expect that the beta functions of the fourth set will also be
implied by the first two sets of beta functions since the
constraints coming from holomorphicity and antiholomorphicity of the
BRST current allows to relate the background fields to some
components of the torsion. This is not too hard to check in the case
of lower dimension, for example, at dimension $5/2$ consider the
beta functions coming from the coupling to $J\Pb^{\bb}$

\be \N^a R_{a\bb} + \N_{(\eb} R_{\bb )\d} P^{\d\eb} + 2
(\N_{\bb} C^{\ab} - R_{\bb\g} P^{\g\ab}) \N_{\ab} \F + 2 C^{\ab}
\N_{\ab} \N_{\bb} \F = 0.\label{JPbbb} \ee 
By using $R_{a\bb} = T_{a\bb}{}^\g \O_\g$
and $R_{\bb\d} = \N_{\bb}\O_\d$, which follow from the definition of
the curvature, and $C^{\bb} = P^{\a\bb}\O_\a$, which follows from
the antiholomorphicity constraints, we find that (\ref{JPbbb}) can be
written as

\be (\N^c T_{c\bb}{}^{\a} -2 \N_{\bb} P^{\a\gb} \N_{\gb} \F +
2 P^{\a\gb} \N_{\gb} \N_{\bb} \F )\O_\a =0,\label{last} \ee 
so, the beta function
(\ref{dalPbbb}) with dimension $2$ implies (\ref{JPbbb}). Similarly we
checked that (\ref{Pialdtbb}) implies (\ref{PialJt}) and that the beta functions
with dimension $5/2$ (\ref{dalPbb}) and (\ref{Piadtbb}) imply respectively the
beta functions with dimension $3$ (\ref{JPbb}) and (\ref{PiaJt}) .

This concludes the study of the beta functions for the type II sigma
model. \\

Another application of the superstring sigma model will be 
presented in the next chapter, based in the heterotic sigma model, in which 
the quantum consistency of the BRST symmetry will be studied.

\chapter{ Yang-Mills Chern-Simons Corrections from the Pure Spinor Superstring}

The BRST properties play a key role when the superstring is coupled to
a generic background. In this chapter it will be shown how these
properties can be computed perturbatively in the inverse of the string
tension, allowing to find expected Yang-Mills Chern-Simons corrections.

\section{Lowest Order Constraints in $\a'$}
In this section we compute the constraints
coming from the nilpotency of the BRST charge and holomorphicity of the BRST
current at tree level.

The action which describes the Heterotic Superstring in a curved 
background can be obtained by adding the massless vertex operators
to the flat action and then covariantizing with respec to to the 
$D=10$ $N =1$ super-reparameterization invariance \cite{BerkovitsUE}, as
discussed in chapter 2 . 
The action is as follows

\be S = {1\over{2\pi\a'}} \int d^2 z ({1\over 2}\Pi^a \Pb^b
\eta_{ab} + {1\over 2}\Pi^A \Pb^B B_{BA}+d_\a \Pb^\a + \Pi^A \Jb^I A_{AI} + d_\a \Jb^I
W_I ^\a \label{actioncurved}\ee 
$$ \l^\a \omega_\b \Jb^I U_{I\a}{}^\b + \l^\a \omega_\b \Pb^C
\O_{C\a}{}^\b) + S_\l + S_{\Jb} + S_{\Phi}, $$
where $\Pi^A = \p Z^M E_{M}^A(Z)$, $\Pb^A = \pb Z^M E_M ^A (Z)$ and $E_M ^A (Z)$ 
is a supervielbein:\\ $ G_{MN} (Z) = E_M ^a E_N ^b \eta_{ba} $. $Z^M$ denote
the coordinates for the $D=10$ $N=1$ superspace $(X^m ,\t^\mu)$ with $m =
0,{\ldots} ,9$ and $\mu = 1,{\ldots}, 16$. $S_\l$ and $S_{\Jb}$, as before, are the 
actions for $\l$ and $\Jb^I = {1\over {2}} {\cal K}_{\cal AB}^{I}
\bar\psi^{\cal A}\bar\psi^{\cal B}$ respectively, with ${\cal A, B }=
0,{\ldots} ,32$. $S_{\Phi}$ is the action for the dilaton coupling to the
worldsheet scalar curvature. The nilpotency of the BRST charge is guaranteed 
in a flat background because of the pure spinor condition. Nevertheless, when the superstring is coupled to the curved
background, the background fields must be constrained in order to maintain this
nilpotency \cite{BerkovitsUE} \cite{ChandiaIX} .  We can find these constrains by performing a tree level
computation. To set that, we perform a background field expansion
\cite{deBoerSkenderis} by splitting every
worldsheet field into a classical and quantum part, where the classical
 part is assumed to satisfy the classical equation of motion and the quantum part
will allow to find propagators and form loops. Specifically, we will use
the following notation for the splitting

\be Z^M = X_0 ^M + Y^M , \,\,\, d_\a = d_{\a 0}+ \dh_\a,
\label{wsfsplitting} \ee 
$$ \l^\a = \l^\a _0 + \hat\l ^\a ,\,\,\, \omega_\a = \omega_{\a 0}+ \hat \omega_\a , \,\,\,
\bar\psi ^{\cal A} = \bar\psi^{\cal A} _0 +\hat{\bar\psi}^{\cal A}. $$
So the expansion for the term ${1\over {2\pi \a'}} \int d^2 z {1\over 2}\p Z^M \pb Z^N G_{NM}$ 
in (\ref{actioncurved}) in second order of the quantum field is

\be {1\over{2\pi\a'}}\int d^2 z({1\over 2}\p Y^a \pb Y^b
\eta_{ab} - {1\over 2}\p Y^a Y^B \Pb^C \Tt_{CB}{}^a - {1\over 2}\pb Y^a Y^B
\Pi^C \Tt_{CB}{}^a  
 +{1\over 4}\p Y^B  Y^C \Pb^a \Tt_{CB}{}^a \label{sndordereta} \ee
 $$ +{1\over 4}\pb Y^B Y^C \Pi^a
\Tt_{CB}{}^a + {1\over 2}Y^B Y^C \Pi^D
\Tt_{DC}{}^a \Pb^E \Tt_{EB}{}^a -{1\over 4}Y^B Y^C \Pi^{(a}\Pb^{D)} \Tt_{DCB}{}^a),$$
where $\Tt$ is the part of the torsion which only contains derivatives of
the vielbein: $\Tt_{MN}{}^A =  \p_{[M} E_{N]}{}^A$ and $\Tt_{DCB}{}^A =  - \Tt_{DC}{}^E \Tt_{EB}{}^A +(-)^{CD}\N_C
\Tt_{DB}{}^A$. Note that $\Tt$ in this chapter is not related
to the one used in the last chapter. Repeated bosonic indices in (\ref{sndordereta}) are assumed to be contracted
with the Minkowski metric. On the other hand, the expansion for ${1\over
{2\pi \a'}}\int d^2 z d_\a \pb
Z^M E_M{} ^\a$ is
\be {1\over{2\pi \a'}} \int d^2 z (\dh_{\a} \pb Y^\a -\dh_\a Y^B 
\Pb^C \Tt_{CB}{}^\a +{1\over 2} (d_{\a 0} + \dh_\a ) \pb Y^B Y^C
\Tt_{CB}{}^\a \label{sndorderdbp} \ee  
$$ -{1\over 2} (d_{\a 0}+ \dh_\a)Y^B \Pb^D Y^C (\p_C 
\Tt_{DB}{}^\a + \Tt_{CD}{}^E 
\Tt_{EB}{}^\a )+{1\over 2}\dh_\a \Pb^D Y^M Y^N \p_N E_{M}{}^B \Tt_{BD}{}^\a )$$
In the subsequent, we will drop off the $0$ subindex. From the first term in the last two expressions we can read the
propagators 

\be Y^a (x,\bar x)Y^b(z,\bar z) \to -\a' \eta^{ab}log|x-z|^2
,\,\,\, \dh_\a (x) Y^\b (z) \to {\a' \d_\a {}^\b\over{x-z}}.
\label{propagators} \ee

\subsection{Nilpotency at tree level}
The propagators (\ref{propagators})  allow to compute the conditions for the 
nilpotency of $Q_{BRST}$ perturbatively in $\a'$. In fact, we can easily
compute a tree level diagram using
the second propagator and the fifth term in (\ref{sndordereta}) expanding $e^{-S}$ 
in a series power, giving as a result

\be  \l^\a d_\a (w) \l^\b d_\b (z) = {1\over 2}\a' {1\over {w-z}}\l^\a \l^\b 
\Pi^c T_{\b\a}{}^c (z). \label{fst} \ee
The expansions for the remaining terms in the expansion of the action
(\ref{actioncurved}) are written in the
appendix. Initially we are interested in computing the tree level
diagrams coming from terms in the expansions with $\pb Y^A Y^B$, since
they will give rise to the same kind of poles as in (\ref{fst}). So, the contributions
to the pole $(w-z)^{-1}$ will be

\be {1\over 2} {\a' \over {w-z}}\l^\a \l^\b \Pi^c (T_{\b\a}{}^c
+H^c {}_{\b\a})(z) +{1\over 2}{\a' \over{w-z}}\l^\a \l^\b \Pi^\g H_{\g \b\a}
\label{polewz} \ee
$$+{\a' \over{w-z}} \l^\a \l^\b d_\g T_{\b\a}{}^\g (z) +{\a'
\over{w-z}}\l^\a \l^\b \l^\g \omega_\d R_{\b\a\g}{}^\d (z). $$
In our notation, the Torsion superfield $T_{\b\a}{}^\g$ is given by
\be T_{\b\a}{}^\g = \Tt_{\b\a}{}^\g - \O_{\b\a}{}^\g -
\O_{\a\b}{}^\g , \label{Torsionfermionic} \ee
while the curvature superfield is given by
\be  R_{\a\b\g}{}^\d = D_\a \O_{\b\g}{}^\d + D_\b \O_{\a\g}{}^\d +
\O_{\a\g}{}^\epsilon \O_{\b\epsilon}{}^\d + \O_{\b\g}{}^\epsilon
\O_{\a\epsilon}{}^\d + \Tt_{\a\b}{}^E
\O_{E\g}{}^\d , \label{Curvature} \ee
where $D_\a$ is the supersymmetric derivative. There also other possible tree level contractions of $\l^\a d_\a (w) \l^\b
d_\b (z)$ with terms including $\p Y^A Y^B$ which will lead to 
\be -{1\over 2} \a' {{\bar w -\bar z}\over {(w-z)^2}} \l^\a \l^\b
\Pb ^c (T_{\b\a}{}^c - H^c {}_{\a\b})(z) +{1\over 2}\a' {{\bar w - \bar
z}\over{(w-z)^2}}\l^\a \l^\b \Pb^\g H_{\g\a\b}(z) \label{polebwbz} \ee
$$ -\a' {{\bar w - \bar
z}\over {(w-z)^2}} \l^\a \l^\b \Jb^I F_{\a\b I}.$$
In our notation the field-strength superfield is given by
\be F_{\a\b I} = D_\a A_{\b I} + D_\b A_{\a I} + f_I
{}^{JK}A_{\a J}A_{\b K} + \Tt_{\a\b}{}^C A_{C I}.\label{fieldstrength} \ee
To compute the tree-level diagrams that give rise to the above result, 
we need to compute the integral

\be \int d^2 x {1\over {(w-x)(x-z)^2}} = -\int d^2 x \pb_x
{(\bar x -\bar w)\over {x-w}} {1\over{(x-z)^2}} = 2\pi {{\bar w- \bar z}\over
{(w-z)^2} } \label{integralddpi} \ee
From (\ref{polewz})\ and (\ref{polebwbz}) we deduce that the conditions for the nilpotency
of $Q_{BRST}$ at the lowest order in $\a'$ are
\be \l^\a \l^\b T_{\a\b}{}^C =0, \,\,\, \l^\a \l^\b
H_{C \a\b} =0, \,\,\, \l^\a\l^\b F_{\a\b I} = 0, \,\,\, \l^\a\l^\b\l^\g
\omega_\d R_{\b\a\g}{}^\d =0. \label{nilpotencyconstraints} \ee
These are the same set of constraints found in \cite{BerkovitsUE} and
\cite{ChandiaIX} .

\subsection{Holomorphicity at tree level}
To compute the conditions for holomorphicity of the BRST current $\pb j =
\pb (\l^\a d_\a ) =0$, we must know the expansion up to first order in $Y^\a$ of
the sigma model action. This expansion for the term ${1\over {2\pi \a'}}
\int d^2 z {1\over 2}\p Z^M \pb Z^N G_{NM}$ is 
\be {1\over{4\pi \a'}}\int d^2 [\Pi^a \pb Y^b \eta_{ab} + \Pb^a \p
Y^b\eta_{ab} + \Pi^b \Pb^D Y^C \Tt_{CD}{}^a \eta_{ab} + \Pi^D \Pb^a Y^C
\Tt_{CD}{}^b \eta_{ab} ]. \label{Gfstorder} \ee
The conditions for holomorphicity will appear as conditions for vanishing
to the independent couplings $\Pi^a \Pb^b$, $\Pi^\a \Pb^b$ and so on.
For example, forming a tree level diagram contracting $\pb d_\a$ in $\pb j$
with the third term in (\ref{Gfstorder}) , we obtain ${1\over 2}\l^\a \Pi^b \Pb^C
\Tt_{C\a}{}^d \eta_{bd}$. Following this procedure with all the terms in the 
expansion written in the appendix up to order $Y$, we arrive at
\be {1\over 2} \l^\a [-\Pi^b \Pb^c (T_{\a b}{}^d \eta_{dc} +
T_{\a c}{}^d \eta_{bd}+ H_{cb\a}) + \Pi^\b \Pb^c (T_{\b\a c} - H_{\b\a c}) +
\Pi^b \Pb^\g (T_{\g\a b} + H_{\g\a b}) \label{holtreelevel} \ee
$$ - \Pi^\b \Pb^\g H_{\g\b\a} 
-2d_\b \Pb^c T_{c\a}{}^\b -2d_\b \Pb^\g T_{\g\a}{}^\b+ 2 \Pi^b \Jb^I F_{b\a
I} +2 \Pi^\b \Jb^I F_{\b\a I} + 2 \l^\b \omega_\g \Pb^d R_{d\a\b}{}^\g 
$$ $$- 2d_\b \Jb^I (D_\a W_I^\b-W_J ^\b A_{\a K}f_I{}^{JK}-U_{I\a}{}^\b ) + 2
\l^\b\omega_\g \Jb^I (\O_{\a\d}{}^\g U_{I\b}{}^\d - \O_{\a\b}{}^\d U_{I\d}{}^\g
+ U_{J\b}{}^\g A_{\a K}f_I^{JK} $$ $$ 
- W_I^\d R_{\d\a\b}{}^\g - \p_\a U_{I\b}{}^\g )] = 0.$$ 
Since $\Pb^\a$ is related to $\Jb^I$ through $\Pb^\a = - \Jb^I W_I^\a$ by
using the equation of motion for the worldsheet field $d_\a$ in
(\ref{actioncurved}) , we arrive at the following set of constraints for the
holomorphicity of the BRST current at the lowest order in $\a'$
\be T_{\a (bc)} = - H_{\a bc} = T_{\a\b}{}^c - H_{\a\b}{}^c = T_{c\a}{}^\b
 = 0, \,\, \l^\a \l^\b R_{d\a\b}{}^\g =0, \,\, F_{\a\b I} = - {1\over 2}W_I^\g
 H_{\g\a\b}, \label{holomorphicityTL} \ee
 $$ F_{\a bI} = - W_I ^\g T_{\g\a b} , \,\,  \N_\a W_I^\b -
 T_{\a\g}{}^\b W_I^\g = U_{I\a}{}^\b , \,\, \l^\a \l^\b (\N_\a U_{I\b}{}^\g +
 R_{\a\d\b}{}^\g W_I ^\d) =0 .$$
This was the same set of constraints found in \cite{BerkovitsUE} and
\cite{ChandiaIX} .

\section{Yang-Mills Chern-Simons Corrections}
In this section we will compute $\a'$ corrections to the nilpotency
constraints (\ref{nilpotencyconstraints}) .  In the first subsection we will explain how to compute all of the
twenty possible contributions to the nilpotency of the BRST charge. In the second
subsection, we will explain how, adding some counter-terms, we can find the
Yang-Mills Chern-Simons $3-$form.

\subsection{One-loop Corrections to the Constraints}
In the expansion for the $\Pi^A \Jb^I A_{A I}$ term,  the following
will play a role in our computation: $\Pi^A Y^B  _0 \Jb^I _2 (\p_B A_{AI}
+ \Tt_{BA}{}^C A_{C I})(x)$ and
$\p Y^A \Jb ^I _2 A_{AI}(y)$. Contracting them with $\l^\a d_\a (w) \l^\b d_\b (z)$ we can form a 1-loop diagram
\be \epsfbox{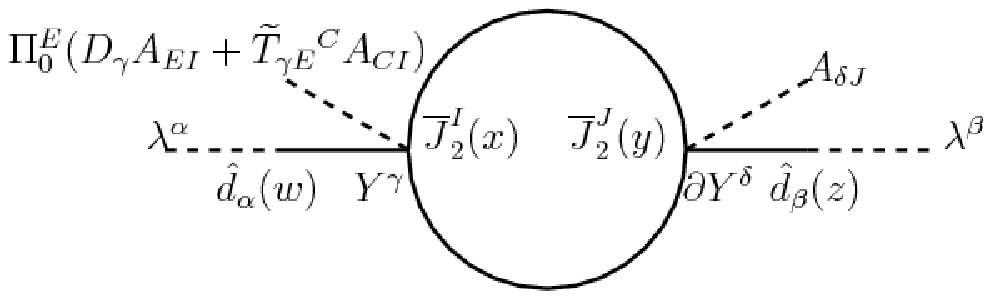}  \label{picI}\ee
The dashed lines denote background fields while the continuous lines denote
the contractions using the propagators. 
So one can compute how these terms contribute
to the nilpotency of $Q_{BRST}$. To determine the coefficient for this
diagram, note that there is an $1/2$ from the expansion of $exp[-S]$ and
there is a factor of $2$ coming from the possible ways to put the
superfields at $x$ or $y$. Denoting the integration over the world-sheet
fields by $\int [Dwsf]$, we find 

\be \l^\a d_\a (w) \l^\b d_\b (z)_{I}  = {1\over {(2\pi\a')^2}}\int [D
wsf]\int d^2 x d^2 y \l^\a \dh_\a (w) \l^\b \dh_\b (z) \label{loopI} \ee 
$$\Pi^E _0 Y^\g (D_\g
A_{E I}+\Tt_{\g E}{}^F A_{F I})(x) \p Y^\d A_{\d J}(y) \Jb^I _2 (x) \Jb_2 ^J (y) $$ 
\be = {{\a'^2 }\over {(2 \pi)^2 }}  \l^\a \l^\b \Pi^C _0 A_{\a
I}(D_\b
A_{C I} + \Tt_{\b C}{}^D A_{D I})(z)
\int d^2 x d^2 y {1\over{(w-x)^2(z-y)}} {1\over{(\bar x - \bar y)^2}}
\label{intermezzoI} \ee
$$
 - {\a'^2 \over {(2\pi )^2}} \l^\a \l^\b \Pi^C _0 A_{\b I}(D_{\a} A_{C I}
 +\Tt_{\a C}{}^D A_{D I}) (z)\int
d^2 x d^2 y {1\over{(w-y)(z-x)^2}} {1\over{(\bar x - \bar y)^2}} ,$$
where $\Jb^I _2 (\bar x) \Jb^J _2 (\bar y) \to {(\a')^2\d^{IJ} \over{(\bar x - \bar y )^2}} $.
The second line in the last equation is obtained from minus the first by
interchanging $\a$ with $\b$ and $w$ with $z$. So, we will just compute one
of the integrals.
\be \int d^2 x d^2 y {1\over {(w-x)^2(z-y)(\bar x - \bar y)^2}} =
\int d^2 x d^2 y {1\over {(w-x)^2(z-y)}}\pb_{\bar y} {1\over
{\bar x -\bar y}} \label{integralI} \ee
$$  = 2\pi \int d^2 x d^2 y {\d^2 (y-z) \over {(w-x)^2(\bar
x - \bar y)}} = 2\pi \int d^2 x {1\over{(w-x)^2 (\bar x -\bar z)}}
,  $$  
where in the second step we integrated by parts with respect to $\bar y$.
In the last integral we can integrate by parts with respect to $x$ to obtain

\be \int d^2 x d^2 y {1\over {(w-x)^2(z-y)}} {1\over {(\bar x -
\bar y)^2}} = -{{(2\pi)^2} \over {w-z}}. \label{integralIR} \ee
Then a first contribution to our check of nilpotency will be
\be \l^\a d_\a (w) \l^\b d_\b (z) _{I} = -2\a'^2  {{\l^\a
\l^\b}\over {w-z}} \Pi^C _0 A_{\b I}(\p_\a A_{C I} +\Tt_{\a C}{}^D A_{D
I})(z). \label{QQI} \ee

A second contribution comes from contracting $\l^\a
d_\a (w) \l^\b d_\b (z)$ with \\$\p Y^\g \Jb^I _2 A_{\g I}(x) \p Y^\d \Jb^J _2 A_{\d
J}(y)$ as shown in the diagram. 
\be \epsfbox{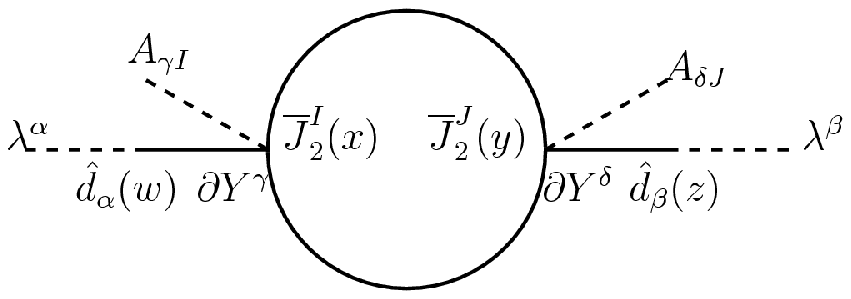}  \label{picII}\ee
To determine the coefficient of this diagram, note that there is an
$1/2$ coming from the Taylor expansion of $exp(-S)$. So we find

\be \l^\a d_\a (w) \l^\b d_\b (z) _{II} = {\a'^2 \over 2 }{{\l^\a \l^\b
(z)}\over{(2\pi)^2}}\int d^2 x d^2 y [{{A_{\a I}(x)A_{\b
I}(y)}\over{(w-x)^2(z-y)^2}} - {{A_{\b I}(x)A_{\a
I}(y)}\over{(w-y)^2(z-x)^2}}] {1\over(\bar x -\bar y)^2} \label{loopII} \ee
The second term in the integrand is obtained from minus the first by
interchanging $w$ with $z$ and $\a$ with $\b$. The integral we are left to
solve is
\be \G = \int d^2 x d^2 y {{A_{\a I}(x)A_{\b
I}(y)}\over{(w-x)^2(z-y)^2 (\bar x - \bar y)^2}} = -\int d^2 x d^2 y
{{\Pb^C \p_C A_{\a I}(x)A_{\b I}(y)}\over{(\bar y - \bar x)(w-x)^2(z-y)^2}}
\label{intermezzoIIa} \ee
$$ + \int d^2 x d^2 y {{A_{\a I}(x)A_{\b I}(y) \p_x \d^2 (x-w)}\over{(\bar y -
\bar x)(z-y)^2}},$$
where we integrated by parts with respect to $\bar x$. The first and second
integral on the right hand side of (\ref{intermezzoIIa}) can be integrated by parts
with respect to $y$ and $x$ to obtain 
\be \G = 2\pi \int d^2 x d^2 y {{\Pb^C \p_C A_{\a I}(x)A_{\b
I}(y)\d^2 (y-x)}\over{(z-y)(w-x)^2}} - 2\pi \int d^2 x d^2 y {{\Pi^C \p_C
A_{\a I}(x)A_{\b I}(y)\d^2 (x-w)}\over{(\bar y -\bar x)(z-y)^2}}.
\label{intermezzoIIb} \ee
Evaluating the superfields in $z$, using (\ref{integralddpi}) in the first integral and integrating by parts with
respect to $y$ in the second, we obtain
\be \G = -(2\pi)^2 {{\bar w -\bar z}\over{(w-z)^2}}\Pb^C \p_C
A_{\a I}A_{\b I}(z) - {{(2\pi)^2}\over{w-z}}\Pi^C \p_C A_{\a I}A_{\b I}(z).
\label{intermezzoIIc} \ee
Then
\be \l^\a d_\a (w) \l^\b d_\b (z) _{II} = -\a'^2 {{\bar w -\bar
z}\over{(w-z)^2}}\l^\a \l^\b \Pb^C \p_C
A_{\a I}A_{\b I}(z) - {\a'^2 \over{w-z}}\l^\a \l^\b \Pi^C \p_C A_{\a I}A_{\b
I}(z) \label{loopIIR} \ee
$$ +\a'^2  {{\bar w -\bar z}\over{(w-z)^2}}\pb \l^\a \l^\b A_{\a
I}A_{\b I} + {\a'^2 \over{w-z}}\p\l^\a \l^\b A_{\a I}A_{\b I}(z) $$

A third contribution to the nilpotency property comes from contractions of
$\Pi_0 ^A \Jb^I _2 A_{A I}$, twice  $\p Y^A \Jb_2
^I A_{AI}$ and $\l^\a d_\a (w) \l^\b d_\b (z)$ giving rise to the diagram
\be \epsfbox{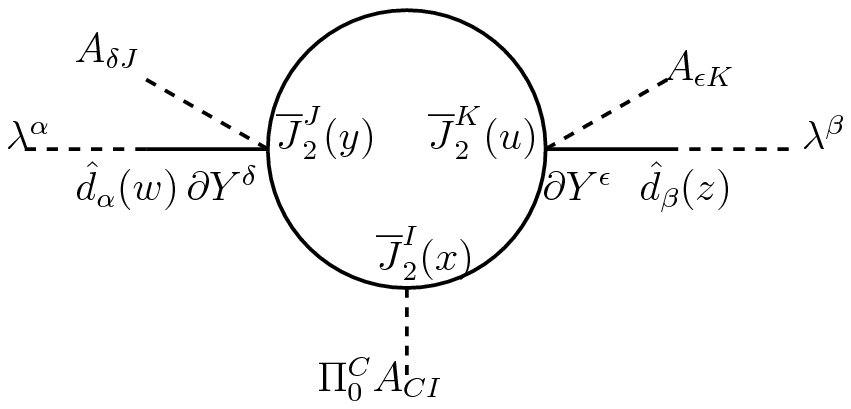}  \label{picIII} \ee
Since we are at order $S^3$ in the expansion of $e^{-S}$, there is an
${1\over{3!}}$ and also a factor of $3$ from the possible ways to put 
the superfields at $x$, $y$ and $u$, so there will be a $- 1/2$
coefficient in front:

\be \l^\a d_\a (w) \l^\b d_\b (z)_{III} = -{1\over{2(2\pi\a')^3}}\int
[Dwsf]\int d^2 x d^2 y d^2 u \l^\a \dh_\a (w) \l^\b \dh_\b (z)
\label{loopIII} \ee
$$\Pi^C _0 \Jb^I
_2 A_{C I}(x) \p Y^D \Jb_2 ^J A_{DJ}(y) \p Y^E \Jb_2 ^K A_{E K}(u). $$

\be =-{1\over{2(2\pi)^3 \a'}} \l^\a \l^\b \Pi^C _0 A_{CI} A_{\g
J}A_{\d K}(z)\int d^2 x d^2 y d^2 u ({\d_\a {}^\g \d_\b {}^\d
\over{(w-y)^2(z-u)^2}} \label{intermezzoIII} \ee 
$$ -{\d_\a {}^\d \d_\b {}^\g
\over{(w-u)^2(z-y)^2}})  \Jb_2 ^I (x) \Jb^J _2 (y) \Jb^K _{2} (u). $$
It is not hard to verify that that 
\be \Jb_2 ^I (x) \Jb^J _2 (y) \Jb^K _{2} (u) = {(\a')^3 f^{IJK} \over{(\bar x -
\bar y)(\bar y -\bar u)(\bar x - \bar u)}} + {\ldots} , \label{JJJ} \ee 
where by ${\ldots} $ means less singular poles which are not important in
this computation. Then the type of integrals we must compute are 
\be  \G_1 =\int d^2 x d^2 y d^2 u {1
\over{(w-y)^2(z-u)^2 (\bar x - \bar y)(\bar y -\bar u)(\bar x - \bar u)}}.
\label{integralIII} \ee
The integral in $x$ gives

\be \int d^2 x {1\over {(\bar x - \bar y)(\bar x - \bar u)}} =
\int d^2 x \partial_x ({{x-y}\over{\bar x - \bar y}}){1\over{\bar x - \bar
u}} =-2\pi {{y-u}\over{\bar y -\bar u}}, \label{integralx} \ee
so (\ref{integralIII}) yields
\be \G_1 = -2\pi \int d^2 y d^2 u \p_y ({1\over w-y}) {y-u \over
(z-u)^2 (\bar y - \bar u )^2}.\label{interalIIIa} \ee
Integrating by parts in $y$, $\bar y$ and then in $u$ we find $\G_1 = (2\pi)^3
/(w-z)$. In this way (\ref{loopII}) gives
\be \l^\a d_\a (w) \l^\b d_\b (z) _{III}= - (\a')^2 {\l^\a \l^\b \over
w-z }f^{IJK} \Pi^C _0 A_{CI} A_{\a J} A_{\b K} (z). \label{loopIIIR} \ee

Note that a fourth loop could be formed with ${1\over 4}\pb Y^\a  Y^\b \Pi^c
(T_{\b\a}{}^c+H^c{}_{\b\a})$, $\dh_\a \Jb^I _2 W_I ^\a$ and $\p Y^\a \Jb^I _2
A_{\a I}$ as shown in the diagram below.
\be \epsfbox{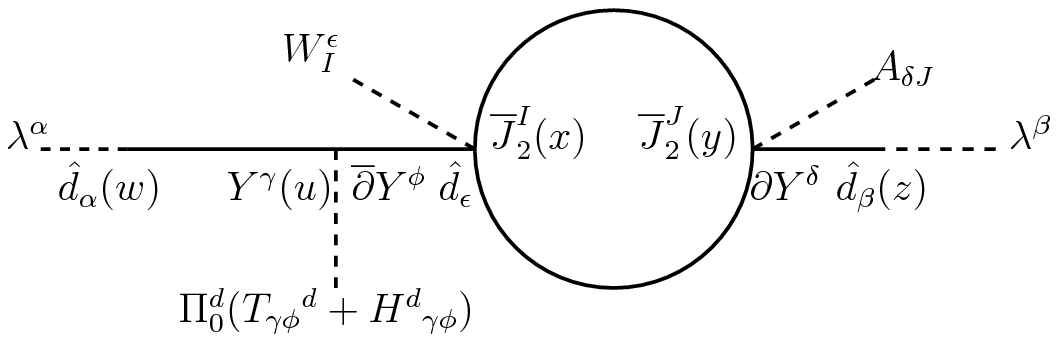} \label{picIV}\ee
In this case, we are also at the order $S^3$, so there is an ${1\over {3!}}$
which is cancelled by the symmetry factor responsible for the localization 
of the superfields, either at $x$, $y$ or $u$. The $1\over 4$ coming from the
coefficient of the term with $\Pi^c$ is cancelled by a symmetry factor of the
possible ways of contraction:

\be \l^\a d_\a (w) \l^\b d_\b (z)_ {IV} = -{{\a'^2}\over{(2\pi)^2} }
\l^\a \l^\b \Pi^c (T_{\d\a}{}^c + H^c {}_{\d\a})W^\d _I A_{\b I}(z)
\label{loopIV} \ee $$\times \int d^2 x
d^2 y d^2 u {{\d^2 (x-w) }\over{(z-u)^2(y-x)(\bar y - \bar u)^2}}$$
Integrating $x$ we have to solve
\be \int d^2 y d^2 u {1\over{(z-u)^2(y-w)(\bar y - \bar
u)^2}} = -2\pi \int d^2 y d^2 u {{\d^2 (y-w) }\over {(\bar u - \bar
y)(z-u)^2}} = -{{(2\pi )^2}\over{w-z}}. \label{intermezzoIVa} \ee

Then 
\be \l^\a d_\a (w) \l^\b d_\b (z)_{IV} = {{\a'^2}\over{w-z}}\l^\a \l^\b
\Pi^c (T_{\a\d}{}^c + H^c {}_{\a\d})W^\d _I A_{\b I}(z) \label{loopIVR} \ee

Considering the same last diagram but with the vertex ${1\over 4} \Pi^\g
H_{\g\b\a}$ instead of ${1\over 4} \Pi^c(T_{\b\a}{}^c + H_{\b\a}{}^c)$,
gives a fifth contribution to the coupling to $\Pi^\g$
\be \l^\a d_\a (w) \l^\b d_\b (z)_{V} = {{\a'^2}\over{w-z}}\l^\a \l^\b
\Pi^\g H_{\g\a\d}W^\d _I A_{\b I}(z) \label{loopVR} \ee

A sixth contribution can be formed with ${1\over 4}\Pi^c \pb Y^A Y^B (\Tt_{BA}{}^c + H^c {}_{BA})$
and twice $\p Y^A \Jb^I _2 A_{AI}$:
\be \epsfbox{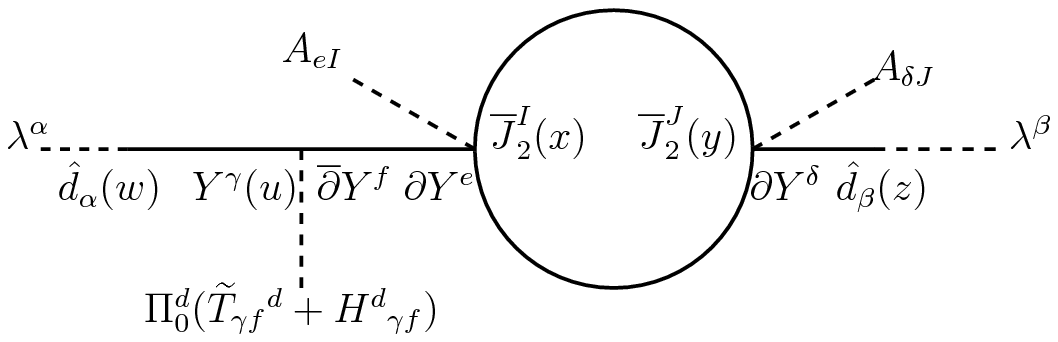} \label{picV} \ee
There are $8$ possible ways of making the
contractions, a $3$ factor from the possible ways to put the superfields at
$x$, $y$ or $u$, an $1/3!$ because we are at $S^3$ in the expansion, and the
factor of $1/4$ of the $\Pi^c$ term gives a one coefficient:

\be \l^\a d_\a (w) \l^\b d_\b (z)_{VI} = -{{\a'^2}\over{(2\pi
)^2}}\l^\a \l^\b \Pi^c(\Tt_{d\a}{}^c + H^c {}_{d\a})A_{d I}A_{\b I}(z)
\times \label{loopVI} \ee 
$$ \int d^2 x d^2 y d^2 u {{\d^2 (x-w) }\over{(y-x)(z-u)^2}} {1\over{(\bar  y
-\bar u)^2}}. $$
The integral is the same as in (\ref{loopIV}) , so the answer is 
\be \l^\a d_\a (w) \l^\b d_\b (z)_{VI} = {{\a'^2}\over{w-z}}\l^\a \l^\b
\Pi^c (\Tt_{d\a}{}^c + H^c {}_{d\a})A_{d I}A_{\b I}(z).
\label{loopVIR} \ee

In the same way, the last diagram but with the vertex ${1\over 4}\Pi^\g
H_{\g BA}$ instead of ${1\over 4} \Pi^c (T_{BA}{}^c + H_{BA}{}^c)$ leads to
a seventh contribution
\be \l^\a d_\a (w) \l^\b d_\b (z)_{VII} = {{\a'^2}\over{w-z}}\l^\a \l^\b
\Pi^\g H_{\g d\a}A_{d I}A_{\b I}(z).\label{looopVIIR} \ee

An eighth contribution can be formed with $-{1\over 2} \pb Y^a Y^\b
\Pi^C \Tt_{C\b}{}^a$ and twice $\p Y^A \Jb^I _2 A_{AI}$:
\be \epsfbox{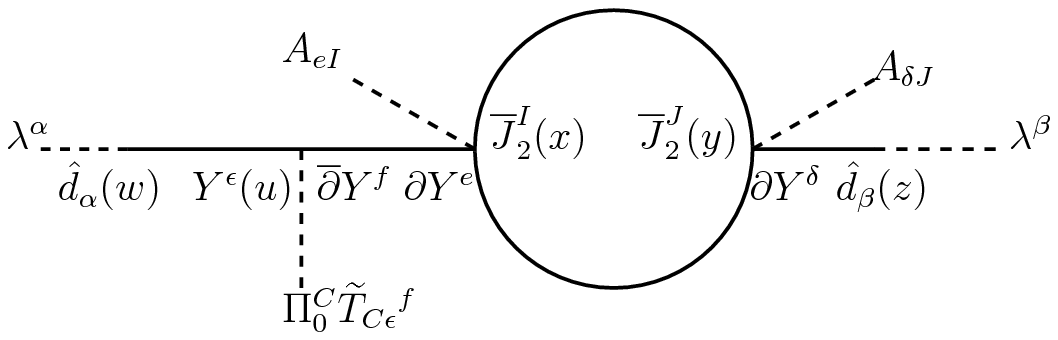}\label{picVI} \ee
There are $4$ possible
ways of making the contractions, a $3$ factor from the possible ways to put
the superfields at $x$, $y$ or $u$, an $1\over 3!$ because we are at $S^3$
order in the expansion and a factor of $1/2$ of the $\Pi^a$ coefficient,
giving at the end a $1$ coefficient:

\be\l^\a d_\a (w) \l^\b d_\b (z)_{VIII} = -{{\a'^2}\over{(2\pi )^3}}
\l^\a \l^\b \Pi^C \Tt_{C\a}{}^d A_{\b I} A_{d I}(z) \times
\label{loopVIII} \ee
$$ \int d^2x d^2 y d^2 u {{-2\pi \d^2 (u-x)}\over{(w-x)(z-y)^2}} {1\over{(\bar u - \bar y)^2
}}.$$
Integrating in $u$, the integral we have to solve is

\be \int d^2 x d^2 y {1\over{(w-x)(z-y)^2(\bar x - \bar y)^2}} = 2\pi \int  d^2 x d^2 y 
{{\d^2 (x-w) }\over{(z-y)^2 (\bar y - \bar x)}}  =  {{(2\pi)^2
}\over{w-z}} ,\label{ingtermezzoVIIIa} \ee
then
\be \l^\a d_\a (w) \l^\b d_\b (z)_{VIII} = {{\a'^2}\over{w-z}}\l^\a
\l^\b \Pi^C \Tt_{C\a}{}^d A_{\b I} A_{d I}(z). \label{loopVIII} \ee

Let's consider the couplings to $\Pb^A$.

A diagram like (\ref{picIV}) can be formed with ${1\over 4} \Pb^c \p Y^A Y^B
 (\Tt_{BA}{}^c - H^c {}_{BA})$, $\p Y^A \Jb^I _2 A_{AI}$ and $\dh_\a \Jb^I _2
W_I ^\a$. There are $4$ possible
ways of making the contractions, a $6$ factor from the possible ways to put
the superfields at $x$, $y$ or $u$, an $1\over 3!$ because we are at $S^3$
order in the expansion and a factor of $1/4$ of the $\Pb^c$ coefficient,
giving at the end a $1$ coefficient to this ninth contribution:

\be \l^\a d_\a (w) \l^\b d_\b (z)_{IX} =  {{\a'^2}\over{(2\pi
)^3}} \l^\a \l^\b \Pb^c (T_{\d\a}{}^c - H^c {}_{\d\a}) W^\d _I A_{\b I}(z)
\times \label{loopIX}\ee  $$ \int
d^2 x d^2 y d^2 u {1\over{(w-x)^2(z-u)^2 (y-x)(\bar y - \bar u)^2}}$$
Integrating $\bar y$ by parts, we are left to solve the integral
\be \int d^2x d^2 y d^2 u {{\d^2 (y-x)}\over{(w-x)^2 (z-u)^2
(\bar u - \bar y)}} = 2\pi \int d^2 x
{1\over{(w-x)(z-x)^2}}.\label{intermezzoIXa} \ee
The right hand side in the last equation is the same as
(\ref{integralddpi}) , so
\be \l^\a d_\a (w) \l^\b d_\b (z)_{IX} = - \a'^2 {{\bar w -
\bar z}\over{(w-z)^2}} \l^\a \l^\b \Pb^c (T_{\d\a}{}^c - H^c
{}_{\d\a})W^\d _I A_{\b I}(z).\label{loopIXR} \ee

In the same way, considering vertex $-{1\over 4} \Pb^\g H_{\g BA}$ instead of
$-{1\over 4} \Pb^c (\Tt_{BA}{}^c - H_{BA}{}^c )$ leads to the tenth contribution

\be \l^\a d_\a (w) \l^\b d_\b (z)_{X} = \a'^2 {{\bar w -
\bar z}\over{(w-z)^2}} \l^\a \l^\b \Pb^\g H_{\g\d\a}W^\d _I A_{\b
I}(z) \label{loopX} \ee

An eleventh contribution comes from a diagram like (\ref{picV}) which can be formed with ${1\over 4} \Pb^c \p Y^A Y^B
 (\Tt_{BA}{}^c - H^c {}_{BA})$ and twice $\p Y^A \p \Jb^I _2 A_{AI}$ . There are
$8$ possible ways of making the contractions, a $3$ factor from the possible ways to put
the superfields at $x$, $y$ or $u$, an $1\over 3!$ because we are at $S^3$
order in the expansion and a factor of $1/4$ of the $\Pb^c$ coefficient,
giving at the end a $1$ coefficient: 

\be \l^\a d_\a (w) \l^\b d_\b (z)_{XI} = {{\a'^2}\over{(2\pi)^3}}
\l^\a \l^\b\Pb^c (\Tt_{d\a}{}^c - H^c {}_{d\a})A_{dI}A_{\b I}(z)
\times \label{loopXI} \ee $$\int d^2 x d^2
y d^2 u {1\over{(w-x)^2(z-u)^2(y-x)(\bar u - \bar y)}}.$$
The last integral is the same as the integral in (\ref{loopIX}) , so the result is
\be \l^\a d_\a (w) \l^\b d_\b (z)_{XI} = -\a'^2 {{\bar w -\bar
z}\over{(w-z)^2}}\l^\a \l^\b \Pb^c (\Tt_{d\a}{}^c - H^c {}_{d\a})A_{d I}A_{\b
I}(z).\label{loopXIR} \ee

In the same way, a twelfth contribution comes from considering the vertex
$-{1\over 4}\Pb^\g H_{\g BA}$ instead of the vertex ${1\over 4} \Pb^c
(\Tt_{BA}{}^c - H_{BA}{}^c)$, leading to 
\be \l^\a d_\a (w) \l^\b d_\b (z)_{XII} = \a'^2 {{\bar w - \bar
z}\over{(w-z)^2}}\l^\a \l^\b \Pb^\g H_{\g d\a}A_{dI}A_{\b
I}(z).\label{loopXII} \ee

Another diagram like (\ref{picVI}) can be formed with $-{1\over 2} \p Y^a Y^\b
\Pb^C \Tt_{C\b}{}^a$, $\p Y^a \Jb^I _2 A_{a I}$ and $\p Y^\a \Jb^I _2 A_{\a
I}$, giving rise to a thirteenth contribution

\be \l^\a d_\a (w) \l^\b d_\b (z)_{XIII} = -\a'^2 {{\bar w -\bar
z}\over{(w-z)^2}} \l^\a \l^\b \Pb^C \Tt_{C\a}{}^d A_{d I}A_{\b
I}(z).\label{loopXIII} \ee

A fourteenth contribution and the last for the couplings to $\Pb^A$ can be formed with $-\dh_\a Y^B \Pb^C
\Tt_{CB}{}^\a$ and twice $\p Y^A \Jb A_{AI}$:
\be \epsfbox{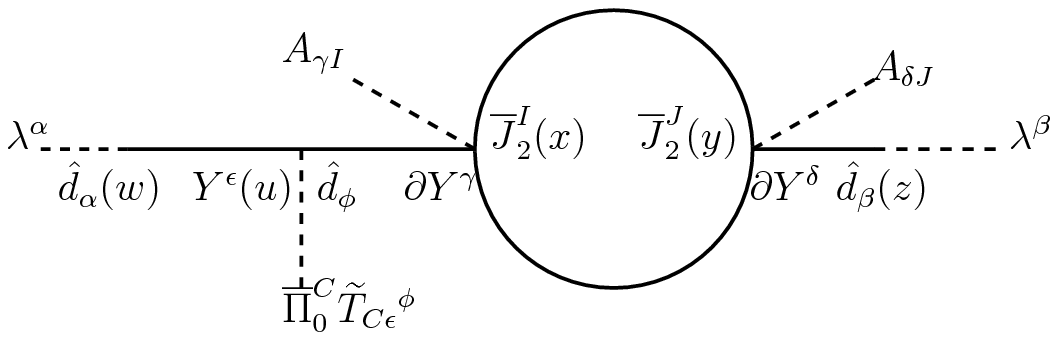} \label{picVII} \ee
giving as result
\be \l^\a d_\a (w) \l^\b d_\b (z) _{XIV} = 2\a'{{\bar w - \bar
z}\over{(w-z)^2}} \l^\a \l^\b \Pb^C A_{\b I}\Tt_{C\a}{}^\g A_{\g
I}\label{loopXIVR} \ee

Let's consider the couplings to $\Jb^I _0$

A fifteenth contribution to the nilpotency will come from a diagram formed with ${1\over 2} \p Y^A Y^B \Jb^I _0
(\p_{[B}A_{A]I} + \Tt_{BA}{}^C A_{C I})$, $\dh_\a \Jb^I _2 W_I ^\a$ and $\p
Y^\a \Jb^I _2 A_{\a I}$:
\be \epsfbox{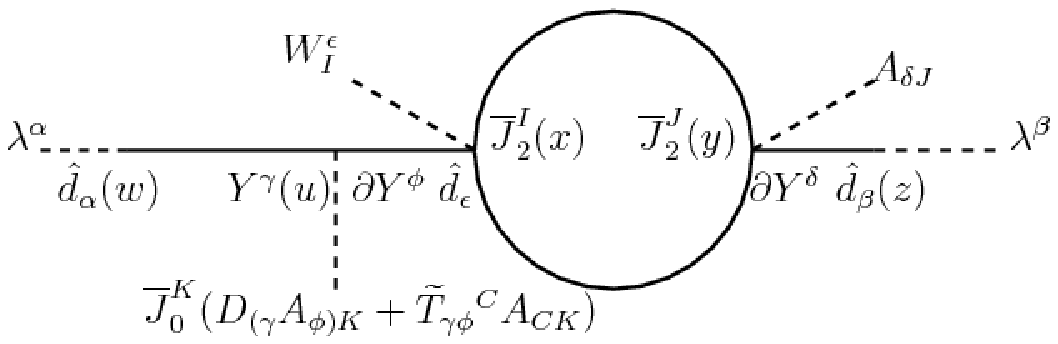} \label{picVIII} \ee
There are $4$
possible ways of making the contractions, a $6$ factor from the possible ways 
to put the superfields at $x$, $y$ or $u$, an $1\over 3!$ because we are at
the $S^3$ order in the expansion and a factor of $1/2$ of the $\Jb_0 ^I$ 
coefficient, giving at the end a 2 factor:
\be \l^\a d_\a (w) \l^\b d_\b (z)_{XV} =
{{2\a'^2}\over{(2\pi)^3}}\l^\a \l^\b \Jb^I _0 (D_{(\g}A_{\a )I} +
\Tt_{\g\a}{}^C A_{C I})W_J ^\g A_{\b J} (z) \times \label{loopXV} \ee  $$ \int d^2 x d^2 y d^2 u
{1\over{(w-x)^2(z-u)^2(y-x)(\bar u - \bar y)^2}}.$$
The last integral is again the same as in (\ref{loopIX}) , so the result is 
\be \l^\a d_\a (w) \l^\b d_\b (z)_{XV} = -2\a'^2 {{\bar w -\bar
z}\over{(w-z)^2}}\l^\a \l^\b \Jb^I _0 (D_{(\g}A_{\a )I} +
\Tt_{\g\a}{}^C A_{C I})W_J ^\g A_{\b J} (z).\label{loopXVR} \ee

A sixteenth contribution can be formed with ${1\over 2} \p Y^A Y^B
\Jb^I _0 (\p_{[B}A_{A]I} + \Tt_{BA}{}^C A_{C I})$ and twice $\p Y^A \Jb^I _2
A_{A I}$:
\be \epsfbox{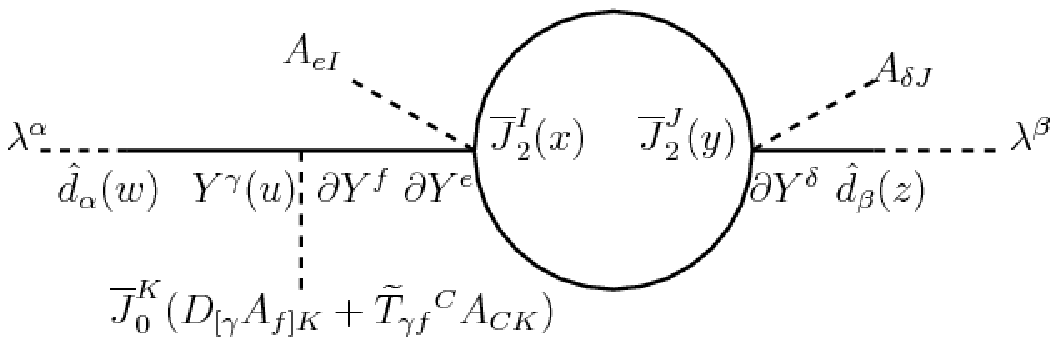} \label{picIX} \ee
There are $8$ possible ways of making the contractions, a $3$ factor 
from the possible ways to put the superfields at $x$, $y$ or $u$, an
$1\over 3!$ because we are at the $S^3$ order in the expansion and a
factor of $1/2$ of the $\Jb^I _0$ coefficient, giving at the end a $2$ coefficient:
\be \l^\a d_\a (w) \l^\b d_\b (z)_{XVI} =
2{{\a'^2}\over{(2\pi)^3}}\l^\a \l^\b \Jb_0 ^I (\p_{[c}A_{\a]I} + \Tt_{c\a}{}^D
A_{D I})A_{c J}A_{\b J} (z)\times \label{loopXVI} \ee $$ \int d^2 x d^2 y d^2 u {1\over{(w-x)^2
(z-u)^2 (y-x)(\bar y - \bar u)^2}},$$
which contains the same integral as before, so the result is  
\be \l^\a d_\a (w) \l^\b d_\b (z)_{XVI} = -2 \a'^2 {{\bar w - \bar
z}\over{(w-z)^2}}\l^\a \l^\b \Jb^I _0 (\p_{[c}A_{\a]I} + \Tt_{c\a}{}^D
A_{D I})A_{c J}A_{\b J} (z). \label{loopXVIR} \ee

Finally, let's consider the couplings to $d_\a$.

A seventeenth contribution can be formed with ${1\over2} d_\a \pb Y^\b Y^\g
\Tt_{\g\b}{}^\a$, 
$\dh_\a \Jb^I _2  W_I ^\a$ and $\p Y^\a \Jb^I _2 A_{\a I}$:
\be \epsfbox{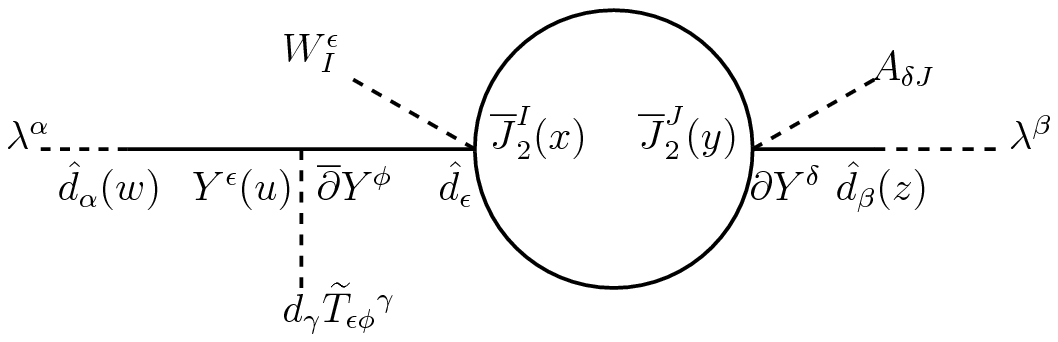}  \label{picX} \ee
There are $4$ possible
ways of making the contractions, a $6$ factor from the possible ways to put the
superfields at $x$, $y$ or $u$, an $1\over 3!$ because we are at the $S^3$ order in the
expansion and a factor of $1/2$ of the $d_\a$ coefficient, giving at the end a $2$
coefficient:

\be \l^\a d_\a (w) \l^\b d_\b (z)_{XVII} =
-2{{\a'^2}\over{(2\pi)^2}}\l^\a \l^\b d_\g \Tt_{\d\a}{}^\g
 W^\d _I A_{\b I} (z) \label{loopXVII} \ee $$ \times \int d^2 x d^2 y d^2 u {{\d^2
 (x-w)}\over{(z-u)^2(y-x)(\bar y - \bar u)^2}} $$
Integrating $x$, the integral we are left to solve is
\be \int d^2 y d^2 u {1\over{(z-u)^2(y-w)(\bar y - \bar
u)^2}} = -2\pi \int d^2 y d^2 u {{\d^2 (y-w)}\over{(\bar u - \bar
y)(z-u)^2}} = -{{(2\pi )^2}\over{w-z}}, \label{intermezzoXVI} \ee
So, 
\be \l^\a d_\a (w) \l^\b d_\b (z)_{XVII} = {{2\a'^2}\over{w-z}}
\l^\a \l^\b d_\g \Tt_{\d\a}{}^\g W_I ^\d A_{\b I}(z). \label{XVIIR} \ee

An eighteenth contribution can be formed with ${1\over2} d_\a \pb Y^B Y^C \Tt_{CB}{}^\a$ and twice
$\p Y^A \Jb^I _2 A_{A I}$:
\be \epsfbox{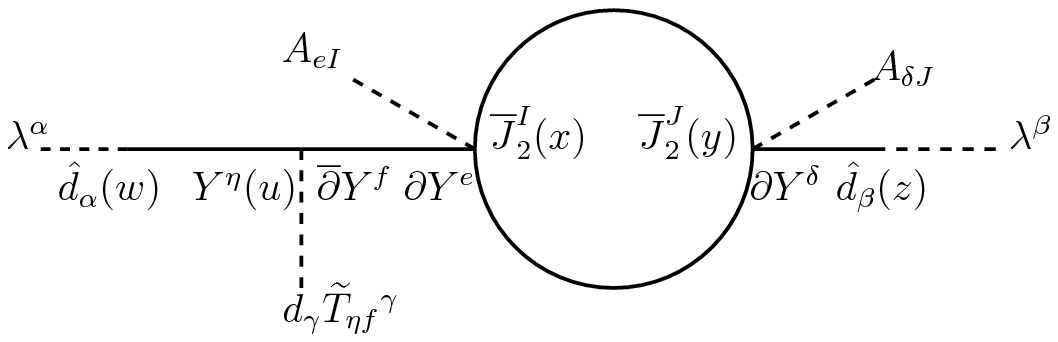}  \label{picXI} \ee
There are $8$ possible ways of making the contractions, a $3$ factor
from the possible ways to put the superfields at $x$, $y$ and $u$, an $1\over 3!$ because we are
at the $S^3$ order in the expansion and a factor of $1/2$ of the $d_\a$ coefficient , giving a $2$ coefficient:

\be \l^\a d_\a (w) \l^\b d_\b (z)_{XVIII} = 2{{\a'^2}\over{(2\pi
)^2}} \l^\a \l^\b d_\g \Tt_{c\a}{}^\g A_{cI}A_{\b I}(z)\times
\label{loopXVIII} \ee $$ \int d^2 x
d^2 y d^2 u {{\d^2 (x-w)}\over{(z-u)^2 (y-x)(\bar y - \bar u)^2}}.$$
This integral is the same as in (\ref{intermezzoXVI}) , so the result is 
\be \l^\a d_\a (w) \l^\b d_\b (z)_{XVIII} = -{{2\a'^2}\over{w-z}}\l^\a \l^\b d_\g 
\Tt_{c\a}{}^\g A_{cI} A_{\b I}(z). \label{loopXVIIIR} \ee

Because of the pure spinor condition, the action is invariant under $\d
\omega_\a = (\L_b \g^b \l)_\a$, so $U_{I\a}{}^\b = U_I \d_\a {}^\b + {1\over
4}U_{I cd}(\g^{cd})_\a {}^\b$. We can form a nineteenth one-loop diagram by
contracting $J \Jb^I_2 U_I (x)$ with $\p Y^\a \Jb^I _2 A_{\a I}$:
\be \epsfbox{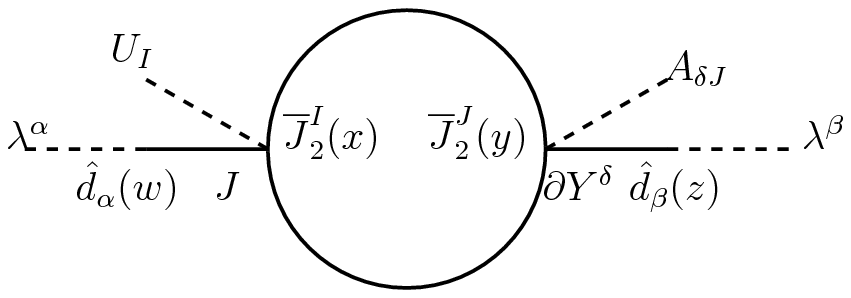} \label{picXII} \ee
giving the contribution 
\be \l^\a d_\a (w) \l^\b d_\b (z) _{XIX} = -2{{\a'^2}\over{w-z}}
\l^\a \l^\b d_\g \d_\a {}^\g A_{\b I} U_I \label{loopXIXR} \ee

Similarly, a diagram like (\ref{picXII}) can be formed contracting ${1\over
2}N^{ab}\Jb^I_2
U_{I ab}(x)$ with $\p Y^\a \Jb^I_2 A_{\a I}$, giving as contribution
\be \l^\a d_\a (w) \l^\b d_\b (z) _{XX} = -{1\over 2}{{\a'^2}\over{w-z}}
\l^\a \l^\b d_\g (\g^{ef})_\a {}^\g U_{I ef}A_{\b I} \label{loopXXR} \ee

Now, let us summarize our results adding the twenty one-loop contributions to the tree level
constraints. Each independent worldsheet coupling will receive corrections,
as indicated below:

Corrections to the the coupling to $\Pi^c$ 
\be {1\over 2} {\a' \over {w-z}}\l^\a \l^\b \Pi^c [(T_{\b\a}{}^c
+H^c {}_{\b\a}) -4\a' A_{\b I}(D_\a A_{c I} +\Tt_{\a c}{}^D A_{D I}) +
2\a'A_{\b I} \p_c A_{\a I} \label{treeplusonelooppic} \ee 
$$ -2\a'f^{IJK} A_{cI} A_{\a J} A_{\b K} +2\a' (T_{\a\d}{}^c +
H^c {}_{\a\d})W^\d _I A_{\b I} +2\a' (T_{d\a}{}^c + T_{c\a}{}^e \eta_{ed}+
H^c {}_{d\a})A_{d I}A_{\b I}](z).$$

Corrections to the coupling to $\Pb^c$ 
\be -{1\over 2} \a' {{\bar w -\bar z}\over {(w-z)^2}} \l^\a \l^\b
\Pb ^c [(T_{\b\a}{}^c - H^c {}_{\a\b}) - 2\a' A_{\b I} \p_c A_{\a I}+2\a'(T_{\d\a}{}^c - H^c
{}_{\d\a})W^\d _I A_{\b I} \label{treeplusonelooppb} \ee 
$$ + 2\a'(T_{d\a}{}^c  + T_{c\a}{}^e \eta_{ed}-
H^c {}_{d\a})A_{d I}A_{\b I} -4\a'A_{\b I}\Tt_{c\a}{}^\g A_{\g I}](z). $$

Corrections to the coupling to $\Pi^\g$ 
\be {1\over 2}{\a' \over{w-z}}\l^\a \l^\b \Pi^\g [H_{\g
\b\a} -4\a' A_{\b I}(D_\a A_{\g I} + \Tt_{\a \g}{}^D A_{D I}) -2\a'A_{\b
I}D_\g A_{\a I} \label{treeplusonelooppig} \ee 
$$ - 2\a'f^{IJK}  A_{\g I} A_{\a J} A_{\b K}+ 2\a' H_{\g\a\d}W^\d _I
A_{\b I} +2\a' (T_{\g\a d} - H_{\g \a d})A_{d I}A_{\b I}](z).$$

Corrections to the coupling to $\Pb^\g$ 
\be {1\over 2}\a' {{\bar w - \bar
z}\over{(w-z)^2}}\l^\a \l^\b \Pb^\g [H_{\g\a\b} -2\a'A_{\b I }D_\g A_{\a I}+ 2\a'H_{\g\d\a}W^\d _I
A_{\b I}-2\a'(H_{\g \a }{}^d + T_{\g\a}{}^d )A_{dI}A_{\b I}
\label{treeplusonelooppbg} \ee $$ + 4\a'A_{\b
I}\Tt_{\g\a}{}^\d A_{\d I}](z).$$

Corrections to the coupling to $d_\g$ 
\be {\a' \over{w-z}} \l^\a \l^\b d_\g [T_{\b\a}{}^\g +
2\a'\Tt_{\d\a}{}^\g W_I^\d A_{\b I} -2 \a'\Tt_{c\a}{}^\g A_{cI}A_{\b I}
-2\a'U_{I\a}{}^\g A_{\b I}]. \label{treeplusoneloopdg} \ee

Corrections to the coupling to $\Jb^I _0$ 
\be -\a' {{\bar w - \bar
z}\over {(w-z)^2}} \l^\a \l^\b \Jb^I [F_{\a\b I} + 2\a' (D_{(\g}A_{\a )I} +
\Tt_{\g\a}{}^C A_{C I})W_J ^\g A_{\b J}  \label{treeplusoneloopJb} \ee $$+ 2\a' (\p_{[c}A_{\a]I} + \Tt_{c\a}{}^D
A_{D I})A_{c J}A_{\b J} ](z). $$

\subsection{Addition of Counter-terms}
Let's now concentrate in finding the Yang-Mills Chern-Simons $3-$form by
adding appropriate counter-terms. Keeping in mind the lowest order in $\a'$ 
holomorphicity constraints $T_{\a bc} + T_{\a cb} = 0 =
H_{\a bc}$; the conditions for nilpotency at one loop look like

From the coupling to $\Pi^c$
\be \l^\a \l^\b [(T_{\b\a}{}^c
+H^c {}_{\b\a}) -4\a' A_{\b I}(D_\a A_{c I} +\Tt_{\a c}{}^D A_{D I}) +
2\a'A_{\b I} \p_c A_{\a I}\label{nilpotencyonelooppi} \ee 
$$ -2\a'f^{IJK} A_{cI} A_{\a J} A_{\b K} +2\a' (T_{\a\d}{}^c +
H^c {}_{\a\d})W^\d _I A_{\b I} ](z) =0.$$

From the coupling to $\Pb^c$
\be \l^\a \l^\b [(T_{\b\a}{}^c - H^c {}_{\a\b}) - 2\a'
A_{\b I} \p_c A_{\a I} +2\a'(T_{\d\a}{}^c - H^c
{}_{\d\a})W^\d _I A_{\b I} -4\a'A_{\b I}\Tt_{c\a}{}^\g A_{\g I}](z) =0
\label{nilpotencyonelooppb} \ee
Adding (\ref{nilpotencyonelooppi}) and (\ref{nilpotencyonelooppb}) gives the condition
\be \l^\a \l^\b [T_{\b\a}{}^c -2\a' A_{\b I}(\p_\a A_{c I} +\Tt_{\a c}{}^D A_{D I})
-\a'f^{IJK} A_{cI} A_{\a J} A_{\b K} +2\a' T_{\a\d}{}^c W^\d _I A_{\b
I} \label{Tcorrection} \ee $$ -2\a'A_{\b I}\Tt_{c\a}{}^\g A_{\g I}]=0  .$$
Subtracting (\ref{nilpotencyonelooppi}) and (\ref{nilpotencyonelooppb}) gives the condition
\be \l^\a \l^\b [H^c {}_{\b\a}
-2\a' A_{\b I}(D_{[\a} A_{c ]I} +\Tt_{\a c}{}^D A_{D I})-\a'f^{IJK} A_{cI}
A_{\a J} A_{\b K} +2\a' H_{\a\d}{}^c W^\d _I A_{\b I} \label{Hcorrection} \ee $$+ 2\a'A_{\b
I}\Tt_{c\a}{}^\g A_{\g I}] =0. $$

Now, suppose that we add a counter-term of the form ${K_1\over{2\pi}}\int d^2 z \p Z^M \pb
Z^N A_{NI}A_{MI}$ to the action, where $K_1$ is a constant to be determined.
This amounts to redefine the space-time metric\cite{HullTownsend}\ $G_{MN} \rightarrow G_{MN}
+ 2\a' K_1 A_{MI} A_{NI}$. The expansion of this counter-term will
contain the terms

\be S_C = {K_1\over {2\pi}} \int d^2 x [\p Y^A \pb Y^B A_{BI}
A_{AI} + \p Y^A \Pb^B A_{BI}Y^C (\p_C A_{AI}+{1\over 2}\Tt_{CA}{}^D A_{DI} )
\label{countertermxpn} \ee $$+
\p Y^A \Pb^B Y^C (\p_C A_{BI}+\Tt_{CB}{}^D A_{DI})A_{AI} + \Pi^A \pb Y^B
A_{BI}Y^C (\p_C A_{AI}+\Tt_{CA}{}^D A_{DI})+ $$ $$\Pi^A \pb Y^B Y^C (\p_C
A_{BI} + {1\over 2}\Tt_{CB}{}^D A_{DI})A_{AI} ] $$
which can be used to compute tree level diagrams contracting with $\l^\a
\dh_\a (w) \l^\b \dh_\b (z)$. However this diagrams will
contribute to the order $\a'^2$, entering at the same foot as the one-loop
diagrams. The result of these tree level diagram is

\be -\a'^2 K_1 {{\bar w-\bar z}\over{(w-z)^2}}\l^\a \l^\b \Pb^C [ A_{C
I}( D_{(\a}A_{\b) I} +\Tt_{\a\b}{}^D A_{DI}) -2 A_{\b I} (D_\a A_{C I}
+ \Tt_{\a C}{}^E A_{EI})](z) \label{fstcc} \ee
$$ \a'^2 K_1 {{\l^\a \l^\b
}\over{w-z}} \Pi^C [A_{C I} (D_{(\a} A_{\b ) I}
+\Tt_{\a\b}{}^D A_{DI})-2A_{\b I} (D_\a A_{C I} + \Tt_{\a C}{}^D A_{DI})](z)$$ $$
+2\a'^2 K_1 {{\bar w - \bar z}\over{(w-z)^2 }} \pb \l^\a \l^\b A_{\a I}A_{\b
I}(z) + 2\a'^2{{K_1 }\over{w-z}} \p\l^\a \l^\b A_{\a I} A_{\b I}(z)$$
Then, (\ref{nilpotencyonelooppi}) and (\ref{nilpotencyonelooppb}) will be modified
respectively to 

\be \l^\a \l^\b [(T_{\b\a}{}^c
+H^c {}_{\b\a}) -4\a' A_{\b I}(D_\a A_{c I} +\Tt_{\a c}{}^D A_{D I}) +
2\a'A_{\b I} \p_c A_{\a I}\label{nilpotencyonelooppic} \ee 
$$ -2\a'f^{IJK} A_{cI} A_{\a J} A_{\b K} +2\a' (T_{\a\d}{}^c +
H^c {}_{\a\d})W^\d _I A_{\b I} + 2\a'K_1 A_{c I}(D_{(\a} A_{\b )I}
+\Tt_{\a\b}{}^D A_{DI}) $$ $$-4\a'K_1
A_{\b I} (D_\a A_{cI} + \Tt_{\a c}{}^D A_{DI})](z) =0.$$

\be \l^\a \l^\b [(T_{\b\a}{}^c - H^c {}_{\a\b}) - 2\a'
A_{\b I} \p_c A_{\a I} +2\a'(T_{\d\a}{}^c - H^c
{}_{\d\a})W^\d _I A_{\b I} \label{nilpotencyonelooppbc} \ee
$$+ 2\a'K_1 A_{c I}(D_{(\a} A_{\b )I} +\Tt_{\a\b}{}^D
A_{DI}) -4\a'K_1
A_{\b I} (D_\a A_{cI} + \Tt_{\a c}{}^D A_{DI}) -4\a'A_{\b I}\Tt_{c\a}{}^\g A_{\g I}](z) =0 $$

We can add (\ref{nilpotencyonelooppic}) with
(\ref{nilpotencyonelooppbc})
to obtain 

\be \l^\a \l^\b [T_{\b\a}{}^c -2\a'A_{\b I}(D_\a A_{c I} +\Tt_{\a
c}{}^D A_{D I}) -\a'f^{IJK} A_{cI} A_{\a J} A_{\b K} +2\a' T_{\a\d}{}^c W^\d
_I A_{\b I} \label{Tresult} \ee 
$$+ 2\a'K_1 A_{c I}(D_{(\a} A_{\b )I} +\Tt_{\a\b}{}^D A_{DI})
-4\a'K_1 A_{\b I} (D_\a A_{cI} + \Tt_{\a c}{}^D A_{DI}) -2\a'A_{\b I}\Tt_{c\a}{}^\g A_{\g I}] =0 .$$
If $K_1 = -1/2$ and using the constriaint $\l^\a \l^\b F_{\a\b I} = 0$ we
arrive at 
\be \l^\a \l^\b [T_{\b\a}{}^c + 2\a' T_{\a\d}{}^c W^\d
_I A_{\b I} -2\a'A_{\b I}\Tt_{c\a}{}^\g A_{\g I}] =0.\label{Tresultb}
\ee
Furthermore, forming a three-level diagram with $\dh_\a Y^\b \Pb^C \Tt_{C
\b}{}^\a $ and $\p Y^\a \pb Y^\b A_{\b I} A_{\a I}$ in
(\ref{countertermxpn}) ,
with precisely this value for $K_1$ we can cancel the term proportional to 
$A_{\b I }\Tt_{c \a}{}^\g A_{\g I}$ in (\ref{Tresultb}) and
(\ref{Hcorrection}) . Also,
with this value for $K_1$, the 
counter-terms in the last line of (\ref{fstcc}) will cancel the contributions
proportional to $\p \l^\a$ and $\pb \l^\a$ in (\ref{loopIIR}) .

Note that we can add a second counter-term of the form ${{K_2}\over{2\pi}}\int
d^2 z d_\a \pb Z^M A_{MI}W_I ^\a $. This amounts to redefine the
supervielbein $E_{M}{}^\a \rightarrow E_{M}{}^\a +\a' K_2 A_{MI}W_I ^\a$.
After expanding this counter-term, we can form a tree-level diagrams contracting 
it with ${1\over 4}\pb Y^\g Y^\d \Pi^c
(T_{\d\g}{}^c + H_{\d\g}{}^c)$:
\be \epsfbox{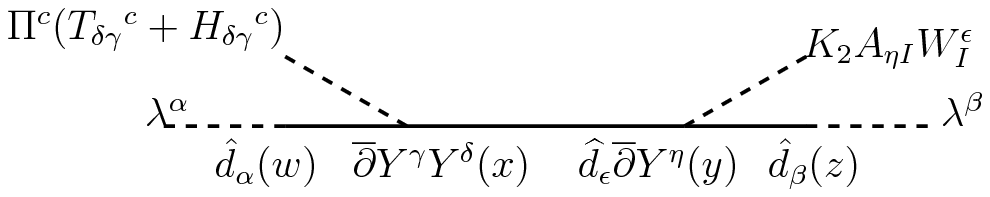} \label{picXIII} \ee
giving a contribution to the nilpotency
\be \a'^2 K_2 {{\l^\a \l^\b }\over{w-z}} \Pi^c (T_{\a\g}{}^c +
H_{\a\g}{}^c )W_I ^\g A_{\b I}(z), \label{KIIa} \ee
while contractions with ${1\over 4}\p Y^\g Y^\d \Pb^c
(T_{\d\g}{}^c - H_{\d\g}{}^c)$ will form the diagram
\be \epsfbox{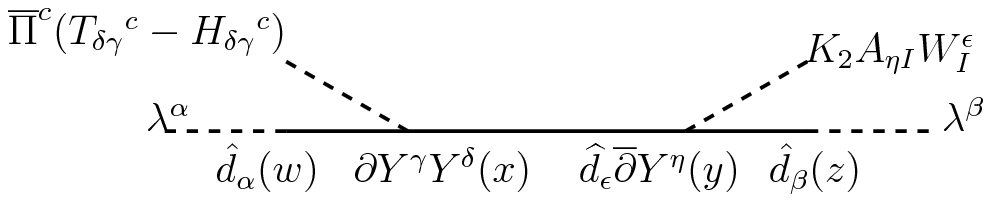} \label{picXIV} \ee
which gives the contribution
\be - \a'^2 K_2 {{\bar w - \bar z}\over{(w-z)^2}} \l^\a \l^\b \Pb^c
(T_{\a\g}{}^c - H_{\a\g}{}^c) W_I ^\g A_{\b I}. \label{KIIb} \ee
It can be easily checked that for $K_2 = -1$, adding (\ref{KIIa}) and
(\ref{KIIb}) to
(\ref{nilpotencyonelooppi}) and (\ref{nilpotencyonelooppb}) respectively; then $\l^\a \l^\b
T_{\a\b}{}^c$ will not receive $\a'$ corrections, i.e. this second counter-term 
cancels the $\a'$ correction in (\ref{Tresultb}); while the corrections for
$H_{\a\b}{}^c$ are
\be \l^\a \l^\b [H^c {}_{\b\a}
-2\a' A_{\b I}(D_{[\a} A_{c ]I} +\Tt_{\a c}{}^D A_{D I})-\a'f^{IJK} A_{cI}
A_{\a J} A_{\b K} ] =0. \label{HcorrectionIMZI} \ee

Now, the couplings to $\Pi^\g$ also receive corrections from the two
counter-terms just introduced. Some of these corrections come from the
coupling to $\Pi^C$ in (\ref{fstcc}) when $C$ is $\g$. Another correction comes
from the tree-level diagram
\be \epsfbox{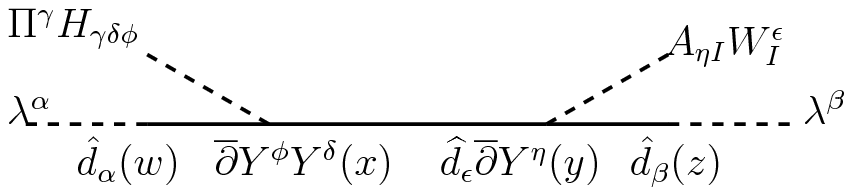} \label{picXV} \ee
Adding those corrections and using the holomorphicity constraint $F_{\a\b I}
= -{1\over 2}W^\g _I H_{\g\a\b}$, we can check that the $\a'$ corrections to
the coupling to $\Pi^\g$ are
\be \l^\a \l^\b [H_{\g\b\a}
-2\a' A_{\b I}(D_{(\a} A_{\g ) I} +\Tt_{\a \g}{}^D A_{D I})-\a'f^{IJK} A_{\g I}
A_{\a J} A_{\b K} ] =0. \label{HcorrectiongIMZI} \ee

Let's now identify the Chern-Simons form. We can use the lowest order constraints in $\a'$ coming from
nilpotency condition $\l^\a \l^\b F_{\a\b I} =0$ to write
(\ref{HcorrectionIMZI}) in the desired form. Since $\l^\a \l^\b = \l^\b \l^\a $

\be \l^\a \l^\b [H^c {}_{\a\b} -\a'Tr
A_{[\a}(D_{\b}A_{c]}+{1\over 2}\Tt_{\b c ]}{}^D A_{D]})
-2\a'f^{IJK}A_{cI} A_{\a J} A_{\b K} ](z) =0 \label{HcorrectionIMZII}
\ee
Since $2f^{IJK}A_{cI} A_{\a J} A_{\b K} = {2\over 3}Tr A_{[c}A_\a A_{\b]}$ then 

\be \l^\a \l^\b [H^c {}_{\a\b} -\a'Tr(A_{[\a}D_{\b}A_{c]} + {2\over
3} A_{[c} A_\a A_{\b ]} +{1\over 2}A_{[\a}\Tt_{\b c]}{}^D A_{D})](z)
=0,\label{HcorrectionIMZII} \ee
which is the desired form.
Similarly, (\ref{HcorrectiongIMZI}) can be written as 
\be \l^\a \l^\b [H_{\a\b\g} -\a'Tr(A_{(\a}D_{\b}A_{\g)} + {2\over
3} A_{(\g} A_\a A_{\b )} +{1\over 2}A_{(\a}\Tt_{\b \g)}{}^D A_{D}
)](z) =0. \label{HcorrectionIMZII}  \ee

Adding a further third counter-term $-{1\over{2\pi}}\int d^2 z \l^\a
\omega_\b \pb
Z^M A_{M I}U_{I \a}{}^\b $, which amounts to redefine $\O_{M\a}{}^\b
\rightarrow \O_{M\a}{}^\b -\a'A_{M I}U_{I\a}{}^\b$;
and thanks also to the other two counter-terms added, can verify that neither 
$\l^\a \l^\b T_{\a\b}{}^\g = 0$ nor $\l^\a \l^\b F_{
\a\b I} =0$ will receive $\a'$ corrections.

There are some similarities between the terms including the gauge
connection and the spin connection in the heterotic sigma model
action. This suggest that a similar computation would help to find 
similar Chern-Simons corrections for the gravity side, which will be
presented in the next chapter.

\chapter{Lorentz Chern-Simons Corrections}
In this chapter we consider the Lorentz Chern-Simons type of corrections 
to the field strength $H$. To achieve this purpose we consider in the first
section the background field expansion of the terms in the action 
(\ref{actioncurved}) that includes the spin connection $\O_{M\a}{}^\b$ and compute
their $\a'$ corrections to the nilpotency of the BRST charge. 

\section{One-loop Correction to the Nilpotency Constraints from Pure Spinors
Lorentz Currents}
Because the pure spinor condition, (\ref{actioncurved}) is
invariant under $\d \omega_\a = (\L_b \g^b \l)_\a$. Then
\be \O_{M\a}{}^\b = \O_M ^{(s)} \d_\a {}^\b + {1\over 4}\O_{Mab}(\g^{ab})_\a
{}^\b ,\label{Omegasplitting} \ee
so the terms including the spin connection can be written
\be \l^\a \omega_\b \Pb^C \O_{C\a}{}^\b = J\Pb^C \O_C ^{(s)} + {1\over 2}N^{ab}\Pb^C
\O_{Cab}, \label{lwpO} \ee
where $J =  \l^\a \omega_\a $ and $N^{ab} = {1\over{2}}(\l \g^{ab}
\omega).$ Because of the splitting (\ref{wsfsplitting}) $J$ and $N^{ab}$ also splits
as 
\be J = J_0 + J_1 +J_2 , \,\,\, N^{ab} = N^{ab}_0 + N^{ab}_1 +
N^{ab}_2 , \label{JNsplitting} \ee 
where the subindex 1 or 2 stands for one or two quantum fields
respectively in each definition. Let us now consider the terms in the
background field expansion of (\ref{lwpO}) that will allow to form
loops. Before that, note that $\l^\a \omega_\b \pb Z^M \O_{M\a}{}^\b$ is analog to
$\p Z^M \Jb^I A_{MI}$, then some diagrams of the Yang-Mills Chern-Simons
corrections will have a Lorentz analog.

In the expansion for $J\Pb^A \O_A ^{(s)}$ there are terms $J_2 \Pb^A Y^\b
(D_{\b}\O_{A} ^{(s)}+\Tt_{\b A}{}^D \O_D ^{(s)})$ and $\pb Y^\a J_2 \O_\a ^{(s)}$ which can be used to form a one-loop
diagram like (\ref{picI}) contributing to the nilpotency of the BRST charge:
\be \epsfbox{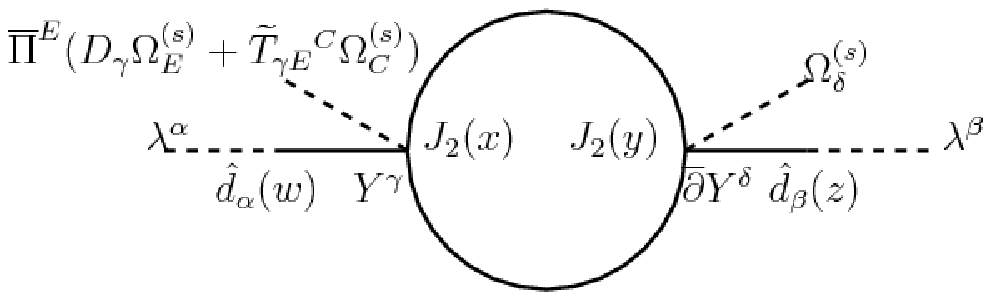}  \label{picXVI} \ee
There is a  $1/2$ coming from the expansion 
at second order or exp(-S), a
factor of $2$ because the different possibilities of putting the superfields
at $x$ or $y$, a factor of $-4$, coming from $J_2 (x) J_2 (y) =-4
(x-y)^{-2}$ and a factor of two because of the symmetries of the diagram, 
giving

\be \l^\a d_\a (w) \l^\b d_\b (z) _{XXI} = {1\over {(2\pi\a')^2}}
\int [Dwsf] \int d^2 x d^2 y \l^\a d_\a (w) \l^\b d_\b (z)
\label{loopXXI} \ee 
$$ \pb Y^\g
\O_\g ^{(s)}(x)
\Pb^E _0 Y^\d (D_{\d} \O_{E} ^{(s)} +\Tt_{\d E}{}^D \O_D ^{(s)})(y) J_2 (y) J_2 (x)$$
\be =-{{8\a'^2}\over{(2\pi)}} \l^\a \l^\b \Pb^E _0 \O_\b ^{(s)}
(D_{\a}\O_{E} ^{(s)}+ \Tt_{\a E}{}^D \O_D ^{(s)})(z) \int d^2 x d^2 y {\d^2 (x-z)
\over{(w-y)(x-y)^2}}, \label{intermezzoIII} \ee
so
\be \l^\a d_\a (w) \l^\b d_\b (z) _{XXI} = -8(\a')^2 {{\bar w -
\bar z} \over {(w-z)^2 }} \l^\a \l^\b \Pb^C \O_\a ^{(s)}(\p_{\b}\O_{C}
^{(s)}+\Tt_{\b
C}{}^D \O_D ^{(s)}) (z). \label{loopXXIR} \ee
This result is analog to (\ref{QQI}).

There is also an one-loop diagram like (\ref{picII}) formed contracting $\l^\a d_\a (w) \l^\b d_\b (z) $
with twice $\pb Y^\g \O_\g ^{(s)} J_2$:
\be \epsfbox{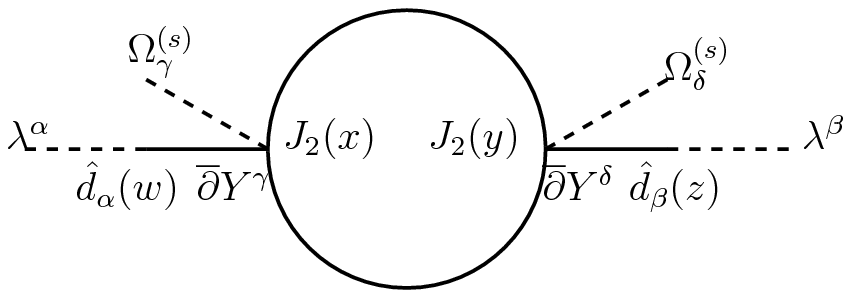} \label{picXVII} \ee
which gives a result analog to (\ref{loopIIR})
\be \l^\a d_\a (w) \l^\b d_\b (z)_{XXII} = -4\a'^2 {{\bar w -
\bar z} \over {(w-z)^2 }} \l^\a \l^\b \Pb^C \p_C \O_\a ^{(s)} \O_\b ^{(s)} (z)
-{{4\a'^2}\over{w-z}}\l^\a \l^\b \Pi^C \p_C \O_\a ^{(s)} \O_\b ^{(s)}
\label{loopXXIIR} \ee 
$$ -4\a'^2 {{\bar w -
\bar z} \over {(w-z)^2 }} \pb \l^\a \l^\b  \O_\a ^{(s)}\O_\b ^{(s)} (z)
-{{4\a'^2}\over{w-z}}\p \l^\a \l^\b \O_\a ^{(s)}\O_\b ^{(s)}$$

There is no contribution $\O_A ^{(s)}\O_B ^{(s)}\O_C ^{(s)}$

There is a similar contribution to (\ref{loopXXIR}) , coming from forming a
diagram with ${1\over 2}N_2 ^{ab}
\Pb^C Y^\d (D_{\d}\O_{C ab}+\Tt_{\d C}{}^E \O_{E ab})$ and ${1\over 2}\pb Y^C
N_2 ^{ab}  \O_{Cab}$ ,
which are in the expansion of ${1\over 2}N^{ab}_2 \Pb^C \O_{Cab}$:
\be \epsfbox{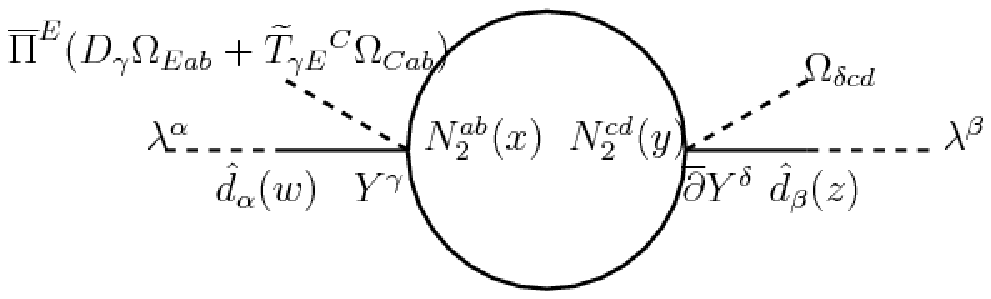} \label{picXVIII} \ee
giving as result
\be \l^\a d_\a (w) \l^\b d_\b (z) _{XXIII} = -3(\a')^2 {{\bar w -
\bar z} \over {(w-z)^2 }} \l^\a \l^\b \Pb^C \O_{\a de}
(D_{\b}\O_{C}{}^{ed}+\Tt_{\b C}{}^F \O_{F}{}^{ed})(z). \label{loopXXIIIR}
\ee
Also there is a similar contribution to (\ref{loopXXIIR})  , making a diagram
contracting \\$\l^\a \dh_\a (w) \l^\b \dh_\b (z)$ with twice ${1\over 2}\pb Y^\g N^{ab}_2 \O_{\b
ab}$:
\be \epsfbox{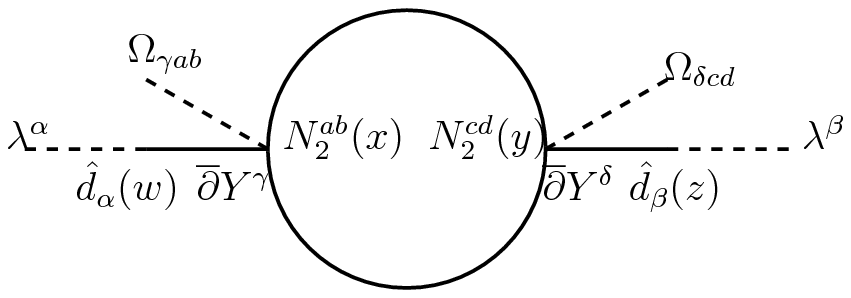} \label{picXIX} \ee
giving as result
\be  \l^\a d_\a (w) \l^\b d_\b (z)_{XXIV} = -{3\over 2}\a'^2 {{\bar w -
\bar z} \over {(w-z)^2 }} \l^\a \l^\b \Pb^C \p_C  \O_{\a ab} \O_\b {}^{ba}(z) 
\label{loopXXIVR} \ee $$ -{3\over 2}{{\a'^2}\over{w-z}}\l^\a \l^\b \Pi^C \p_C \O_{\a ab}\O_\b {}^{ba}
  -{3\over 2}\a'^2 {{\bar w -\bar z} \over {(w-z)^2 }} \pb \l^\a \l^\b  \O_{\a ab} \O_\b {}^{ba} (z)
-{3\over 2}{{\a'^2}\over{w-z}}\p \l^\a \l^\b \O_{\a ab} \O_\b {}^{ba}. $$

There is a diagram like (\ref{picIII}) which gives a cubic contribution in $\O$,
coming from contracting ${1\over 2}N_2 ^{ab} \Pb^C \O_{C ab}$ and
a product of two ${1\over 2}\pb Y^\a N^{bc}_2 \O_{\a bc}$:
\be \epsfbox{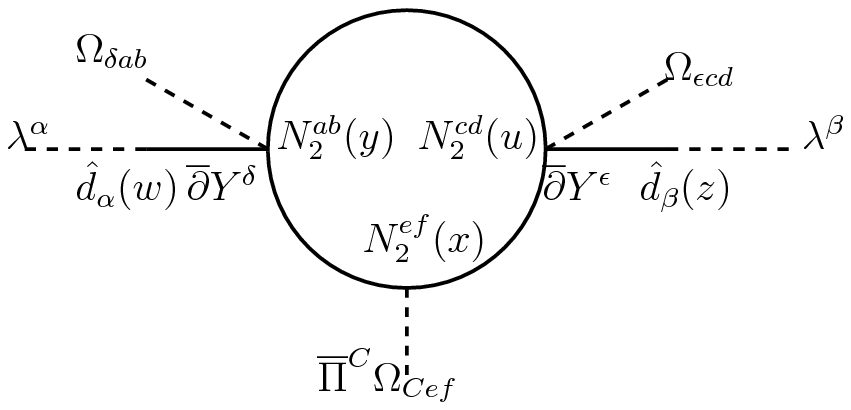} \label{picXX} \ee
To determine the
coefficient of this diagram, note that there is a ${1\over 3!}$ coming from
the expansion of $exp(-S)$ at third order, a factor of $3$ because the
different ways to put the superfields at $x$, $y$ or $u$, an ${1\over 8}$
coming from the one halves in each of the three terms and a factor of $2$
because of the possible ways of contracting, giving

\be \l^\a d_\a (w) \l^\b d_\b (z) _{XXV} = -{1\over{8(2\pi)^3 \a'}}
\l^\a \l^\b \Pb^C \O_{C ab} \O_{\a cd
}\O_{\b ef}(z) \times \label{loopXXV} \ee 
$$\int d^2 x d^2 y d^2 u (2\pi)^2 \d^2 (x-w) \d^2 (y-z)N_2 ^{ab}
(x) N^{cd} _2 (y) N^{ef} _2 (u),$$
It is not hard to compute
\be N_2 ^{ab}(x)N_2 ^{cd}(y) N_2 ^{ef}(u) = {-3(\a')^3(\eta^{bc} \eta^{a[f} \eta^{e]d} - \eta^{ac} \eta^{b[f}
\eta^{e]d} - \eta^{bd} \eta^{a[f} \eta^{e]c} + \eta^{ad} \eta^{b[f} \eta^{e]c}) \over
{(x-y)(y-u)(x-u)}} + {\ldots} , \label{NNN} \ee
where by ${\ldots} $ is meant less singular terms which are not of
importance in this computation. Then
\be \l^\a d_\a (w) \l^\b d_\b (z) _{XXV} = {{3\a'^2 }\over{2\pi}}
\l^\a \l^\b \Pi^C _0 \O_{C ab} \O_{\a}{}^{be}
\O_{\b e}{}^a(z) \times \int d^2 u {1\over{(w-z)(z-u)(w-u)}}.
\label{intermezzoloopXXV} \ee
The type of the last integral was already solved and has the form of
(\ref{integralx}) , then we arrive to an answer analog to
(\ref{loopIIIR})
\be \l^\a d_\a (w) \l^\b d_\b (z)_{XXV} = -3 (\a')^2 {\bar w -\bar z
\over (w-z)^2}\l^\a \l^\b \Pb^C \O_{Cde}\O_{\a}{}^{ef}\O_{\b f}{}^d
(z) \label{loopXXVR} \ee
With the computations of the present section, the $\a'$ corrections to the
nilpotency condition for the couplings to $\Pi^c$ and $\Pb^c$ are respectively

\be \l^\a \l^\b [(T_{\b\a}{}^c + H_{\b\a}{}^c) + 8\a' \O_\b ^{(s)} \p_c \O_\a
^{(s)} +3 \a' \O_{\b}{}^{ba}\p_c \O_{\a ab} ] = 0 \label{LCTplusH} \ee

\be \l^\a \l^\b [(T_{\b\a}{}^c - H_{\b\a}{}^c) +16\a' \O_\b ^{(s)}(D_\a
\O_c  ^{(s)}+ \Tt_{\a c}{}^D \O_D ^{(s)})-8 \a' \O_\b ^{(s)}\p_c \O_\a
^{(s)} \label{LCTminusH} \ee $$ +6\a'\O_{\b ab}(D_\a \O_c
{}^{ba} + \Tt_{\a c}{}^D \O_{D}{}^{ba})  - 3 \a'\O_{\b}{}^{ba}\p_c \O_{\a
ab} +6\a'\O_{c ab}\O_{\a}{}^{bd}\O_{\b d}{}^a]  =0,$$
so adding and subtracting (\ref{LCTplusH}) and (\ref{LCTminusH}) we obtain respectively

\be \l^\a \l^\b [T_{\b\a}{}^c + 8\a'\O_\b ^{(s)} (D_{\a} \O_{c} ^{(s)}  +
\Tt_{\a c}{}^E \O_E ^{(s)}) + 3\a'\O_{\b ab}(D_{\a} \O_{c}{}^{ba}+\Tt_{\a c}{}^E
\O_E{}^{ba}) \label{LCT} \ee $$ + 3\a' \O_{cab}\O_{\a}{}^{bd}\O_{\b d}{}^a ]
=0 $$

\be \l^\a \l^\b [H_{\b\a}{}^c - 8\a'\O_\b ^{(s)} (D_{[\a} \O_{c]}^{(s)}  +
\Tt_{\a c}{}^E \O_E ^{(s)}) - 3\a'\O_{\b ab}(D_{[\a} \O_{c]}{}^{ba}+\Tt_{\a c}{}^E
\O_E{}^{ba}) \label{LCH} \ee $$ - 3\a' \O_{cab}\O_{\a}{}^{bd}\O_{\b d}{}^a ]
=0 $$

Using the lowest order in $\a'$ constraint $\l^\a \l^\b
R_{\b\a cd} =0$, we can write (\ref{LCH}) as
\be \l^\a \l^\b [H_{\b\a}{}^c - 8\a'\O_\b ^{(s)}(D_{[\a}
\O_{c]}^{(s)}   +
\Tt_{\a c}{}^E \O_E ^{(s)}) - {3\over 2}\a' ( \O_{\b a}{}^b(D_{[\a}
\O_{c]b}{}^a +\Tt_{\a c}{}^E
\O_{E b}{}^a) \label{LCHintermezzoI} \ee 
$$ + \O_{\a a}{}^b(D_{[\b} \O_{c ]b}{}^a +\Tt_{\b c}{}^E
\O_{E b}{}^a) +\O_{c a}{}^b(D_{(\a } \O_{\b )b}{}^a +\Tt_{\a\b }{}^E
\O_{E b}{}^a )  + 4 \O_{c a}{}^b \O_{\a b}{}^d\O_{\b d}{}^a ) ] =0.$$
To use the same notation as in the gauge case, let's use the same
representation as in that case. Let's write $\O_{Ab}{}^c = \O_{AI} (T^I )_b
{}^c $, where $(T^I T^J - T^J T^I)_b {}^c = f^{IJ}{}_K (T^K )_b {}^c $ and
$(T^I )_b {}^c (T^J)_c{}^b = 2\d^{IJ}$. Using this notation
(\ref{LCHintermezzoI})
can be written as 
\be \l^\a \l^\b [H_{\b\a}{}^c - 4\a'\O_{[\b }^{(s)}(D_{\a}
\O_{c]}^{(s)}   +
{1\over 2}\Tt_{\a c}{}^E \O_E ^{(s)}) - 3\a' Tr( \O_{[\b }(D_{\a} \O_{c]}
+{1\over 2}\Tt_{\a c ]}{}^E
\O_E) \label{LCHintermezzoII} \ee $$+ {2\over 3} \O_{[c}\O_{\a}\O_{\b ] } )
] =0. $$

Which gives the desired form of the Lorentz Chern-Simons. 


Summarizing, the Yang-Mills and Lorentz Chern-Simons corrections are
\be \l^\a \l^\b[H_{\b\a c} -\a'Tr (A_{[\a}D_{\b}A_{c]} + {2\over
3} A_{[\a} A_\b A_{c]} +{1\over 2}A_{[\a}\Tt_{\b c]}{}^D A_{D})
\label{YMLCH} \ee
$$
- 3 \a' Tr (\O_{[\a}D_{(\b}\O_{c ]} + {2\over 3}
\O_{[\a}\O_{\b}\O_{c]}+ {1\over 2}\O_{[\a}\Tt_{\b c]}{}^D 
  \O_{D} ) - 4\a'\O_{[\b }^{(s)}(D_{\a}
\O_{c]}^{(s)}   +
{1\over 2}\Tt_{\a c}{}^E \O_E ^{(s)})] =0$$

There are further one-loop diagrams that can be formed with 
terms in the expansion containing three quantum fields. It's 
computation constitutes work in progress.
\chapter{Conclusions}

This thesis covered two applications of the non-linear sigma models, namely
the computation of equations of motion for the background fields coupled to the
bosonic and type II superstring and also the appearance of the Yang-Mills
Chern-Simons three-form for the heterotic superstring.\\

The first application was explained in detail for the bosonic string and for 
the type II superstring
using the pure spinor formalism. Both of them are conformally invariant in a
flat space, so when they are coupled to a generic background, which
has a direct correspondence with the massless states in each case, the 
conformal invariance
must be checked. The background field method, useful to
obtain a covariant expansion was discussed in detail for the bosonic string 
computations. A version adapted to
superspace \cite{deBoerSkenderis} of the background field method was
used to obtain the expansions for the type II and heterotic string. 
The result of expanding an action using this method allowed to form Feynman 
diagrams at one loop, contributing to the possible lack of
invariance of conformal symmetry at the quantum level. When all those
diagrams were computed, giving contributions to the beta functions, it was 
shown that for the bosonic sigma model, these beta
functions can be set consistently to zero. In the introduction it was also
presented a spacetime action from which the conditions for conformal
invariance can be obtained as equations of motion by a simple variation of
this action in space-time. The necessity to
use another formalism to make the computations for the superstring is
supported because neither the RNS nor the GS sigma model can not be 
covariantly quantized and at the same time include all the background fields. 
The pure spinor formalism was briefly discussed in the introduction, and the
non-linear sigma model for the heterotic and type II superstring were
discussed in more detail. It was explained how the properties of 
nilpotency of the pure spinor BRST charge and the conservation of its
corresponding current allows to find constraints on the background fields at
the lowest order in $\a'$: super $N=1$ $D=10$ Yang-Mills/supergravity for
the heterotic and $N=2$ $D=10$ for the type II superstring.
For the heterotic string, it was explained by two different methods how
to arrive to those constraints: defining canonical momenta and using Poisson
brackets, as explained in chapter 2, or by performing a tree level
computation as explained in chapter 4.\\
In the one-loop computation of the beta functions for the type II
superstring it was necessary to introduce a scale $\Lambda$ to
regulate the diagrams. By studying the conditions under which the
theory does not depend on that scale, a set of equations was computed
in chapter 3, corresponding to all the independent couplings to
products of worldsheet fields. Because of the background field expansion used,
the result of the one-loop computation has super-Poincar\'e
symmetry. With the help of some Bianchi identities and gauge
invariances of the sigma model action, some components of the torsion
where gauge fixed. Also, the scales connections $\Omega_\a$ and
$\widetilde \Omega_{\bar\alpha}$ where related to the derivatives of the
dilaton $\nabla_\a \Phi$ and $\nabla_{\bar\a} \Phi$ respectively.
It was verified for the lowest dimension equations of motion that the lowest order $\a'$ type II supergravity constraints can set the beta-functions to zero,
implying in this way in conformal invariance. This is a straightforward,
although non-trivial task, whose level of difficulty increases as one
considers equations of motion with higher dimension.\\

The second application concerns the quantum regime of the BRST symmetry for
the heterotic string sigma in the pure spinor formalism. A similar background
field expansion as the one used for the pure spinor type II sigma model was
included in the appendix, where the gauge and spin connections appear
explicitly. One-loop diagrams were formed as a result of considering the 
product of two BRST charges evaluated in different points. The result of
computing these diagrams has poles structure as the two points are
approached. This pole structure are of two types: double poles and
$(\bar w -\bar z)(w-z)^{-2}$ poles, coupling to independent worldsheet fields.
 From the set of equations obtained by imposing the vanishing of the poles
coefficients, corrections of the order $\a'$ are obtained for the
classical nilpotency conditions. Chapter 4 was focused
in the computations including the gauge fields, which allowed to find the
Yang-Mills Chern-Simons three-forms correction of $\a'$ order to the field
strength of the two-form superpotential, or Kalb-Ramond superfield: $H_{MNP}
\to H_{MNP} - \a' \omega_{MNP} $. These
corrections are known since the studies of $N=1$ super Yang-Mills coupled to
$N=1$ supergravity in $10$ dimensions \cite{ChaplineManton}, \cite{GS}, \cite{AtickDharRatra}. Other interesting redefinition of the fields
of $\a'$ order were found, such as a redefinition of the metric superfield
$G_{MN} \to G_{MN} - \a' Tr A_M A_N$, also known since the preservation of
the $N=1$ supersymmetry at the quantum level in the RNS sigma model
\cite{HullTownsend}. A field redefinition not known until now was found
$E_{M}{}^\a \rightarrow E_{M}{}^\a -\a'  A_{MI}W_I ^\a$, since this component
of the super-vielbein does not appear neither in the RNS nor GS sigma
models.\\

A perspective of the present thesis is to compute the Lorentz Chern-Simons
corrections. Partial results were presented in chapter 5, in which similar
diagrams to the gauge side were computed. Using the Lorentz connection
$\Omega_{Ma}{}^b$ a Lorentz Chern-Simons three-form can be identified, although
the role of a Lorentz Chern-Simons three form formed with the scale
connection $\Omega^{(s)}$ is not yet understood. The results presented
in chapter 5 concerned only computations involving the pure spinors
Lorentz currents $N^{ab} = {1\over 2}\l \g^{ab}\omega$ and pure
spinor ghost current $J = \l^\a d_\a$.
Nevertheless, because of the holomorphicity
and nilpotency constraints at the lowest order in $\a'$ and using some
symmetries of the action, some components of the torsion can be gauged fixed 
to zero, allowing to write $\Tt_{c\a}{}^\b = \O_{c\a}{}^\b$ and $\Tt_{\g\a}{}^\b =
\O_{\g\a}{}^\b + \O_{\a\g}{}^\b$. Furthermore, the spin connection can
be written in terms of a Lorentz and scale connections. Considering these 
facts in the background field expansion for the term ${1\over
{2\pi\a'}}\int d_\a \pb Z^M E_{M}{}^\a$, written in equation (\ref{sndorderdbp}) 
and denoting by $M^{ab} = {1\over 2}d_\a (\g^{ab})^\a{}_\b Y^\b$, this
expansion can be written as follows 

\be {1\over{2\pi \a'}}\int d^2 z [- dY \Pb^d \O^{(s)}_d +
{1\over 2}M^{ab} \Pb^d \O_{dab} - dY \pb Y^\g \O^{(s)}_\g + {1\over 2} 
M^{ab}\pb Y^\g \O_{\g ab} - {1\over 2}\pb d_\a Y^\a Y^\b \O^{(s)}_\b
\label{torsionconection} \ee $$ + {1\over 8}(\pb d
\g^{ab} Y)Y^\b \O_{\b ab} -{1\over 2} dY Y^\b \Pb^d \p_d \O^{(s)}_\b + {1\over
4}M^{ab} Y^\b \Pb^d \p_d \O_{\b ab} -{1\over 2} dY \Pb^d Y^\g (\p_\g
\O^{(s)}_d +
\Tt_{\g d}{}^E \O^{(s)}_E) $$ $$ +{1\over 4}M^{ab} \Pb^d Y^\g (\p_\g \O_{d ab}+
\Tt_{\g d}{}^E \O_{E ab}) -{1\over 2} d_\a Y^\b \Pb^d Y^\g \Tt_{\g d}{}^\a
\O^{(s)}_\b - {1\over 8}d_\a Y^\b \Pb^d Y^\g \Tt_{\g d}{}^\epsilon
\O_{\b ab}(\g^{ab})_\epsilon
{}^\a $$ $$ + {{(-)^{d(b+p)}}\over 2}\dh_\a \Pb^C Y^B Y^D E_{B}{}^P E_{D}{}^N \p_N
E_{P}{}^\a,]$$ 
where $dY$ denotes the current $dY = d_\a Y^\a$ and it satisfy 
\be dY (y) dY (z) \rightarrow 16 \a'^2 (y-z)^{-2}, \ee
while the Lorentz currents $M^{ab}$ have the following OPE

\be M^{ab}(y) M^{cd}(z) \rightarrow \a'{{\eta^{c[b} M^{a]d}(z) - \eta^{d[b}
M^{a]c}(z)}\over{y-z}} +4\a'^2{{\eta^{a[d}\eta^{c]b}}\over{(y-z)^2}},
\label{MM} \ee
which is not surprising since this is the Lorentz current algebra of
Siegel approach to the Green-Schwarz superstring. It is expected that
including the one-loop diagrams contributing to the nilpotency of the
BRST charge, formed with the terms in (\ref{MM}), the $-3$ coefficient in front of the Lorentz Chern-Simons
three-form (\ref{YMLCH}) will turn into $+1$ and also the relative
coefficient between the Chern-Simons three-forms constructed with the
scale connection and with the Lorentz connection can be understood.\\
Some one-loop diagrams have been computed formed with the terms in the 
expansion (\ref{torsionconection}), from which some of them give finite 
results, while others give divergent results which have no analog in the gauge side. It
would be very interesting to understand if those diagrams giving
infinite result cancel among themselves or the Fradkin-Tseytlin term
will play a role in the cancellation of divergences. It would also be 
interesting to check if further $\a'$ field redefinitions are
necessary. One field redefinition which one could find is $G_{MN} \to
G_{MN} - K \a' Tr \O_M \O_N$, where $K$ is some number, which is an analog of the redefinition in
the gauge side. In that case the torsion component $T_{\a\b}{}^c$ did
not receive $\a'$ corrections, so it will be necessary to check if
this component of the torsion receives or not corrections. There is no
direct analog to the $\a'$ redefinition of $E_M {}^\a$ found on the
gauge side, so it will be interesting to check if the gravity side
computations suggest a field redefinition for this component of the
supervielbein. \\
Having found all the $\a'$ corrections to the classical constraints,
the next thing to do is to tray to relate them to those found in the
literature, see Gates et al.
\cite{BellucciGates} , \cite{BellucciDepireuxGates} and 
\cite{GatesKissMerrell} and Bonora et al.
\cite{AuriaFreRaciti}, \cite{RacitiTivaZanon},
\cite{BonoraPastiTonin}, \cite{BonoraBregolaLechnerPastiTonin} 
\cite{Bonoraetal} 
in which the two groups have given answers which could not be realted
among them. Recently Lechner and Tonin \cite{LechnerTonin} have proposed
a new set of $N=1 D =10$ supergravity constraints. Those authors also claim
that in this new formulation of $N=1 D=10$ supergravity the apparently
not conciliated set of constraints can be related. So, a perspective
of the work presented in this thesis will also be to relate the
constraints coming from the pure spinor computation with the set
recently proposed.

\begin{appendix}
\chapter{Appendix}
In this appendix we present the results of the background field
expansions of the terms in the pure spinor heterotic sigma model.
\section{Background Field Expansions}
From the expansion of the term ${1\over 2}\p Z^M \pb Z^N B_{NM}$
\be {1\over {2\pi \a'}}\int d^2 z [{1\over 2}\Pi^B \Pb^A Y^C H_{CAB} + {1\over 4}Y^A \p Y^B \Pb^C
H_{CBA}-{1\over 4}Y^A
\pb Y^B \Pi^C H_{CBA} \label{expansionB} \ee $$+ {1\over 4}Y^A Y^B \Pi^C \Pb^D H_{DCBA}],$$

where $H_{ABC} = (-)^{a(b+n)+(c+p)(a+b)}3 E_C ^P E_B ^N E_A ^M
\p_{[M}B_{NP]}$,
\be \p_{[M}B_{NP]} = {1\over 3 }(\p_M B_{NP} + (-)^{m(n+p)}\p_N B_{PM} +
(-)^{p(m+n)}\p_p B_{MN}) \label{defHMNP} \ee 
and $H_{DCBA} = (-)^{B(C+D)}\N_B H_{DCA} - (-)^{BC}T_{DB}{}^E H_{ECA} +
(-)^{D(B+C)}T_{CB}{}^E H_{EDA}.$

From the expantion of $ \p Z^M \Jb^I A_{M I}$
\be {1\over {2\pi\a'}}\int d^2 z [(\Jb^I_0 + \Jb^I_1 + \Jb^I_2)(\p Y^A A_{AI} + \Pi^A
 Y^B (\p_B A_{AI} + \Tt_{BA}{}^C A_{CI})+\Pi^A A_{AI}
 \label{expansionA} \ee $$  + {1\over 2} \p Y^A  Y^B
(\p_{[B}A_{A] I} + \Tt_{BA}{}^C A_{CI}) +{1\over 2}  Y^A Y^B \Pi^C
\Tt_{CB}{}^D (\p_{D}A_{A I} + \Tt_{DA}{}^E A_{E I}) $$ $$ -{{(-)^{BC}}\over
2} Y^A Y^B \Pi^C \p_B (\p_{C}A_{A I} +\Tt_{CA}{}^D
A_{DI})  $$

From the expansion of $d_\a \pb Z^M E_{M}^\a$
\be {1\over{2\pi\a'}}\int d^2 z [(d_{\a 0} + \dh_\a)(\pb Y^\a +
\Pb^B Y^C \Tt_{CB}{}^\a)],\label{expansionE} \ee
where the terms quadratic in $Y$ were written in (\ref{sndorderdbp}) .

From the expansion of $d_\a \Jb^I W_I ^\a$
\be {1\over {2\pi\a'}}\int d^2 z [(d_{\a 0} + \dh_{\a})(\Jb^I_0
+ \Jb^I _1 + \Jb^I_2 )({1\over 2} Y^B
Y^C \p_C \p_B W_I ^\a  +Y^C \p_C W_I ^\a+ W_I ^\a ).
\label{expansionW} \ee

From the expansion of $\l^\a \omega_\b \Pb^C \O_{C\a}{}^\b$
\be {1\over {2\pi\a'}}\int d^2 z [ (\lh^\a \omega_\b +
\l^\a \oh_\b + \lh^\a \oh_\b) ({1\over 2}\pb Y^D Y^C (\p_{[C} \O_{D]\a}{}^\b +
\Tt_{CD}{}^E \O_{E\a}{}^\b) +  \Pb^C \O_{C\a}{}^\b \label{expansionO}
\ee $$+{1\over 2}Y^C Y^D
\Pb^E \Tt_{ED}{}^F
(\p_{F} \O_{C \a}{}^\b + \Tt_{FC}{}^G \O_{G\a}{}^\b ) + \pb Y^C
\O_{C\a}{}^\b + \Pb^C Y^D (\p_D \O_{C\a}{}^\b + \Tt_{DC}{}^E \O_{E\a}{}^\b)$$
$$ -{1\over 2}(-)^{DE} Y^C Y^D \Pb^E \p_D (\p_{E}\O_{C\a}{}^\b + \Tt_{EC}{}^F
\O_{F\a}{}^\b))].$$ 

From the expansion of $\l^\a \omega_\b \Jb^I U_{I\a}{}^\b$
\be {1\over {2\pi\a'}}\int d^2 z [(\l^\a \omega_\b + \lh^\a \omega_\b +
\l^\a \oh_\b + \lh^\a \oh_\b )(\Jb^I_0 + \Jb^I_1 + \Jb^I_2)({1\over 2}
Y^C Y^D \p_D \p_C U_{I\a}{}^\b \label{expansionU} \ee $$+ Y^C \p_C
U_{I\a}{}^\b + U_{I\a}{}^\b )]. $$




\end{appendix}

%
%
%
%
%

\end{document}